\documentclass{article}

\usepackage{amssymb}
\usepackage{amsmath}
\usepackage{wasysym}
\usepackage{graphicx}
\usepackage{booktabs}
\usepackage{array}
\usepackage{adjustbox}
\usepackage[figuresleft]{rotating} 
\usepackage{enumerate}
\usepackage{dsfont}

\def\etal{et al.}
\def\micron{$\mu$m}

\def\amp{\&}
\def\aap{A\&A}
\def\araa{A\&A Ann. Rev.}

\def\apj{ApJ}

\def\apss{ApSS}

\def\mnras{MNRAS}
\newcommand{\me}{\mathrm{e}}
\newcommand{\pasp}{Publications of the ASP} 
\newcommand{\josaa}{J.\ Opt.\ Soc.\ Am.\ A} 

\begin{document}

\title{Interferometry concepts} 
%
\newcommand{\folder}{.}


%
\author{F. Millour\footnote{Laboratoire Lagrange, UMR7293, Universit\'e de Nice Sophia-Antipolis, CNRS, Observatoire de la C\^ote d'Azur, Bd. de l'Observatoire, 06304 Nice, France. email: \texttt{fmillour@oca.eu}}}

\date{EAS publication series, vol. 69, 2014}

\begin{abstract}
This paper serves as an introduction to the current book. It provides
the basic notions of long-baseline optical/infrared interferometry prior
to reading all the subsequent chapters, and is not an extended
introduction to the field.
\end{abstract}
\maketitle
\section{Introduction}

Long-baseline interferometry in the optical and infrared wavelengths
is living a ``golden age'' which indicates its maturity as an observing
technique. I chose here not to develop the history of interferometry
as it has already been extensively presented in numerous reviews
(e.g. Shao \& Colavita \cite{1992ARAandA..30..457S},
Lawson \cite{Lawson2000b}, Jankov \cite{Jankov2010}, and including in
this book: L\'ena \cite{Lena2015}).

I would also suggest reading the excellent book on optical
interferometry from A. Glindemann (\cite{2011psi..book.....G}), where
all the notions which are rapidly explained here, are detailed.

I will rather try to get into more details of new ideas and
now-commonly understood aspects which have been developed in the years
after the publication of Millour (\cite{2008NewAR..52..177M}), namely
the breakthrough of spectrally-dispersed interferometry and its
consequences, how to cope with chromatic datasets, how to make a model
of such data, and imaging techniques. I will also try to present what
makes a good interferometer.

\section[Why HAR]{Why high-angular resolution?}

The resolution power of an optical system, given its optical elements
are perfect, is only related to its size (diameter). This property was
noted by Lord Rayleigh, which gave his name to the so-called empiric
Rayleigh criterion $\theta$:

\begin{equation}
\theta = 1.22\frac{\lambda}{D}
\end{equation}

with $D$ the telescope diameter, and $\lambda$ the wavelength of
observation. This relation comes from an approximate estimate of the
radius of the first zero in the Airy function, which is involved in
the description of the diffraction pattern of a round pupil (see
later).

The consequence is that, even making abstraction of all practical
problems affecting an instrument, there is a fundamental limit in its
resolution power, directly linked to its diameter and the
wave-properties of light. If one takes the simple example of our Sun,
which has an approximate diameter of 30'', an instrument with a pupil
smaller than $\approx 145\mu$m will not be able to resolve it in the
visible (i.e. at $\lambda=555$\,nm, see
Defr\`ere \etal\ \cite{Defrere2014b}). As an illustration of this
effect, most insects, whose eyes are composed of tiny ommatidia
($\leq50\mu$m) see the Sun as a point source, whereas men, whose pupil
is $\approx 1$\,mm, can resolve it (with the use of an adequate
filter, of course). To resolve one of the biggest star in the sky,
Betelgeuse with a diameter of 44\,mas (Michelson \&
Pease, \cite{1921ApJ....53..249M}, Haubois \etal\ \cite{Haubois2009}),
the needed telescope diameter would be $\approx3.2$\,m, i.e. slightly
larger than the 100 inches (2.5\,m) of the Hooker telescope used by
Michelson \& Pease (\cite{1921ApJ....53..249M}) to resolve it (hence
the installation of a boom supporting mirrors to enlarge the available
aperture). To resolve a dwarf star similar to the Sun located at
10\,pc (i.e. a star with an angular diameter of 0.9 milli-arcsecond),
one would need to build a 150\,m diameter telescope, which is simply
unfeasible with the current techniques (see e.g. Monnet \&
Gilmozzi, \cite{2006IAUS..232..429M}).


The way to go to get finer details on stars is interferometry,
i.e. combining several telescopes into a ``virtual telescope'' the
diameter of the utmost-separated apertures.


\section{PSF and $(u,v)$ plane}

To understand what an interferometer does, one needs to understand what
a Point Spread Function (PSF) is. I recall here the introduction of
Millour (\cite{2008NewAR..52..177M}).

\subsection{Single-aperture PSF and U-V patch}

The light propagating from the astrophysical source to the observer
has come a long way. Let us represent it by the classical
electromagnetic wave:

\begin{eqnarray}
  \vec{E}(\vec{z},t) & = &
  \vec{E_0}(\vec{z}) \me^{\imath \omega t}\\
  \vec{B}(\vec{z},t) & = &
  \vec{B_0}(\vec{z}) \me^{\imath \omega t}
  \nonumber
  \label{eq:onde_electromagnetique}
\end{eqnarray}

Here, $\vec{E}$ represents the electric field, $\vec{B}$ the magnetic
field, which form a plane perpendicular to the propagation direction,
$\vec{z}$ is the position in space, $t$ is the time and $\omega$ the
light pulsation, related to the wavelength $\lambda$ and the speed of
light $c$ by $\omega = 2 \pi c / \lambda$.

\begin{figure}[htbp]
  \centering
  \includegraphics[width=0.9\textwidth,angle=-0]{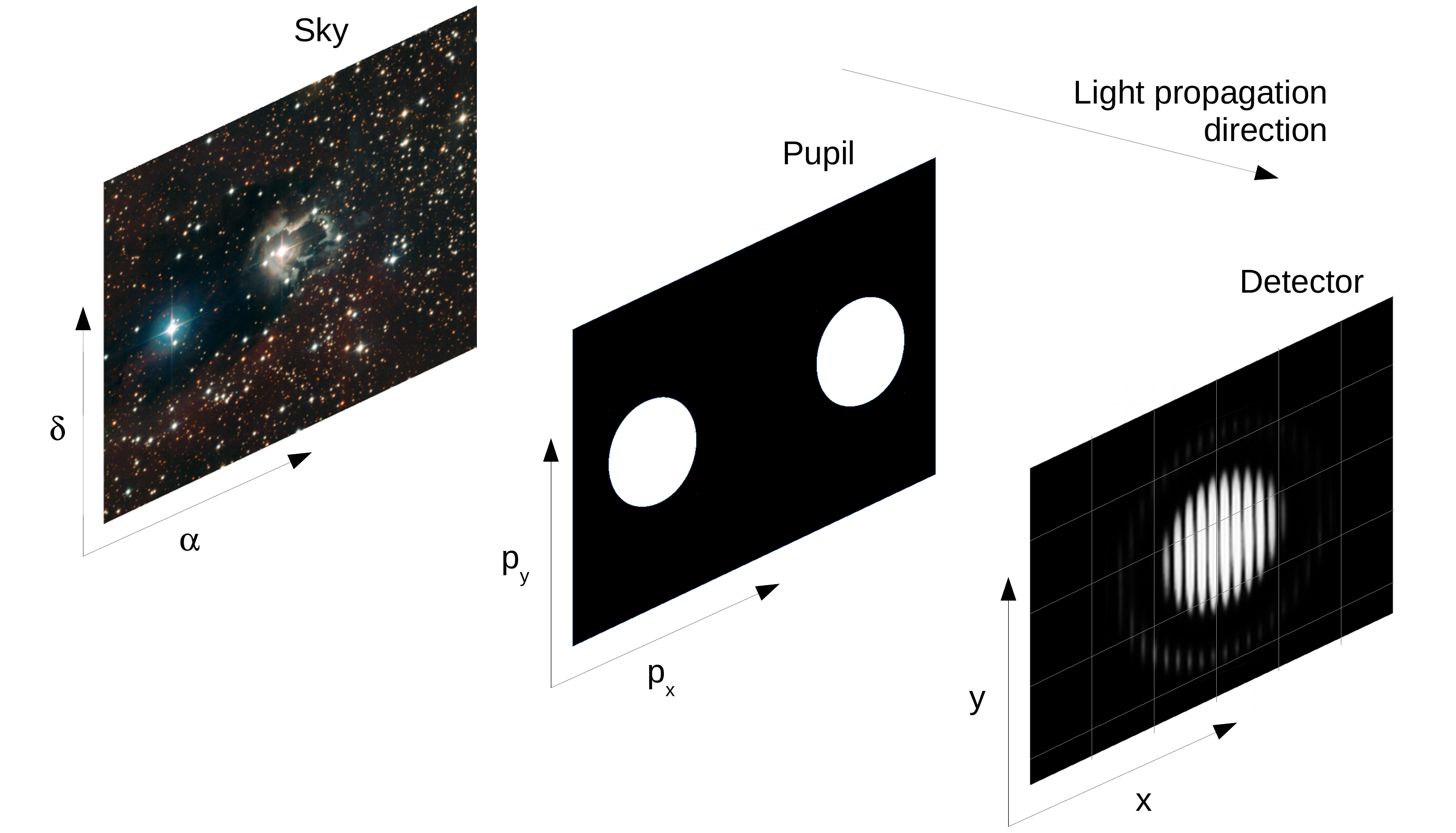}
  \caption{Notations used in this paper. The light propagates from the
    sky to the detector through the instrument pupil. Each plane of
    interest has its own coordinate system.}
  \label{fig:lightPropagation}
\end{figure}

The light intensity at the focus of the instrument (see
Fig.~\ref{fig:lightPropagation} for details) is the result of the
superposition of many electromagnetic waves coming from the pupil of
the instrument:

\begin{eqnarray}
\label{eq:intensity}
  I(\vec{x}) & = &
  \left< \left\| \vec{E}(\vec{x},t) \right\|^2 \right>_t\\
\label{eq:intensity2}
  & = &
  \left< \left\| \sum_i \vec{E}(\vec{p_i},t - \tau_i)
  \right\|^2 \right>_t
\end{eqnarray}

The $i$ index represent a number of arbitrarily chosen points in the
plane of interest. $\vec{x}$ is the 2D coordinate vector onto the
focal plane, screen or detector. For example, $\vec{p_i}$ is the
coordinate vector onto the pupil plane. $\tau_i$ represents the
propagation delay between the different incoming electromagnetic
waves.

When the pupil is split like in Fig.~\ref{fig:lightPropagation}, it is
convenient to define $\vec{B}$ the separation vector between the
sub-pupils. This vector, or its length, is often called ``baseline''.

If one considers a point-source light emitter (i.e. the wavefront at the
entrance pupil is a plane), this expression can be integrated onto the
pupil, instead of summed as in eq.~\ref{eq:intensity2} to see what the
shape of the intensity in the image plane is.

\begin{figure}[htbp]
  \centering
  \begin{tabular}{ccccc}
    \includegraphics[width=0.17\textwidth]{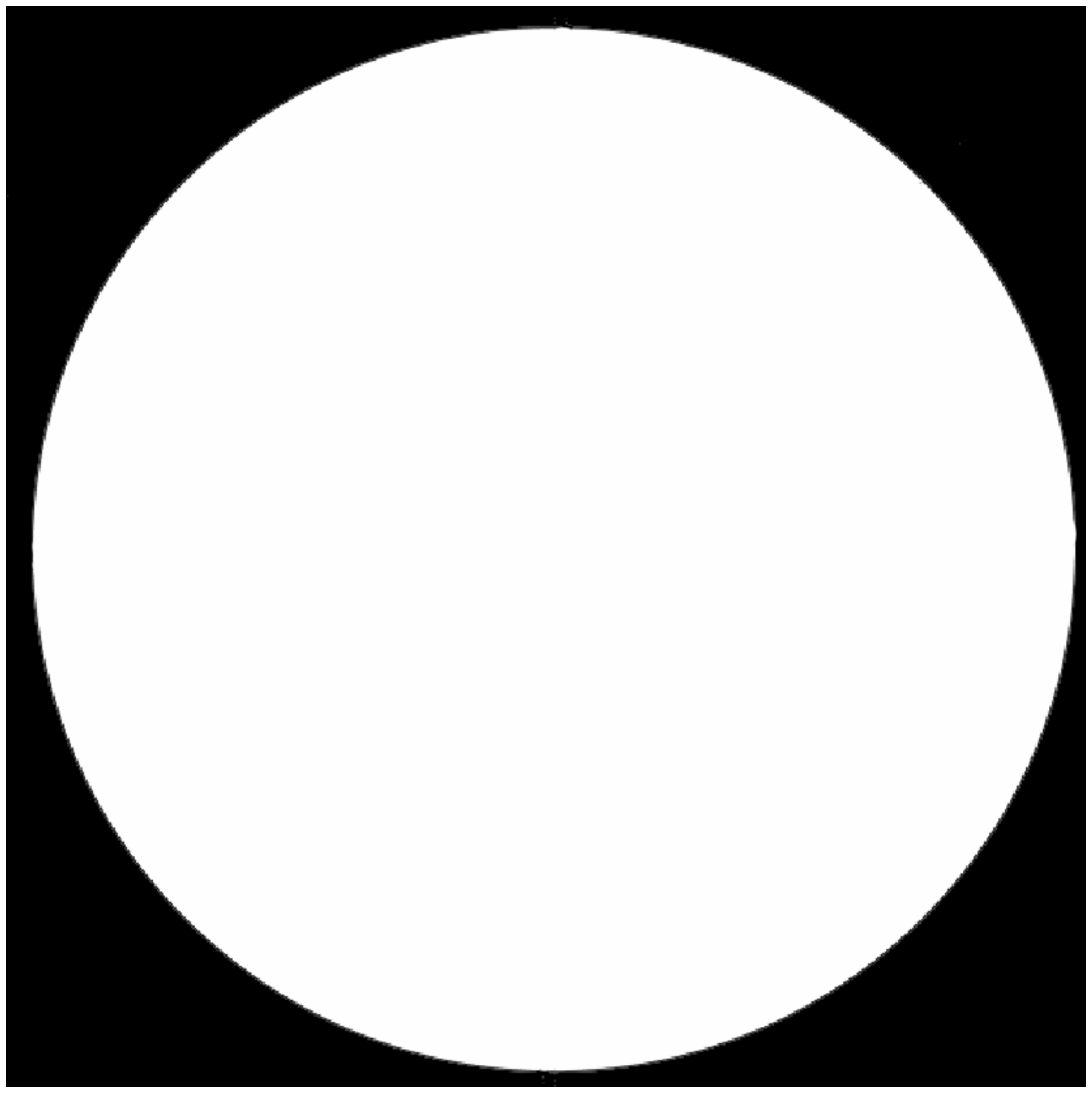} &
    \includegraphics[width=0.17\textwidth]{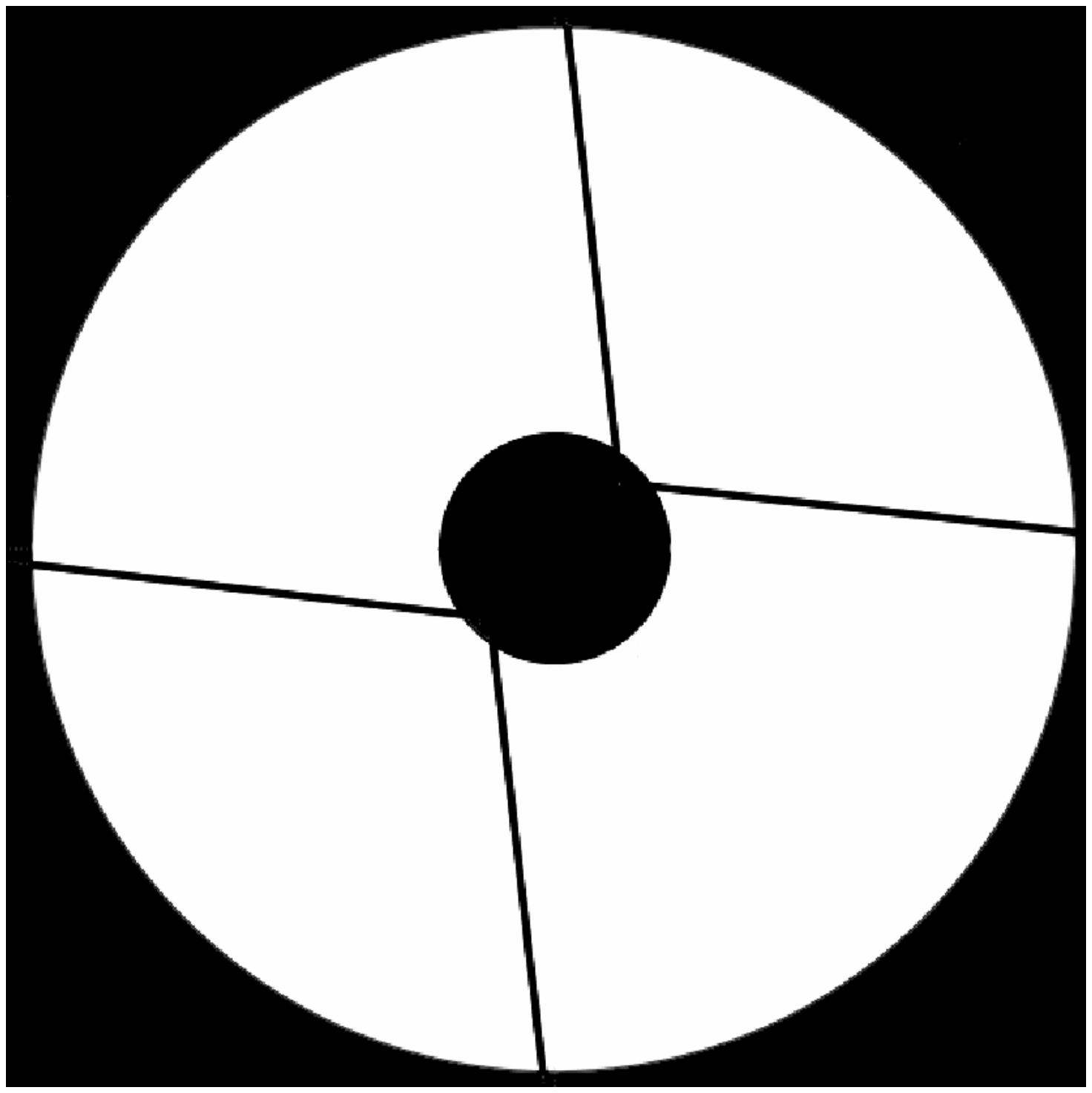} &
    \includegraphics[width=0.17\textwidth]{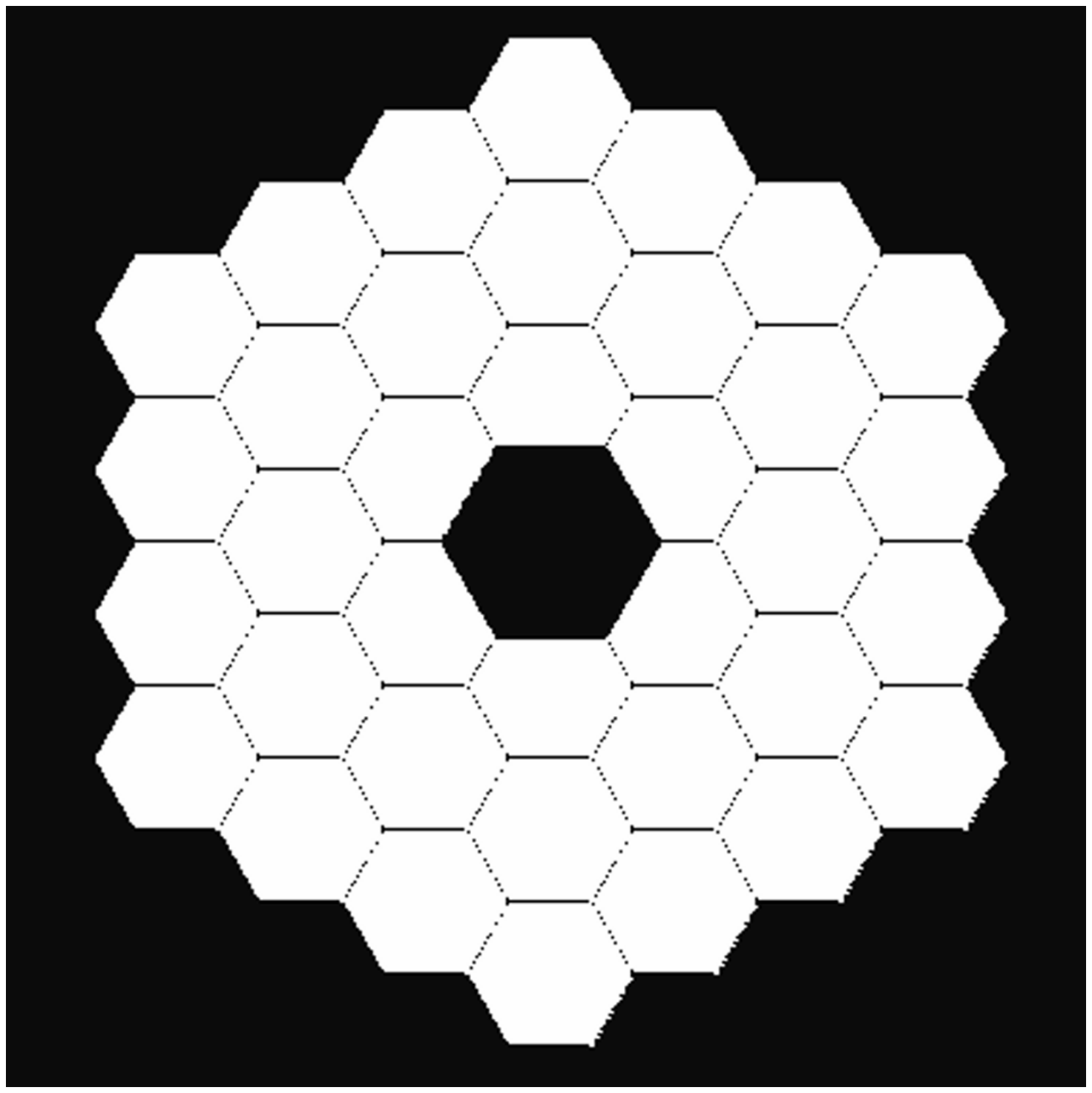} &
    \includegraphics[width=0.17\textwidth]{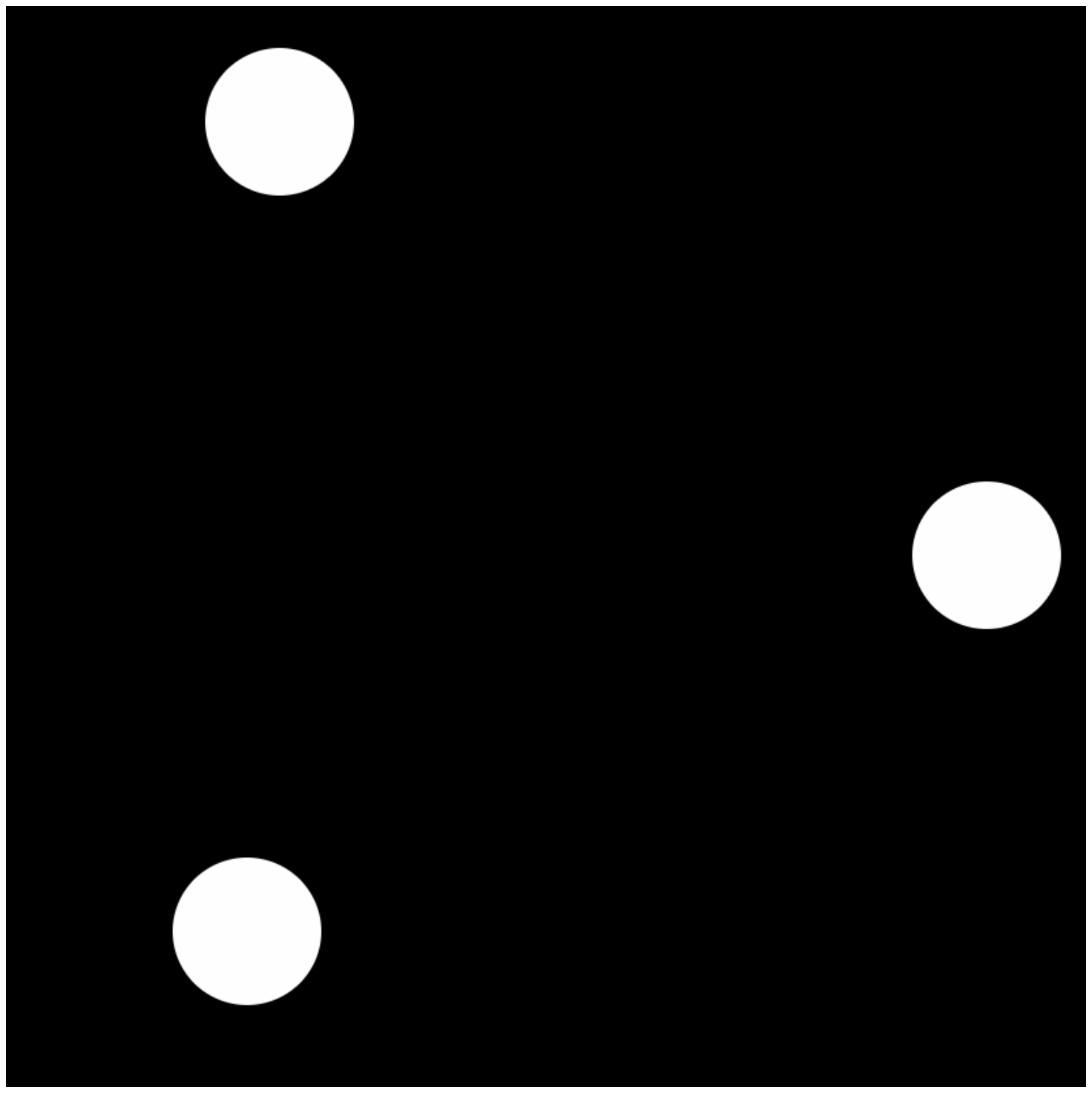} &
    \includegraphics[width=0.17\textwidth]{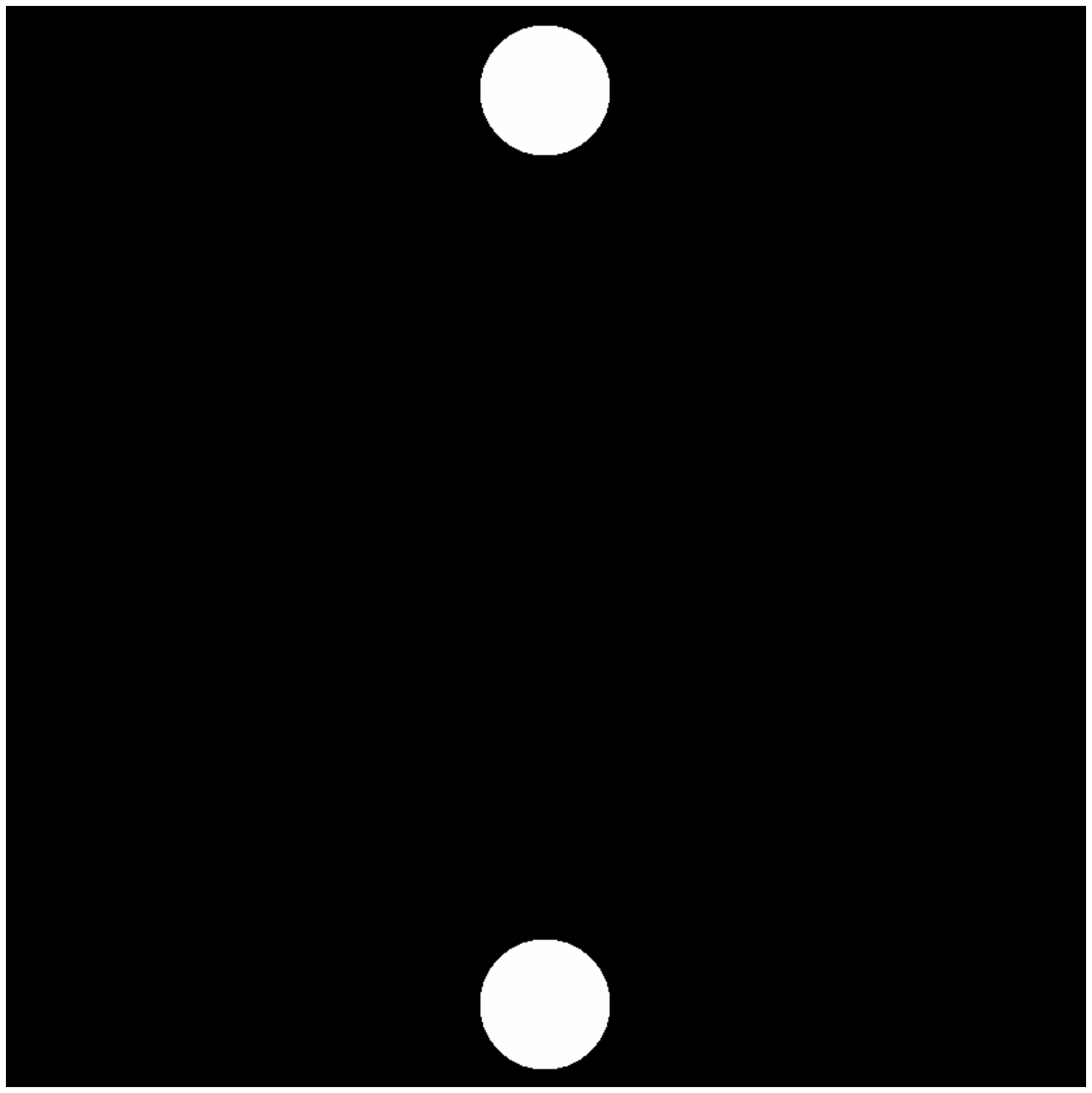} \\
    \includegraphics[width=0.17\textwidth]{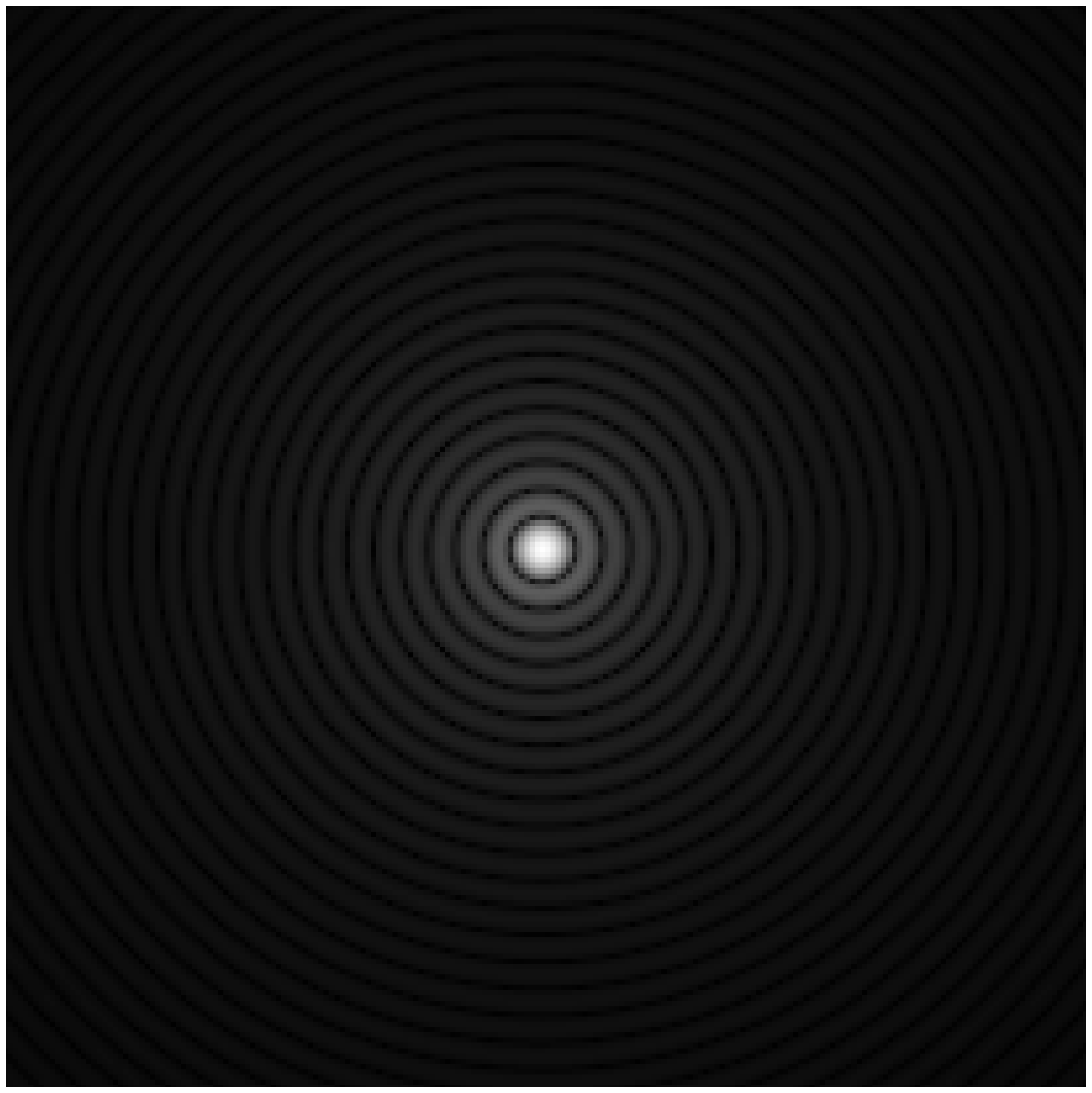} &
    \includegraphics[width=0.17\textwidth]{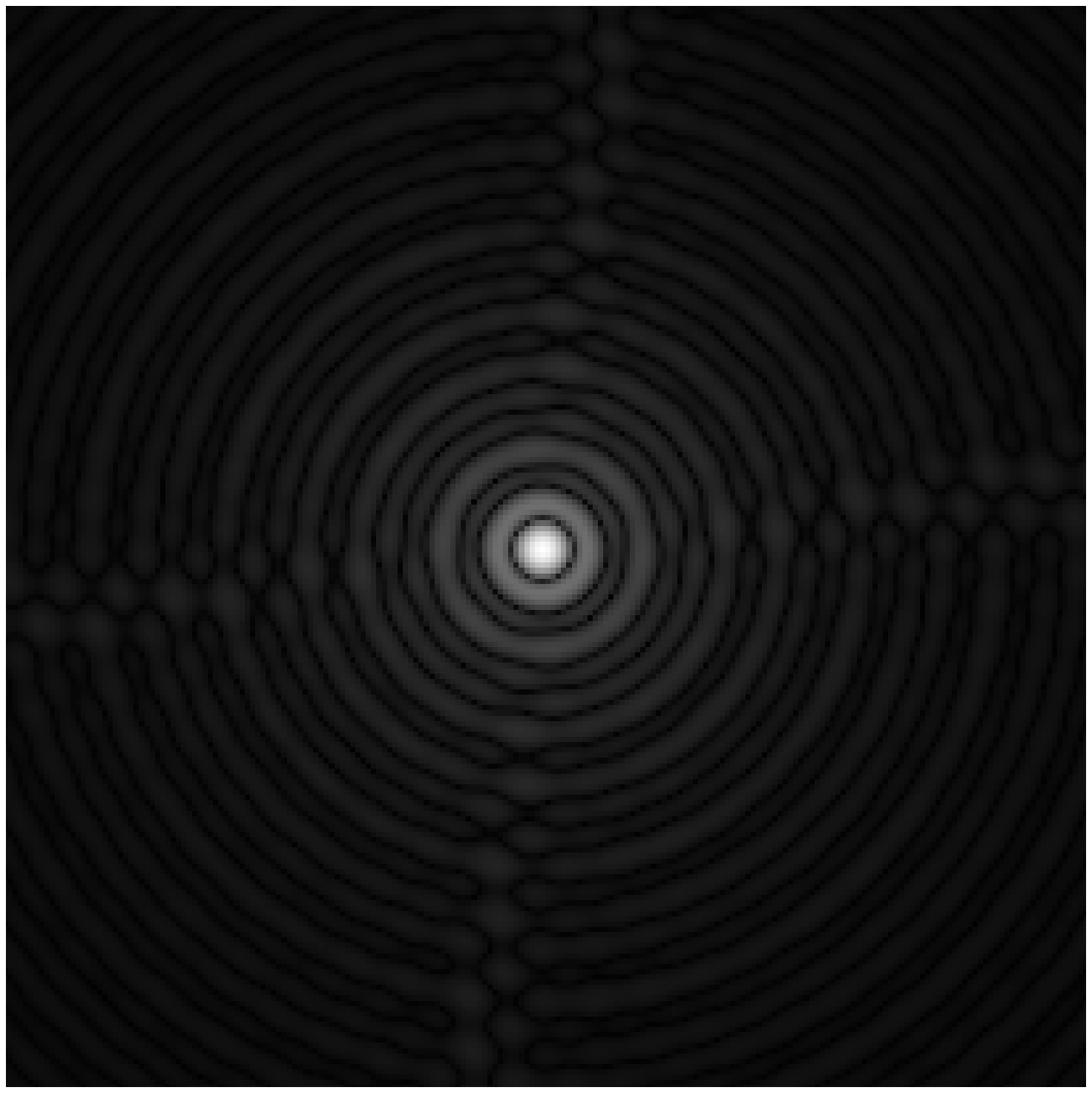} &
    \includegraphics[width=0.17\textwidth]{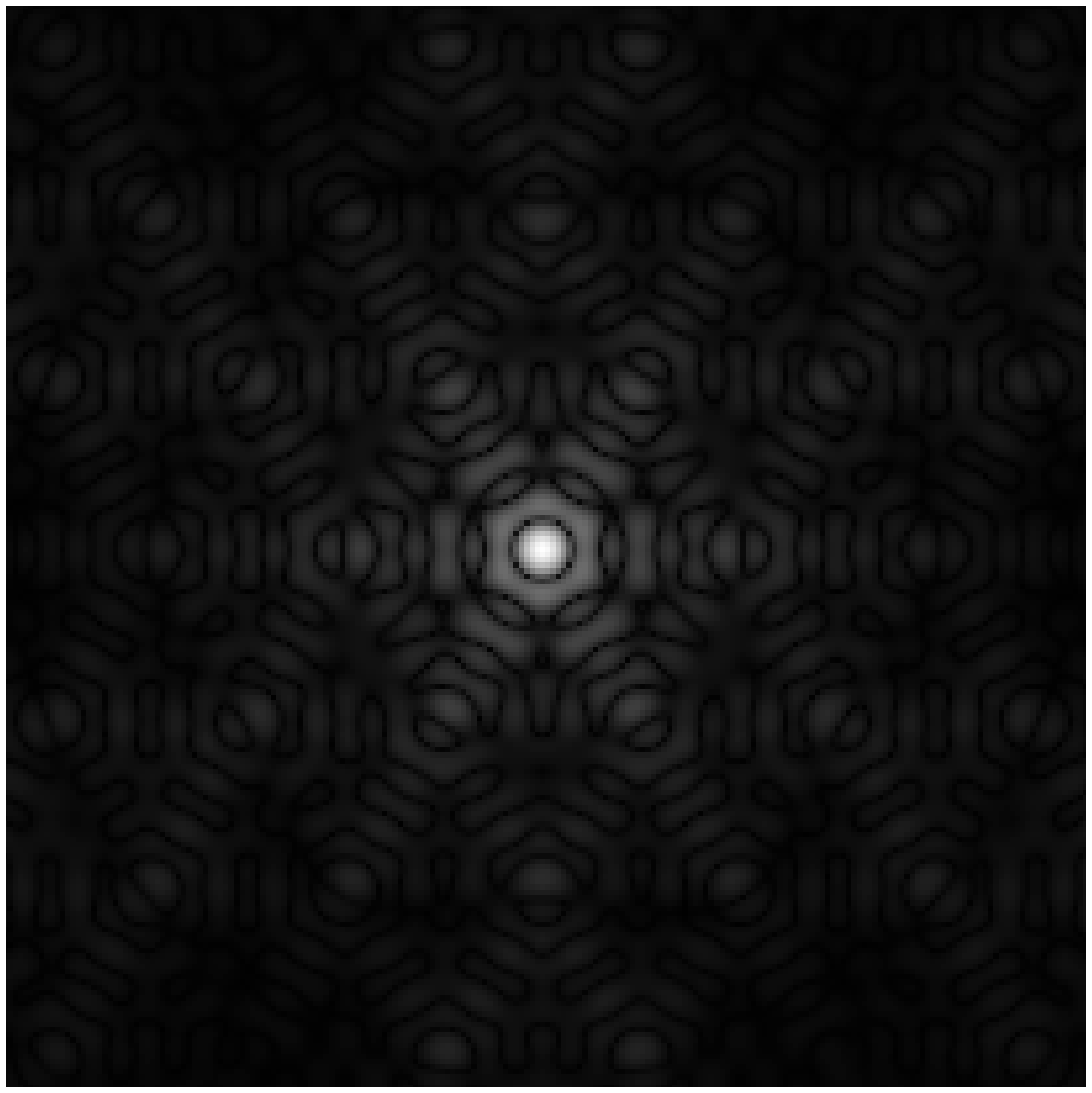} &
    \includegraphics[width=0.17\textwidth]{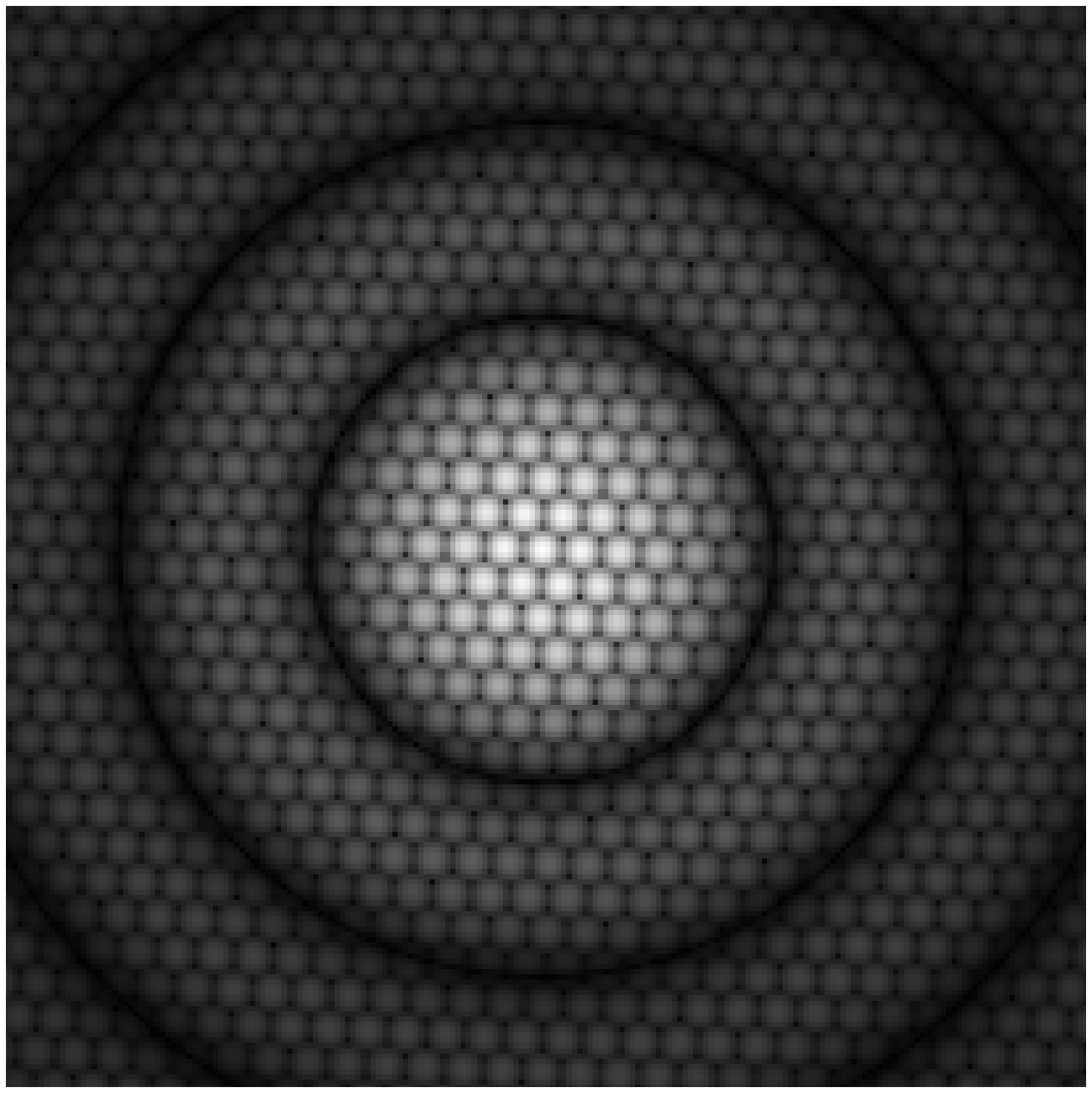} &
    \includegraphics[width=0.17\textwidth]{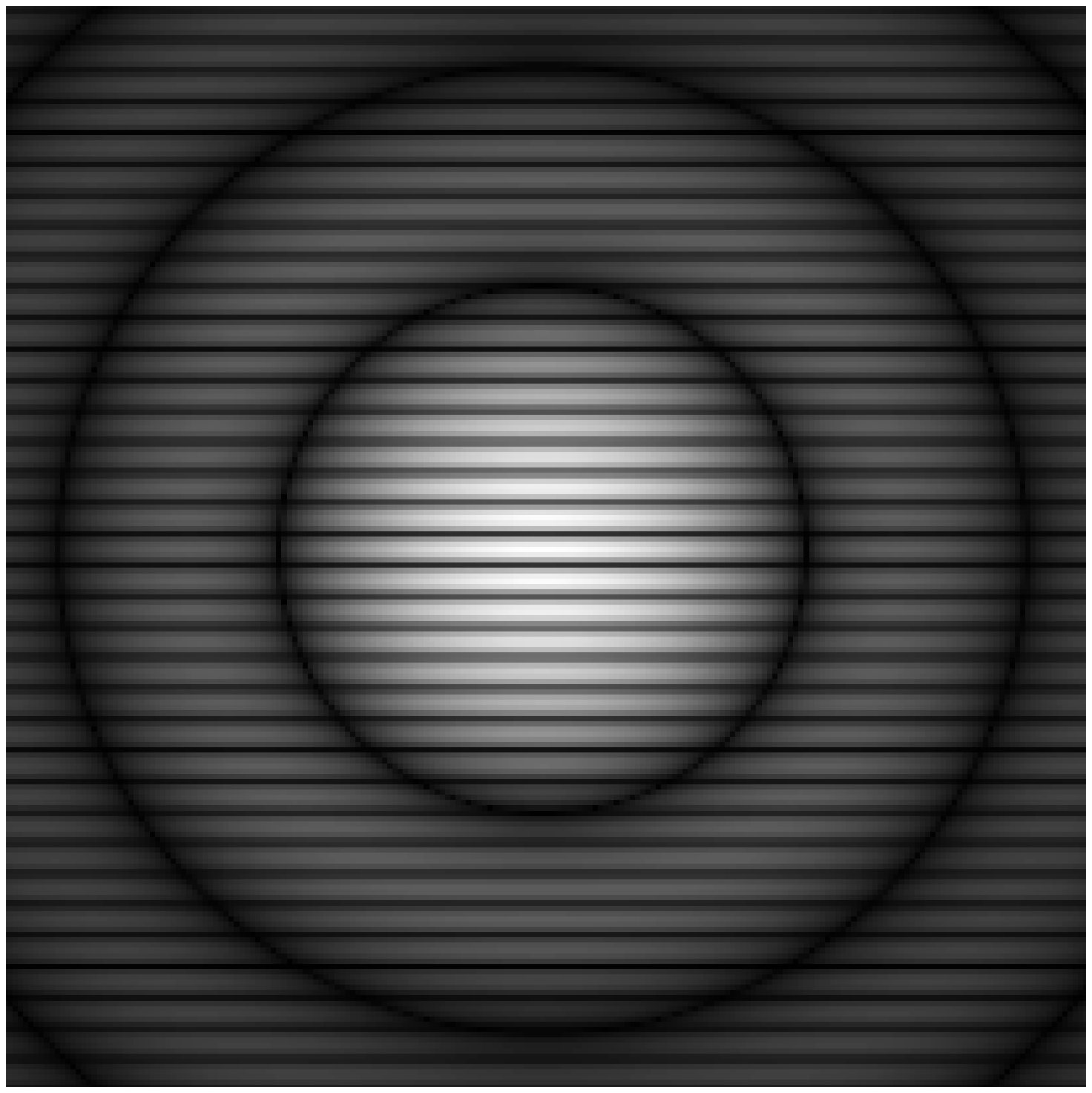} \\
  \end{tabular}
  \caption{{\bf Top, from left to right:} Simulated apertures for
    different instruments with similar angular resolution ; round
    pupil ; VLT pupil ; Keck pupil ; 3 telescopes interferometer ; 2
    telescopes interferometer. {\bf Bottom, from left to right:} The
    corresponding PSF }
  \label{fig:pupilPSF}
\end{figure}

For example, in the case of a round pupil of diameter $D$, the light
intensity will follow an Airy pattern (see the demonstration
  in Perez \cite{Perez1988} page 288), which writes:

\begin{eqnarray}
I(\rho) & = & \left(\frac{\pi D^2}{4}\right)^2 \left[\frac{2
    J_1(\pi \rho D)}{\pi \rho D}\right]^2
\label{eq:airy}
\end{eqnarray}

with $\rho = \|\vec{x}\|$ and $J_1$ the 1$^{\rm st}$ order Bessel
function. An illustration of different pupils and the associated PSF
is shown in Figure~\ref{fig:pupilPSF}. The consequence is that a
point-source does not appear as a point source through a telescope or
instrument, owing to the Rayleigh criterion (the factor 1.22 comes
from the first zero of the Bessel function). An instrument is
therefore limited in angular resolution by the diameter (or maximum
baseline) of its aperture.

\subsection{Diluted or masked-aperture PSF and U-V plane}

An interferometer, or a pupil-masking instrument, is a set of multiple
telescopes (or apertures $D$) which are combined together to form
interference patterns on a given common source, represented by its
sky-brightness distribution $S_\lambda(\alpha,\delta)$. For
convenience, one will always considers that the pupils are infinitely
small, and they are identified by their indices $i$ or $j$.

The interferometer is sensitive to the incoming light coherence,
measured by the mutual coherence function. This function is defined as
the correlation between two incident wavefronts $\vec{E}$ coming from
positions $\vec{p_i}$ and $\vec{p_j}$. The two beams of light from
each positions have a delay $\tau$:

\begin{equation}
  \Gamma_{i,j}(\tau) =  \left<
  \vec{E}(\vec{p_i},t - \tau)
  \vec{E}^{*}(\vec{p_j},t) \right>_t
  \label{eq:coherence_mutuelle}
\end{equation}

The light intensity can then be developed as a function of $\Gamma$
from eq~\ref{eq:intensity}:

\begin{equation}
  I(\vec{x}) = \sum_i \Gamma_{i,i}(0) + 
  \sum_{i,j} \Gamma_{i,j}(\tau_j - \tau_i) + {\Gamma_{i,j}}^{*}(\tau_j - \tau_i)
  \label{eq:intensite}
\end{equation}

One can note here that the terms $\Gamma_{i,i}(0)$ are just the light
intensity $I_i(\vec{x})$, as if there was only one unperturbed source
of light. The delays $\tau_i$ are set as a function of the origin of
the two wavefronts, and of the configuration of the instrument (used
optics, focal length, etc.) and, in the focal plane of the instrument,
both depend only on the coordinates in that plane. Let us pose $\tau_j
- \tau_i = \tau$. When dealing with 2 wavefronts, just like in an
interferometer, the equation \ref{eq:intensite} simplifies in:

\begin{eqnarray}
  I(\vec{x}) & = & I_1(\vec{x}) + I_2(\vec{x}) +
  \Gamma_{1,2}(\tau) + {\Gamma_{1,2}}^{*}(\tau)
  \label{eq:intensite2tel}
\end{eqnarray}

If one normalises the term $\Gamma_{i,j}$ by the total flux, this
defines the complex coherence degree $\gamma_{1,2}(\tau)$:

\begin{equation}
  \gamma_{1,2}(\tau) =
  \frac{\Gamma_{1,2}(\tau)}{\Gamma_{1,1}(0) + \Gamma_{2,2}(0)}
\end{equation}

When considering a 1D-interferogram with abscissa $x$ (for example
when one axis is anamorphosed in order to feed it into a
spectrograph), equation \ref{eq:intensite2tel} becomes:

\begin{eqnarray}
  I(x) & = & \left[ I_1(x) + I_2(x) \right] \left[ 1 + \Re \left(
    \gamma_{1,2}(\tau) \right) \right]\\ & = & \left[ I_1(x) + I_2(x)
    \right] \left[ 1 + \mu^{\rm obj}_{1,2} \cos \left( \frac{2 \pi x}{\lambda} + \phi^{\rm obj}_{1,2}
    \right) \right]
  \label{eq:interferogramme}
\end{eqnarray}

$\mu$ being the modulus of $\gamma_{1,2}(0)$ and $\phi$ its phase
($\gamma_{1,2}(0) = \mu \me^{\i \phi}$). The cosine modulation
corresponds to the intensity fringes that an optical interferometer
measures. Eq.~\ref{eq:interferogramme} and its variants is often
referred as \emph{``the interferometric equation''}, and describes the
intensity interference pattern (or interferogram) as seen on a screen
or detector. $\mu$ and $\phi$ are often called the
\emph{visibility} or \emph{contrast}, and \emph{phase} of the
interferogram, respectively. An illustration of this equation can be
seen in Figure~\ref{fig:fringePattern}, left. The $x$ variable is a length
corresponding to the delay difference between the two recombined
beams. It can be directly projected on the detector, as is done in a
\emph{multiaxial} instrument, or a time-modulated variation of $x =
v_{\rm mod} \times t$ can be introduced as is often done in a
\emph{coaxial} instrument (see e.g. Berger \etal\ \cite{1999ASPC..194..264B}, or the paper in this book: Berger \cite{Berger2015} for more
  details).

This equation, with minor modifications (due to the flux envelope of
the slits) also drives the well-known Young's two-slit experiment.

\section{Light source and light coherence}

With the Young's experiment, a simple test to do is to change the
physical size of the source by e.g. putting a varying-size diaphragm
in front of it. When the source's size changes, one can observe that
the fringe contrast also changes, and there are specific sizes at
which the fringes completely wash out. We saw in the previous sections
the intensity function of an interferometer in the case of a point
source. Here I will detail a little what happens when the source is
\emph{resolved} by the instrument or interferometer.

This is where the Zernicke and van Cittert (ZVC) theorem comes into
light, linking the value of $\gamma_{1,2}(0)= \mu^{\rm obj}_{1,2} \me^{\i
  \phi^{\rm obj}_{1,2}}$ to the object's shape projected onto the plane of
sky:

\emph{For a non-coherent and almost monochromatic extended source, the
  complex visibility is the normalised Fourier transform (hereafter
  FT) of the brightness distribution of the source.}

Or written in a mathematical way:

\begin{eqnarray}
  \gamma_{1,2}(0) & = & \frac{\iint_{-\infty}^{\infty} S(\alpha,\delta) \me^{-2i\pi (u\alpha +
      v\delta)} \,d \alpha\, d \delta}{\iint_{-\infty}^{\infty} S(\alpha,\delta) \,d \alpha\, d \delta} \\
  & = & \frac{FT(S)}{S^{\rm tot}}
  \label{eq:ZVC}
\end{eqnarray}

with here $S(\alpha,\delta)$ is the brightness distribution of the
source at angular coordinates $\alpha$ and $\delta$, $u$ and $v$ are
the spatial frequencies at which the Fourier Transform is
computed. The demonstration of this theorem can e.g. be found in
Born \& Wolf (\cite{BornandWolf1999}). And here is why
Fourier-transforms are so important to interferometry!

\begin{figure}[htbp]
  \centering
  \begin{tabular}{rc|ccc}
&& \multicolumn{3}{c}{Object}\\
 & & point source & small & large \\
\hline
&\vspace{0.01cm}\\
& &
    \includegraphics[width=0.17\textwidth]{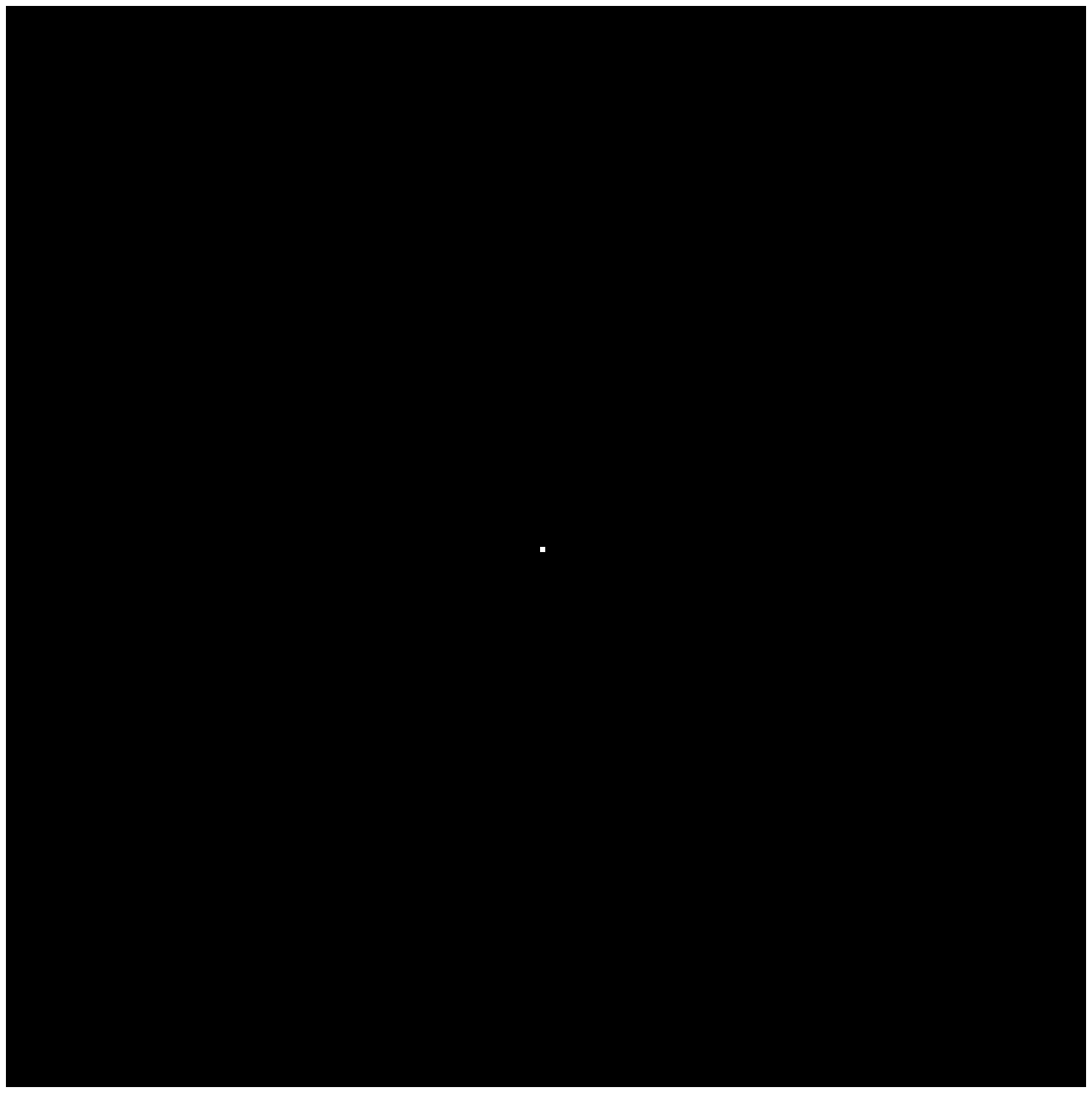} &
    \includegraphics[width=0.17\textwidth]{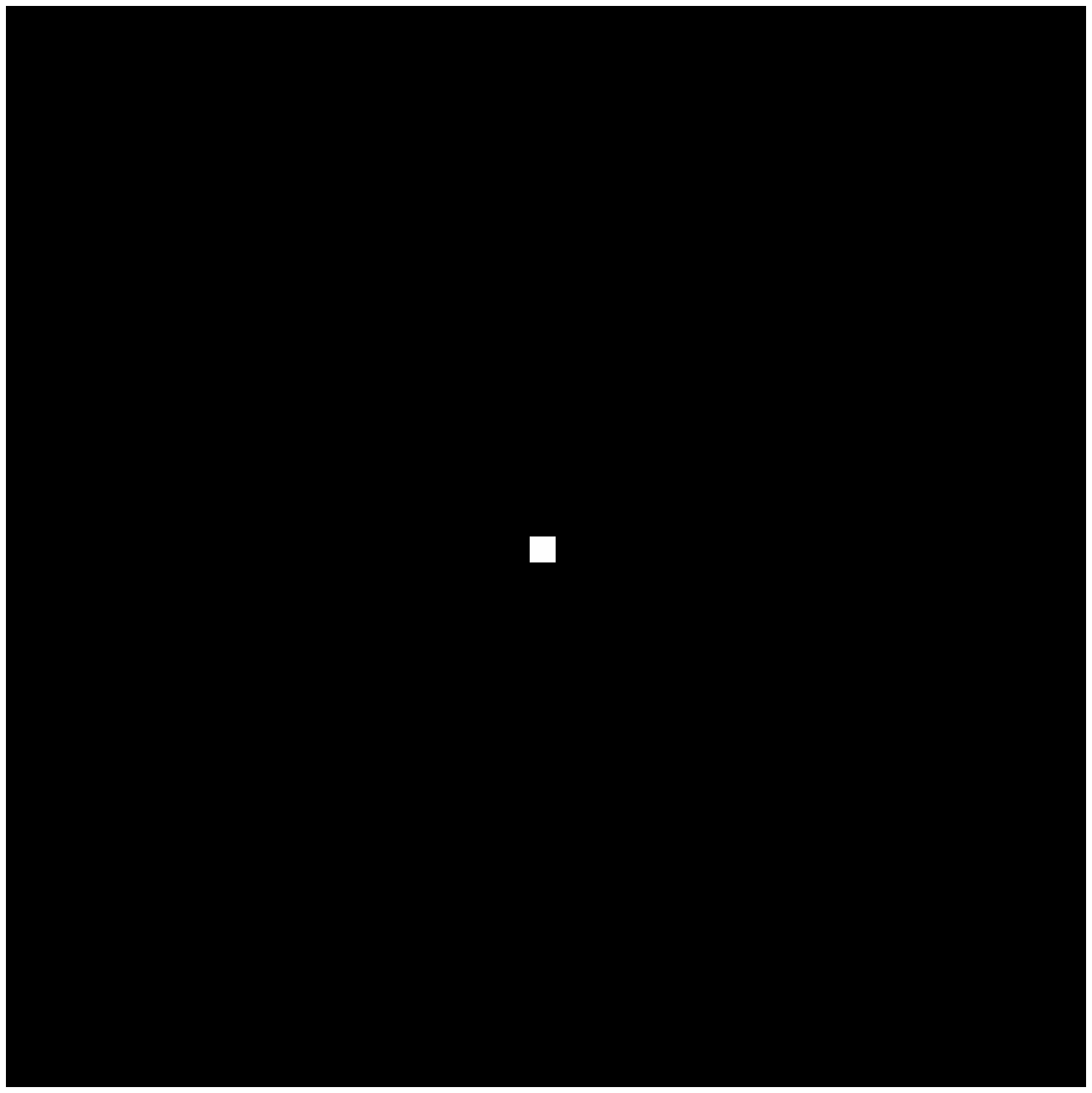} &
    \includegraphics[width=0.17\textwidth]{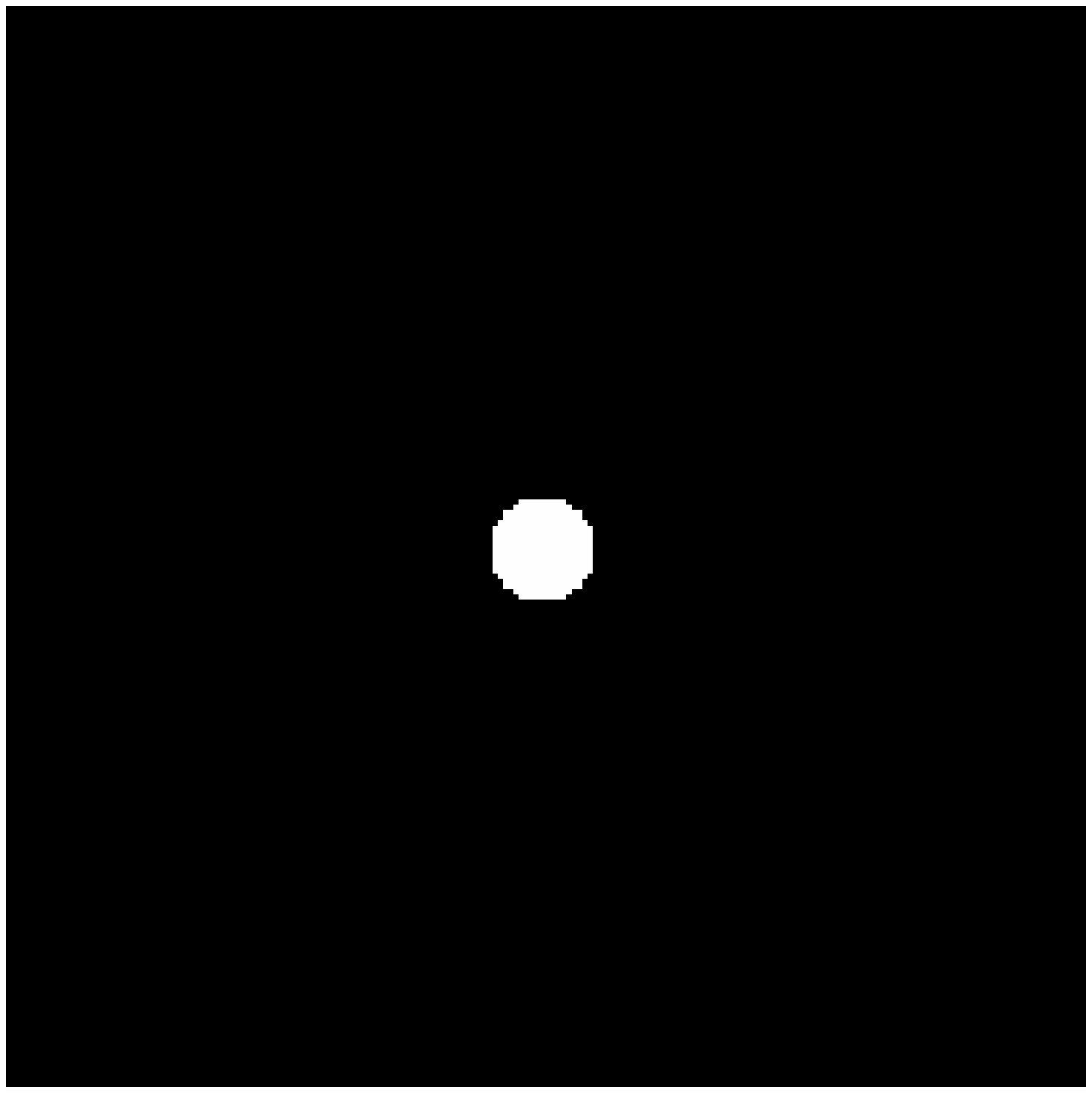} \\
&\vspace{0.2cm}\\
&& \multicolumn{3}{c}{Image}\\
\hline
&\vspace{0.1cm}\\
\raisebox{1cm}{Interferometer} &     \includegraphics[width=0.17\textwidth]{\folder/PUP_interf2T} &
    \includegraphics[width=0.17\textwidth]{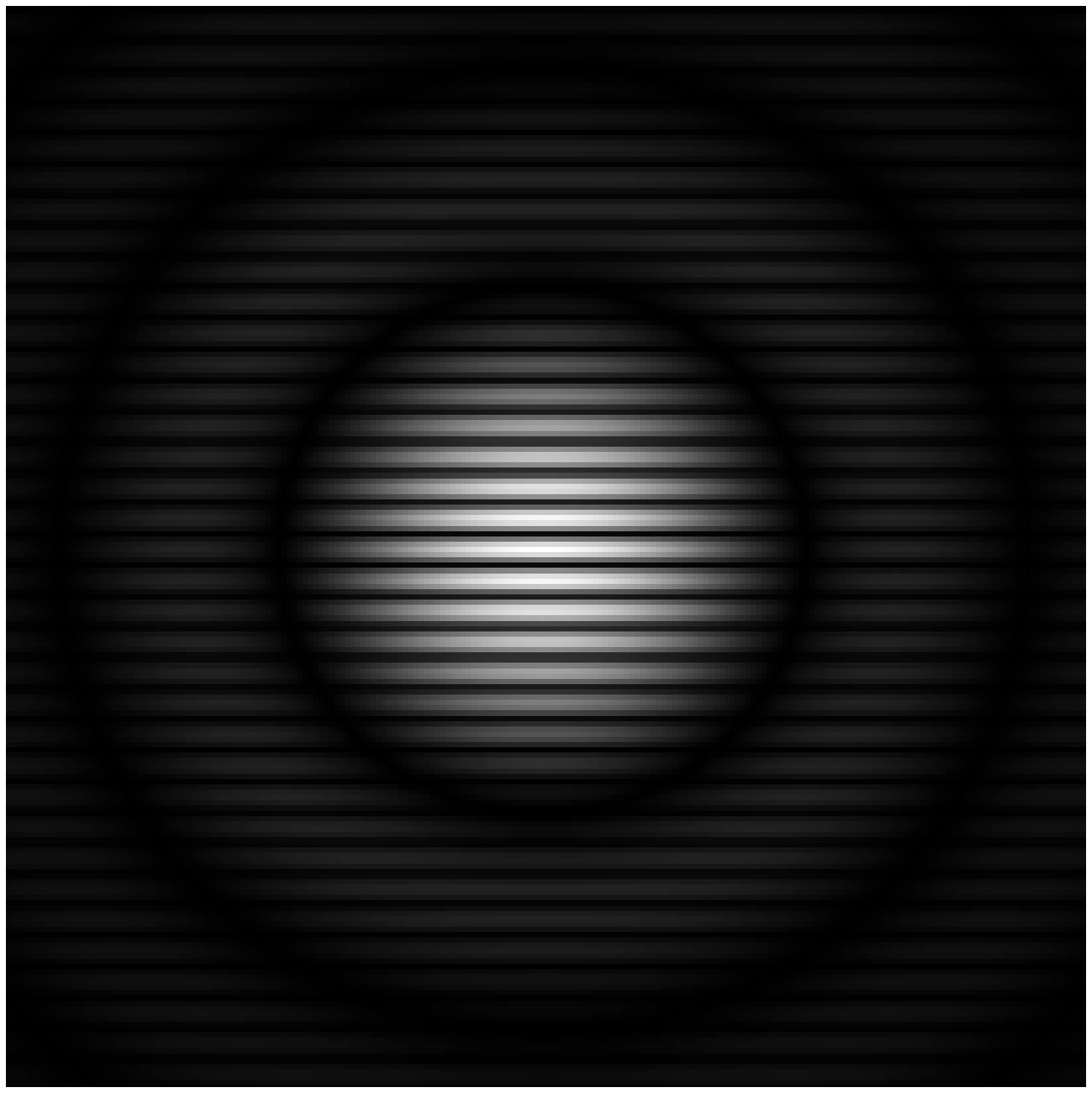} &
    \includegraphics[width=0.17\textwidth]{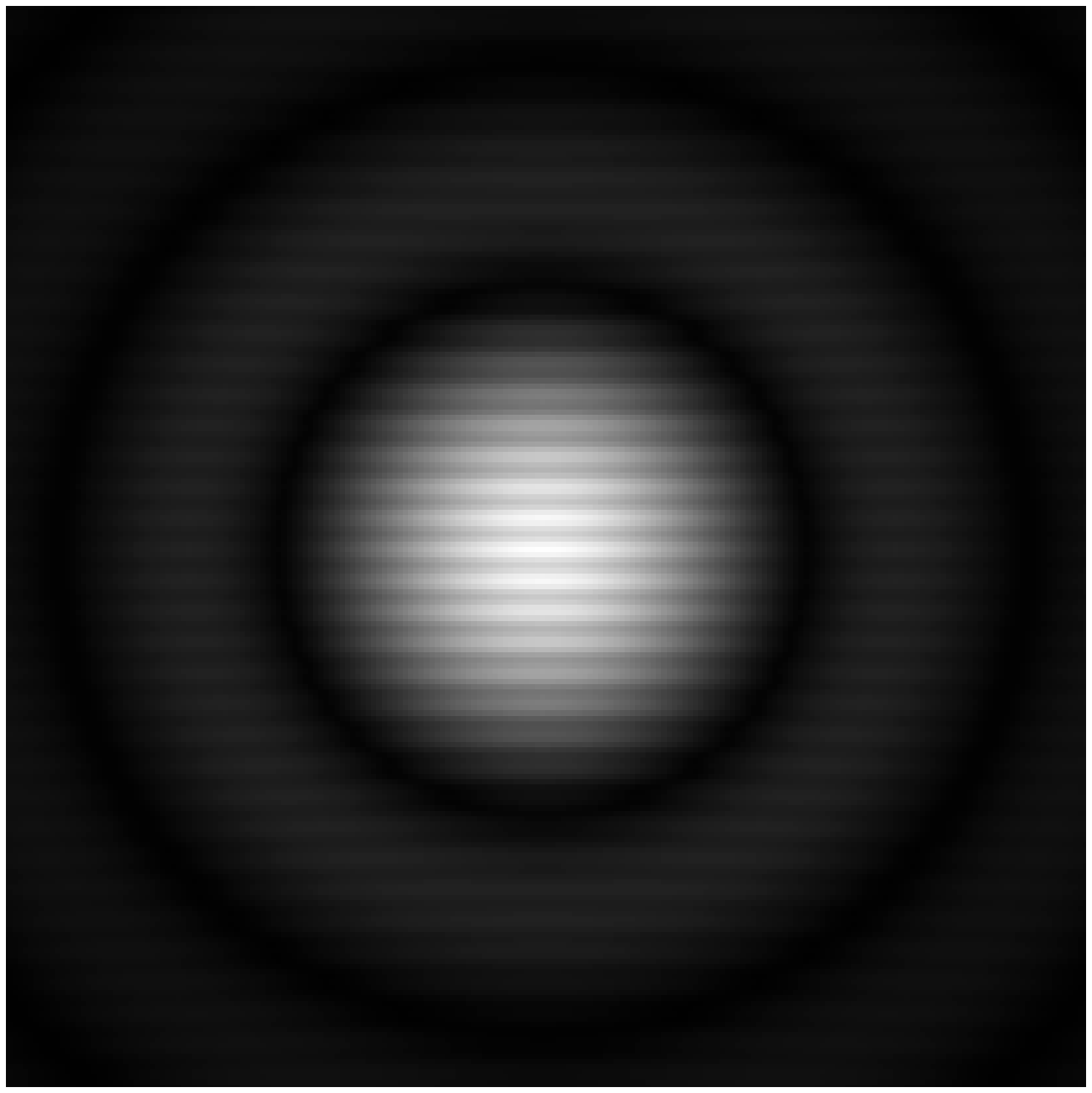} &
    \includegraphics[width=0.17\textwidth]{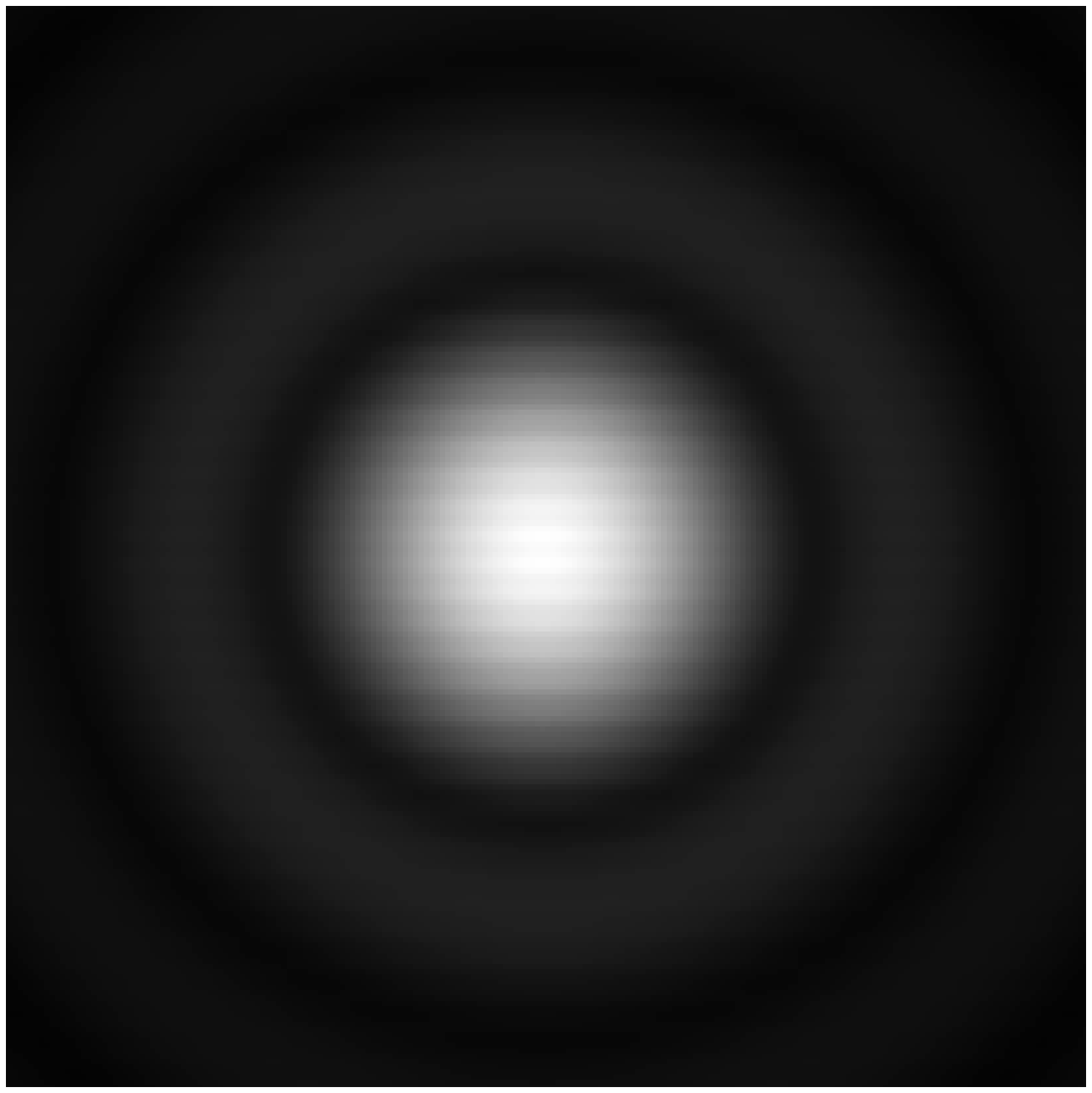} \\
\raisebox{1cm}{Full pupil} &    \includegraphics[width=0.17\textwidth]{\folder/PUP_ideal} &
    \includegraphics[width=0.17\textwidth]{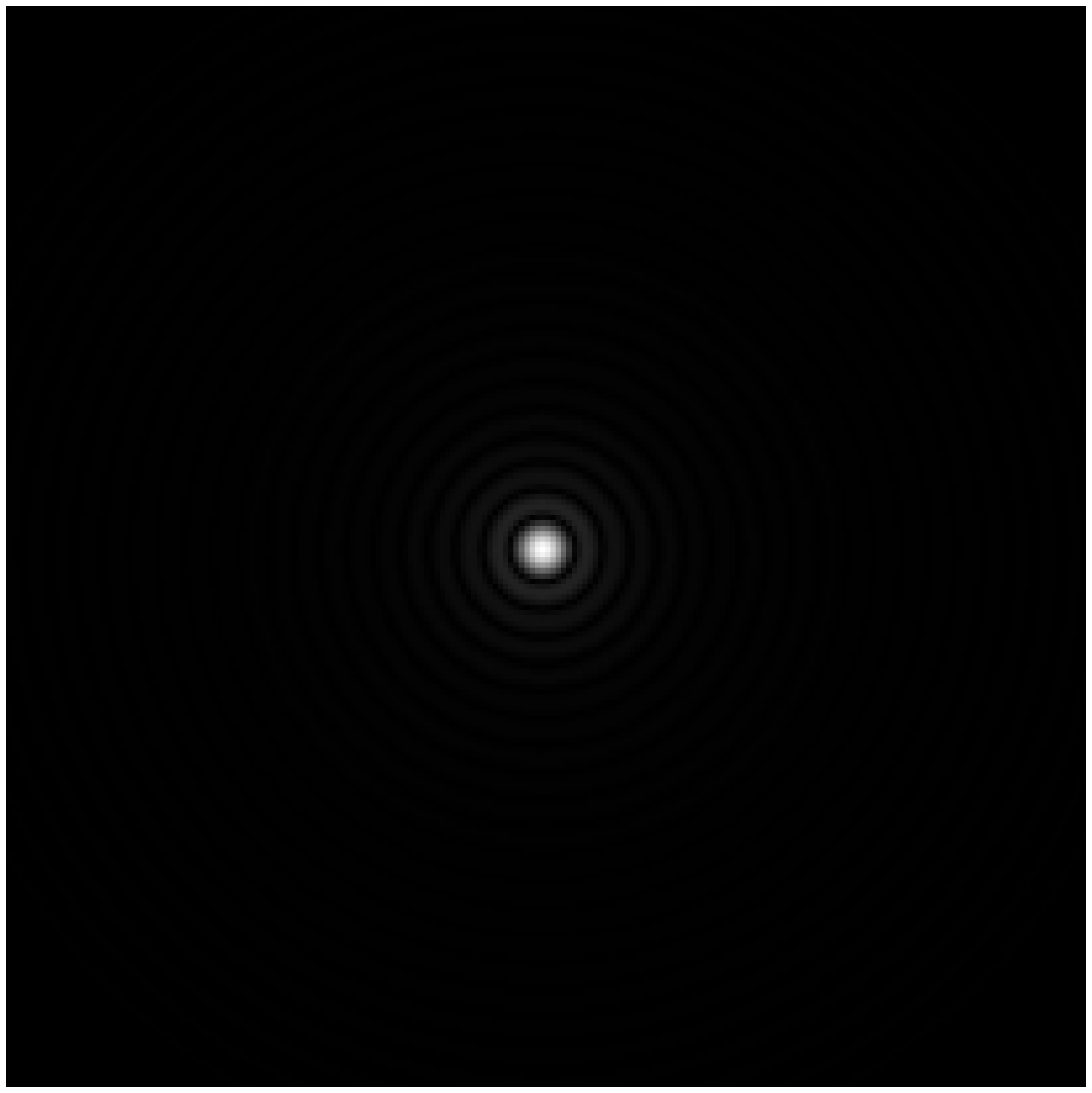} &
    \includegraphics[width=0.17\textwidth]{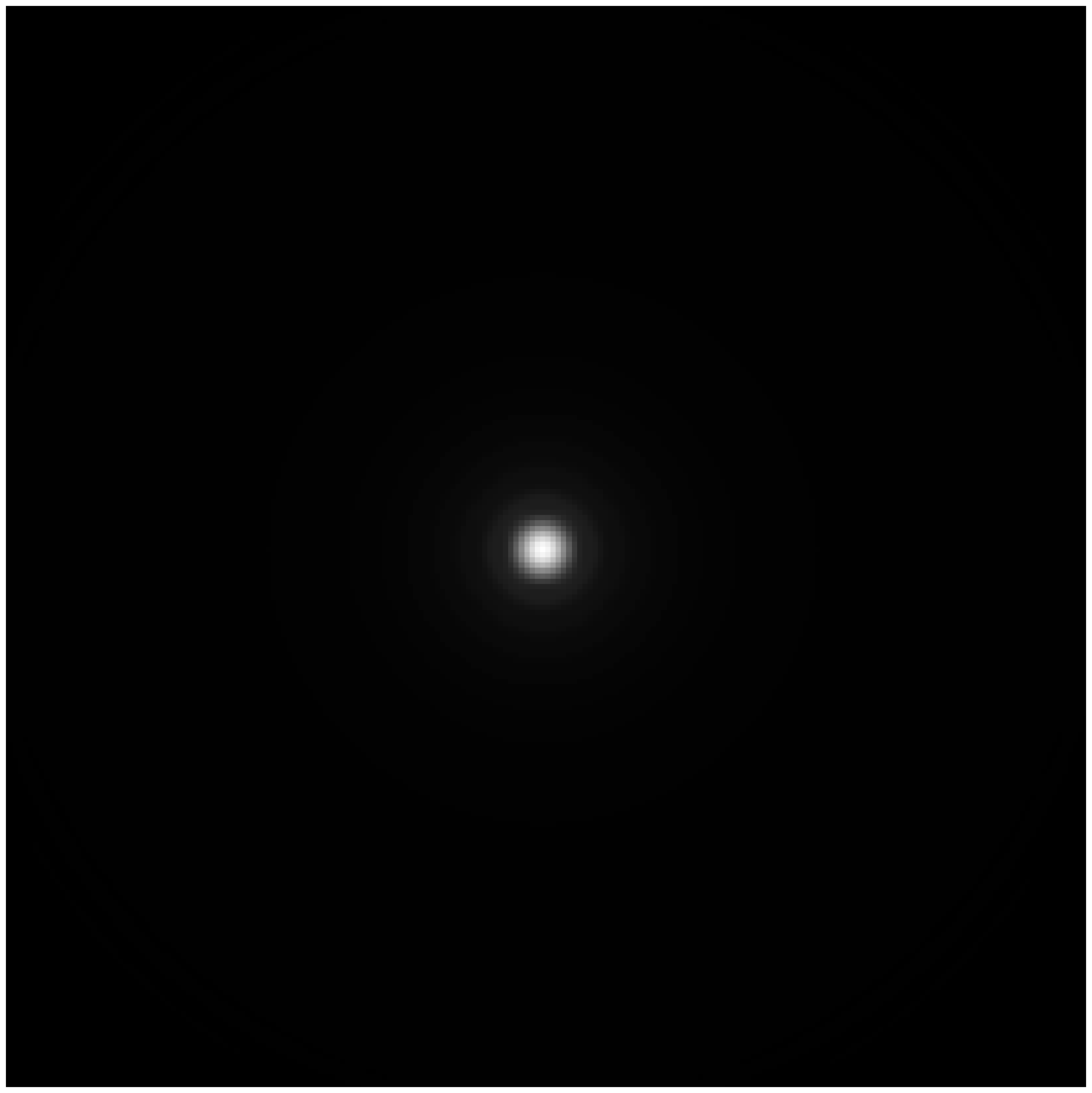} &
    \includegraphics[width=0.17\textwidth]{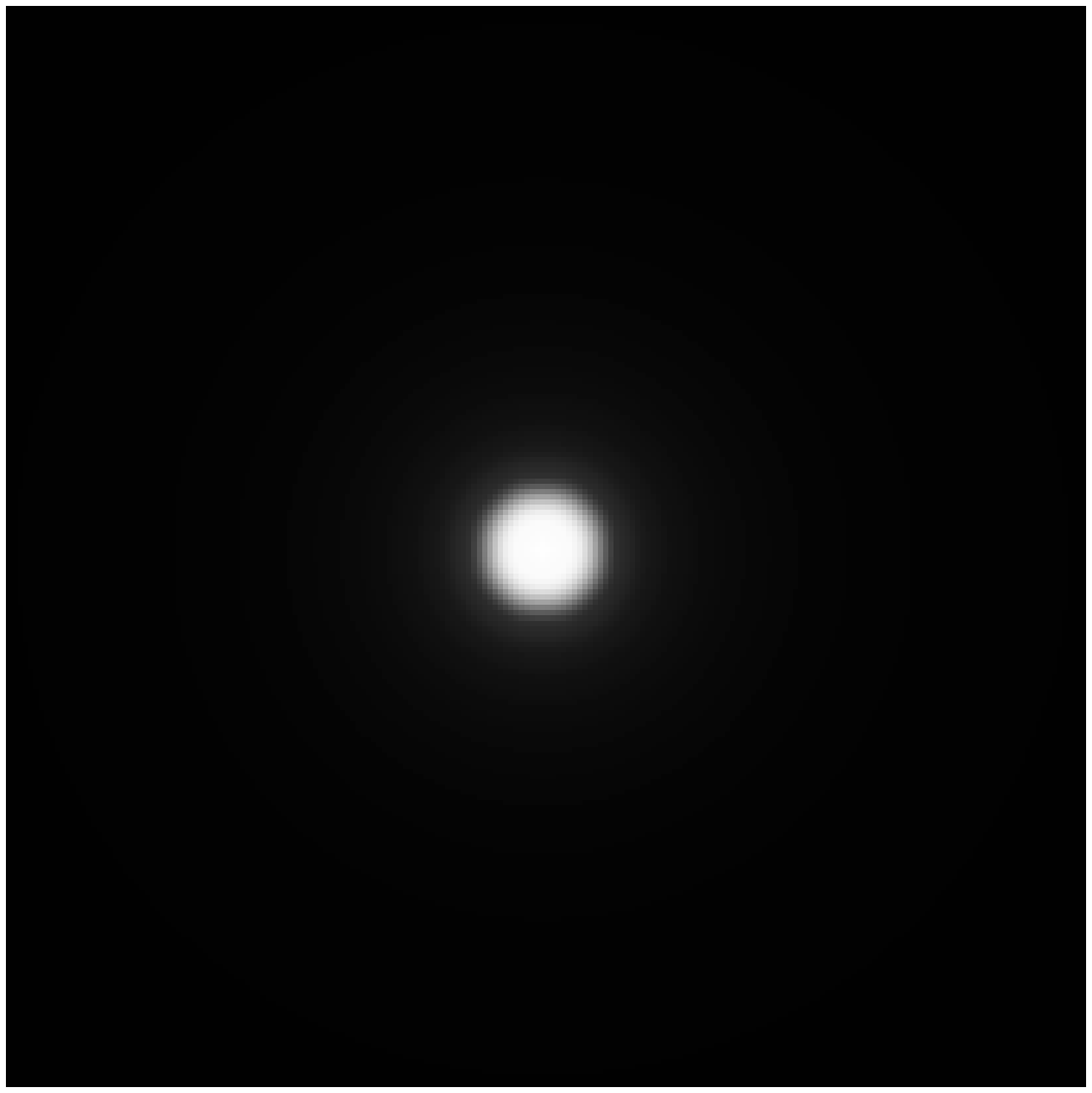} \\
  \end{tabular}
  \caption{{\bf Top, from left to right:} Simulated round objects of
    different diameters. {\bf Middle-left:} Reproduction of the 2
    telescope interferometer pupil. {\bf Middle:} Fringe pattern for
    the different object sizes. Note the fringe contrast
    change. {\bf Bottom-left:} Reproduction of the ideal round
    pupil. {\bf Bottom:} corresponding image. Note the disappearance
    of the Airy rings when the object is resolved. }
  \label{fig:resolveSource}
\end{figure}

The direct consequence of this theorem is that the fringe contrast and
phase are related to Fourier Transforms: the larger the object, the
lower will be the contrast (for a ``regular'' object). An illustration
of this effect is shown in Figure~\ref{fig:resolveSource}.

\subsection{Coherent flux}

As has been seen, the interferometer is sensitive in theory to the
degree of coherence of light $\gamma_{i,j}(0) = \mu^{\rm obj}_{1,2}\times
\me^{\i\phi^{\rm obj}_{1,2}}$, or complex visibility, which is given by the
Zernicke and van Cittert theorem.

\begin{figure}[htbp]
\vspace{-2cm}
  \centering \begin{tabular}{lr} \includegraphics[width=0.48\textwidth]{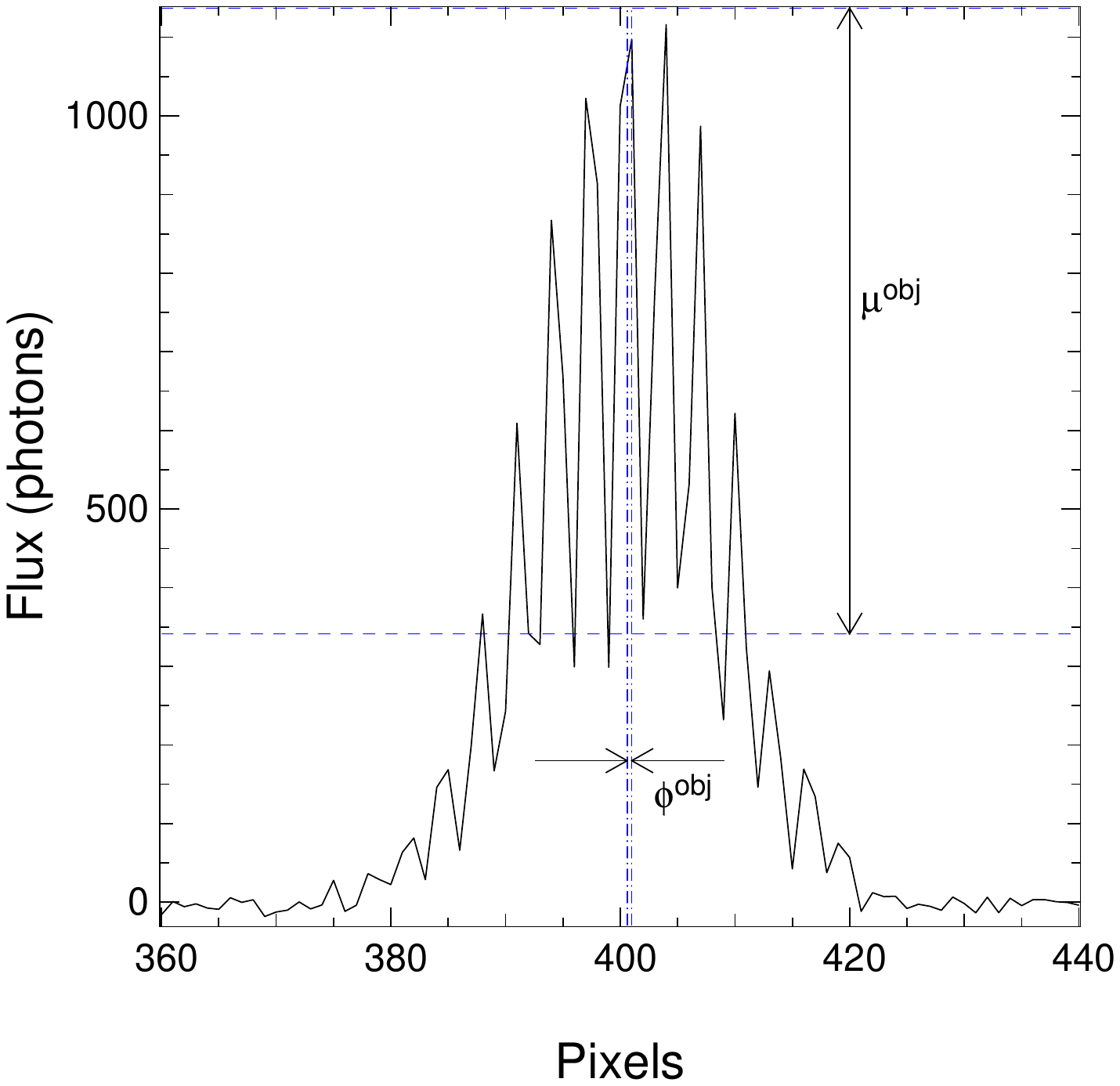}
  & \includegraphics[width=0.48\textwidth]{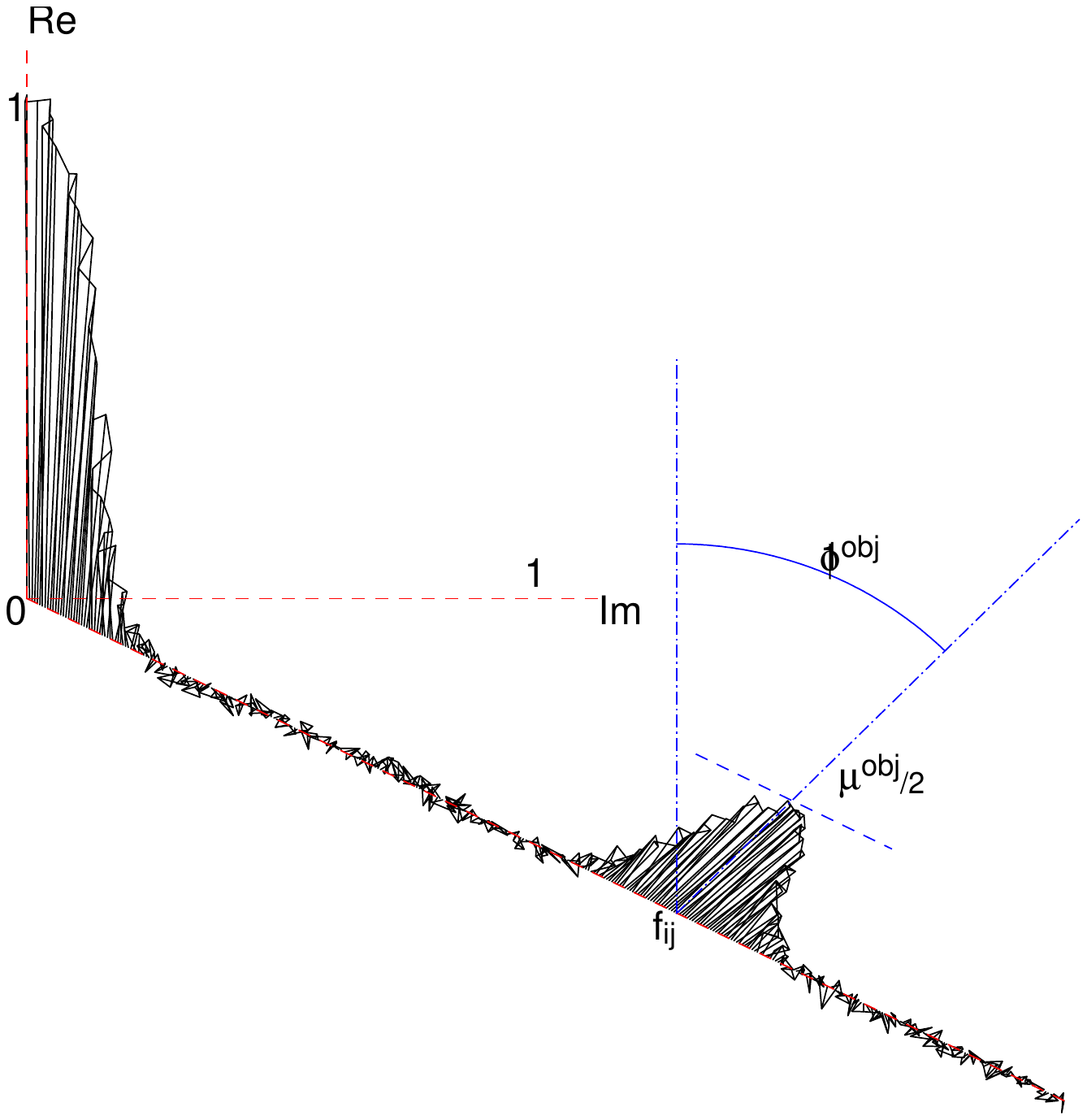} \\ \end{tabular}
\vspace{-3cm} \caption{{\bf
  Left:} A simulated fringe pattern for a typical multiaxial
  interferometer, with the representation of where the fringe contrast
  and phase can be measured. {\bf Right:} The Fourier Transform of
  that fringe pattern exhibits two peaks: one at zero frequency,
  representing the total flux, and one at frequency $f_{i,j}$
  representing the fringe contrast. Note that the white noise from the
  data appears as ``grass'' in the Fourier
  Transform.}  \label{fig:fringePattern}
\end{figure}

One needs to calculate the visibility values from the interferogram
signal. As this signal has a cosine modulation, one way to extract its
amplitude and phase is to apply a Fourier Transform and calculate the
power at the modulation frequency $f_{\rm i,j}$ (See
Fig.~\ref{fig:fringePattern}). The \emph{observed} $\gamma_{i,j}(0)$,
noted with ``$\widetilde{\gamma_{i,j}(0)}$'' is:

\begin{equation}
  \widetilde{\gamma_{i,j}(0)} = \frac{N_{\rm bases}{\rm FT}_{f_{\rm
        i,j}}\left[I(x,\lambda) \right]}{{\rm
      FT}_{0}\left[I(x,\lambda) \right]}
\label{eq:coherentdegree}
\end{equation}

with $N_{\rm bases} = \frac{N_{\rm tel} (N_{\rm tel}-1)}{2}$. This
equation contains the approximation that the two fluxes $I_1(\vec{x})$
and $I_2(\vec{x})$ are equal. The case where $I_1(\vec{x})$ and
$I_2(\vec{x})$ are not equal is treated later in this book
(ten Brummelaar \cite{Brummelaar2015}).

This method is often called the \emph{Fourier} method, but note that
it has nothing to do with the ZVC theorem (eq.~\ref{eq:ZVC}), as it is
just a way to actually measure the visibility.

\begin{figure}[htbp]
  \centering
    \includegraphics[width=0.48\textwidth, angle=90]{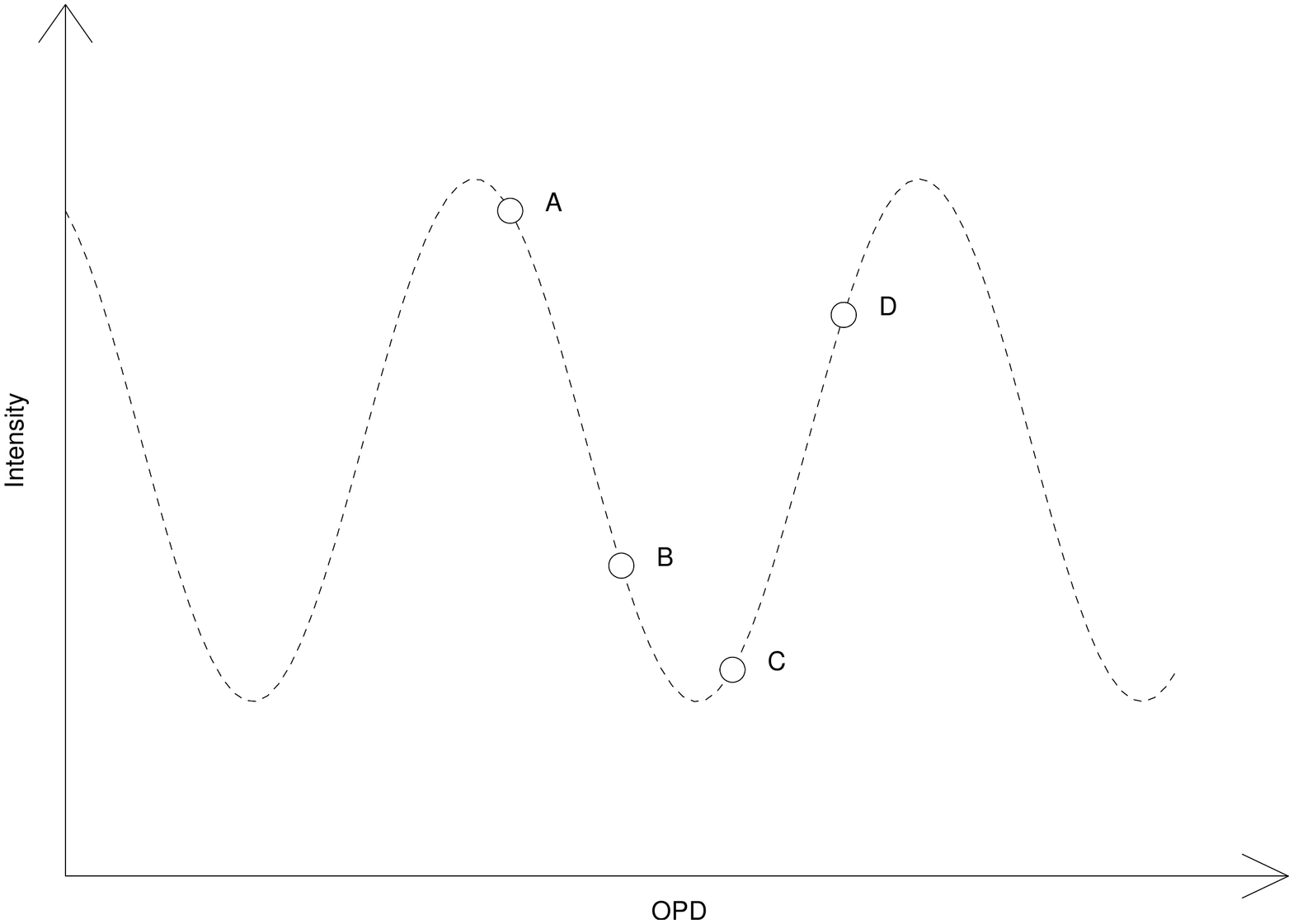} 
  \caption{Principle of the ABCD method: 4 measurements are made onto
    one single fringe, dephased by $\pi/2$ each, owing to the fringe
    contrast and phase.}
  \label{fig:ABCD}
\end{figure}

Another way is to measure the amplitude of the fringes in the image
space, for example, one can measure one fringe at 4 different points
$A,B,C,D$ each one separated to each other in phase by $\pi/2$ (see
Fig.~\ref{fig:ABCD}). The visibility amplitude can be computed this
way:

\begin{equation}
  \widetilde{\mu} = \frac{\sqrt{(I_A - I_C)^2 + (I_B - I_D)^2}}{2
    \sum_j I_j}
\end{equation}

and the visibility phase:

\begin{equation}
  \widetilde{\phi} = \arctan\left(\frac{I_A-I_C}{I_B-I_D}\right)
\end{equation}

One needs to note here that the ABCD method relies on the knowledge of
the shape of the fringes (a cosine function) and on the fact that the
A,B,C and D samples are exactly offset by $\pi/2$. A generalisation of
that method, called {\sc p2vm} (Millour \etal\ \cite{2004SPIE.5491.1222M},
  Tatulli \etal\ \cite{2007AandA...464...29T}), was proposed and implemented on the {\sc
  amber} instrument (Petrov \etal\ \cite{2007AandA...464....1P}). The basic idea is to
use the \emph{a priori} information of the fringes shape to adjust a
model to the data in order to obtain the visibilities. The method is
thoroughly described in Tatulli \etal\ (\cite{2007AandA...464...29T}).

To get an overview and understanding of how to reduce data for a
specific instrument, it is always better to read the corresponding
paper. For example Mourard \etal\ (\cite{Mourard2011a}) for the {\sc
chara}/{\sc vega} instrument, Petrov \etal\ (\cite{2007AandA...464...29T}) for the {\sc vlti}/{\sc amber} instrument,
ten Brummelaar (\cite{Brummelaar2015}) for {\sc chara}/{\sc classic},
Perrin (\cite{2003AandA...398..385P}, \cite{2003AandA...400.1173P})
for {\sc chara}/{\sc fluor} and {\sc vlti}/{\sc vinci}, etc.

\subsection{The $(u,v)$ problem}

One very specific problem of optical long-baseline interferometry is
called the ``$(u,v)$ problem''. It is related to the sparsity of
measurements the interferometers can provide. Indeed, contrary to
classical imaging, a two telescopes interferometer measurement samples
only one point in the frequency domain of equation~\ref{eq:ZVC},
usually noted $(u,v)$ plane. More details are given in
Millour (\cite{2008NewAR..52..177M}), therefore I will just recall the
different ways to fill the $(u,v)$ plane:
\begin{description}
\item[Supersynthesis:] The rotation of Earth relative to the celestial
     sphere makes the baseline change with time. The $(u,v)$ tracks
     are on an arc of ellipse. The exact expression of the $(u,v)$
     tracks is given in Segransan (\cite{2007NewAR..51..597S}) and
     recalled in Millour (\cite{2008NewAR..52..177M}).
\item[Add more telescopes:] The number of $(u,v)$ points for one measurement
     is equal to the number of baselines, roughly proportionnal to the
     square of the number of telescopes, following the relation
     $N_{\rm bases} = \frac{N_{\rm tel} (N_{\rm tel}-1)}{2}$
\item[Make use of wavelength:] The spatial frequencies are
  proportionnal to the wave number $\sigma = 1/\lambda$ and trace
  radial lines in the $(u,v)$ plane.
\end{description}

An illustration of these is shown in Table~\ref{tab:uvfilling} for the
future {\sc matisse}/{\sc vlti} instrument (Lopez \etal\ \cite{2006SPIE.6268E..31L}) in the L
band (3$\mu$m).

\begin{sidewaystable}
\caption{The different ways of filling the $(u,v)$ plane,
  illustrated by {\sc matisse} observation in the L band (3.5--4.1$\mu$m) of
  the star $\gamma^2$~Velorum (declination $-47^o$). ``2T'' means ``2
  Telescopes'', ``4T'' means ``4 Telescopes'' and ``4T+conf'' means
  ``4 Telescopes and change of configurations'', i.e. the best
  situation at {\sc vlti} today}
\label{tab:uvfilling}
 \centering
  \begin{tabular}{|m{1.cm}|c|c|c|c|}
\toprule
     & Single measurement &  Wavelength coverage& Supersynthesis & Supersynthesis  \\
     &  &  &  & + Wavelength coverage \\
\midrule
    2 T &
 \adjustbox{valign=m}{\includegraphics[width=0.18\textwidth]{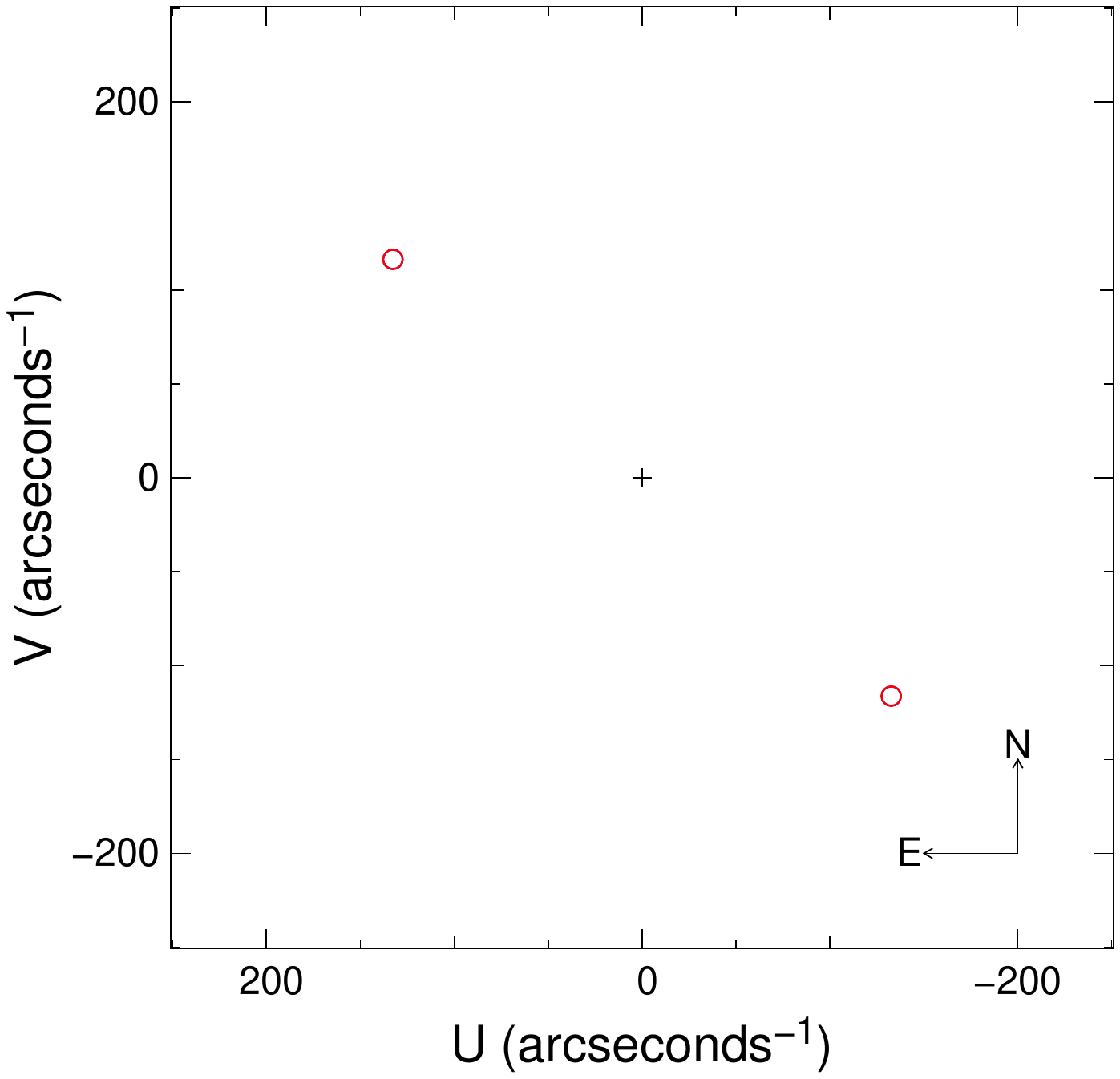}} &
\adjustbox{valign=m}{ \includegraphics[width=0.18\textwidth]{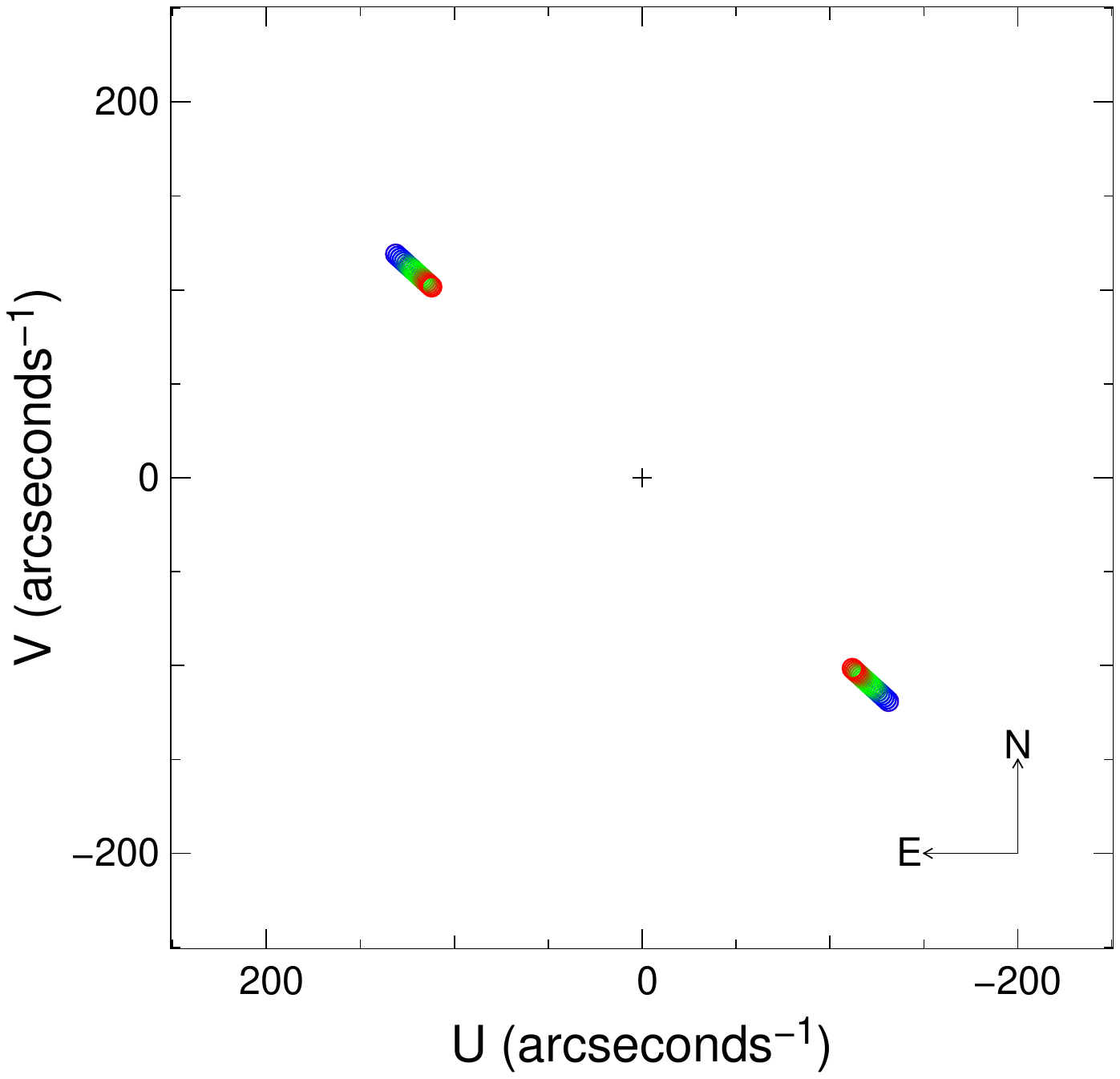}} &
 \adjustbox{valign=m}{\includegraphics[width=0.18\textwidth]{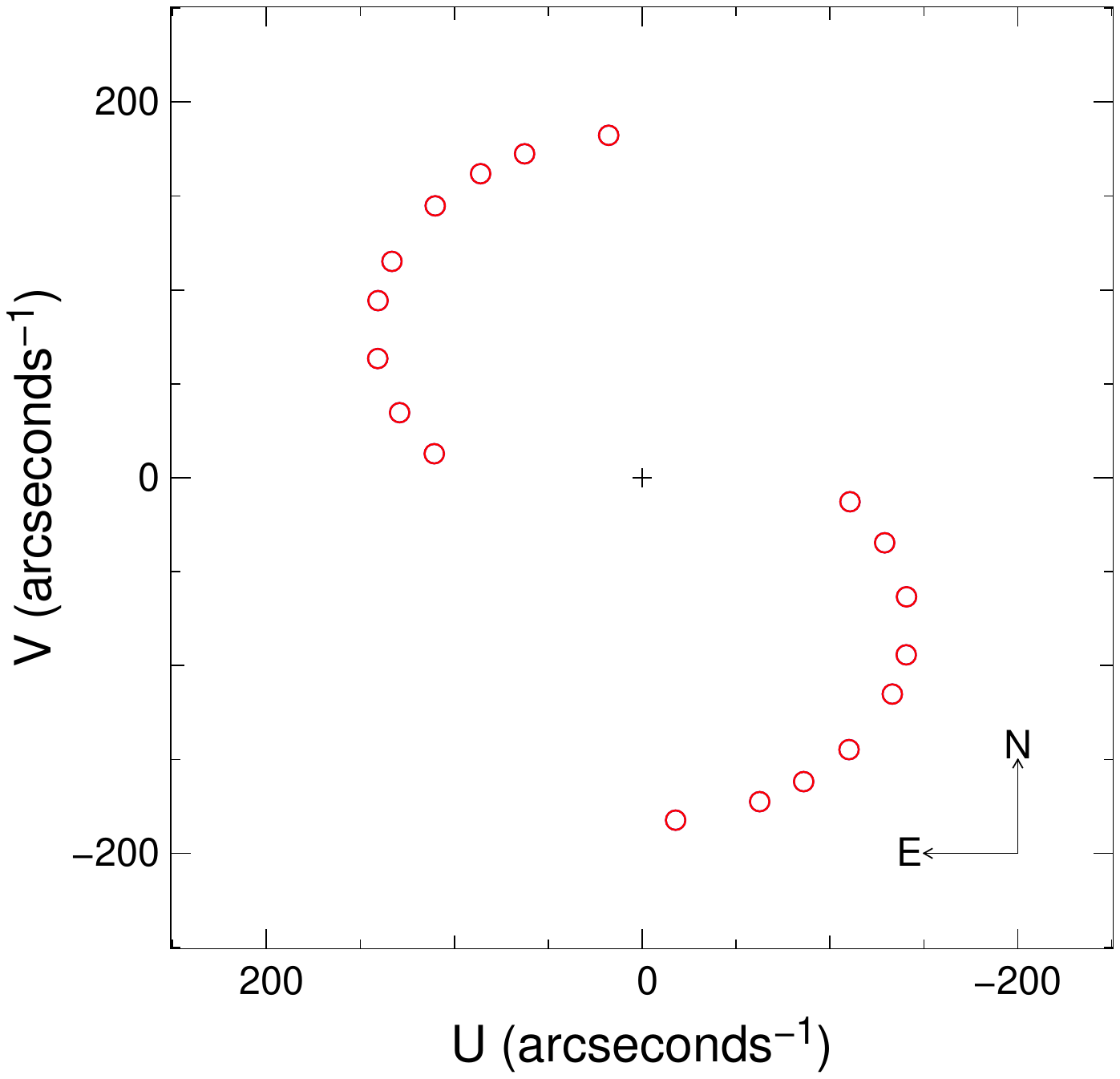}} &
 \adjustbox{valign=m}{\includegraphics[width=0.18\textwidth]{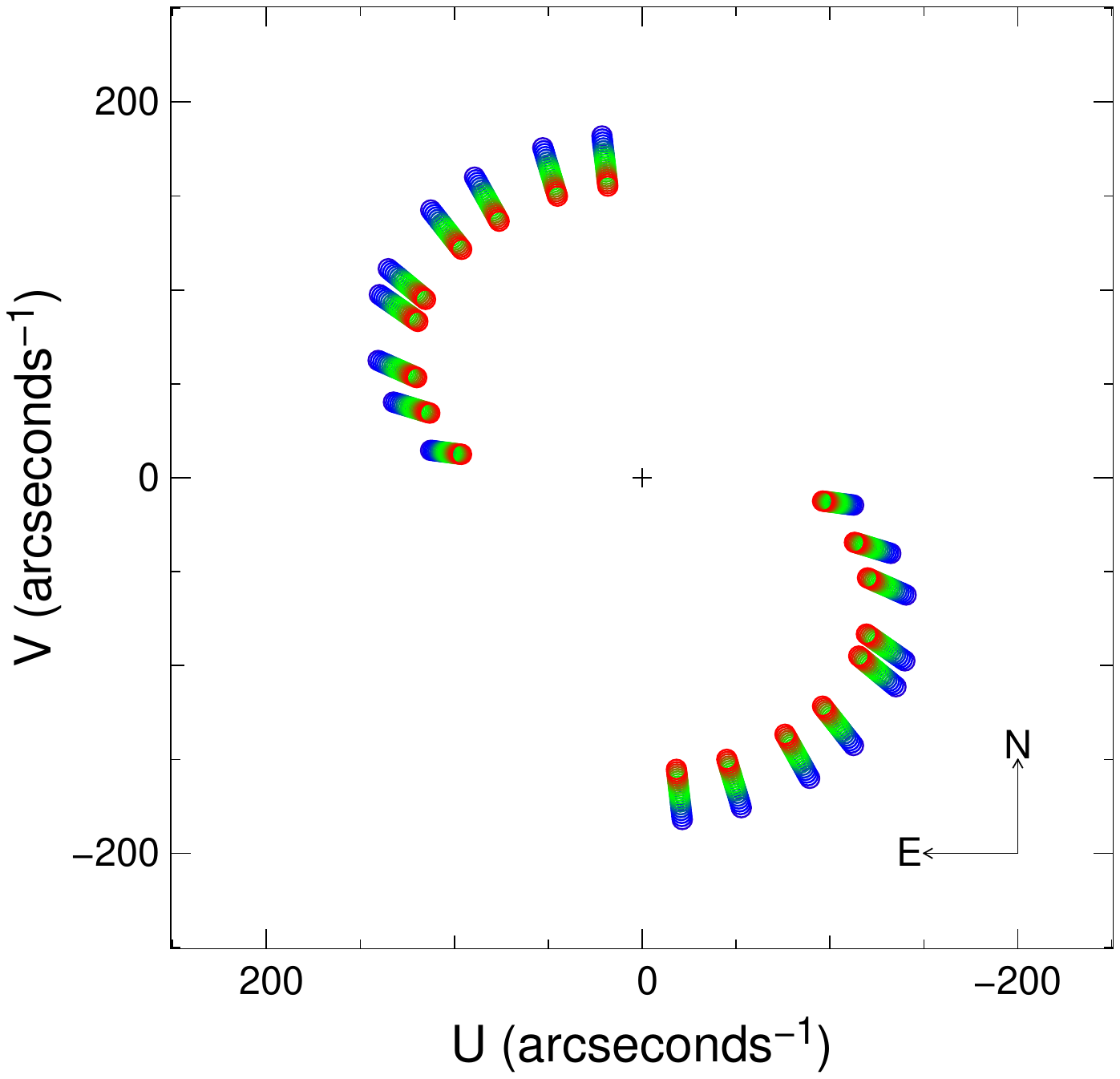}}\\
\midrule
    4 T &
 \adjustbox{valign=m}{\includegraphics[width=0.18\textwidth]{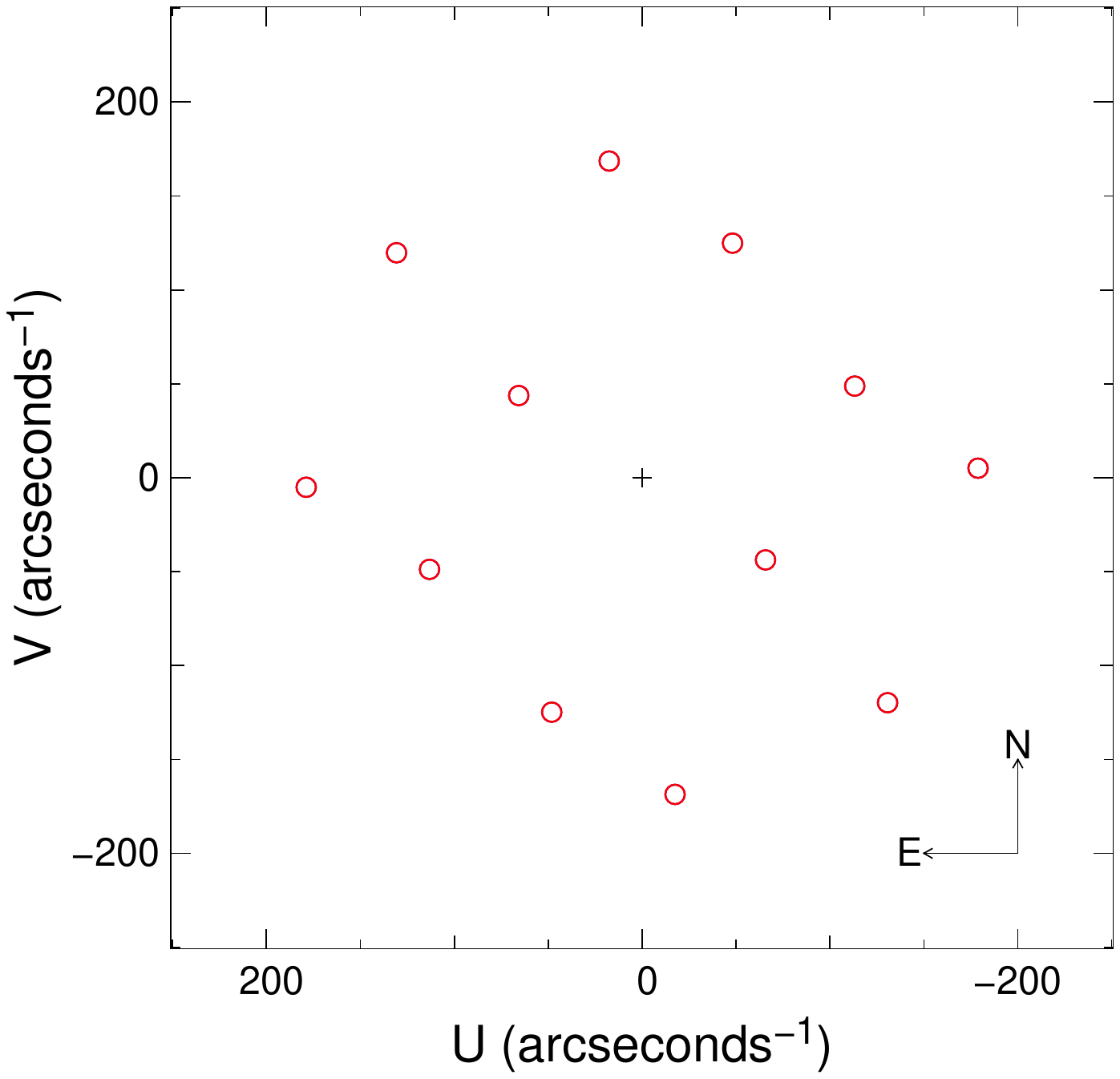}} &
\adjustbox{valign=m}{ \includegraphics[width=0.18\textwidth]{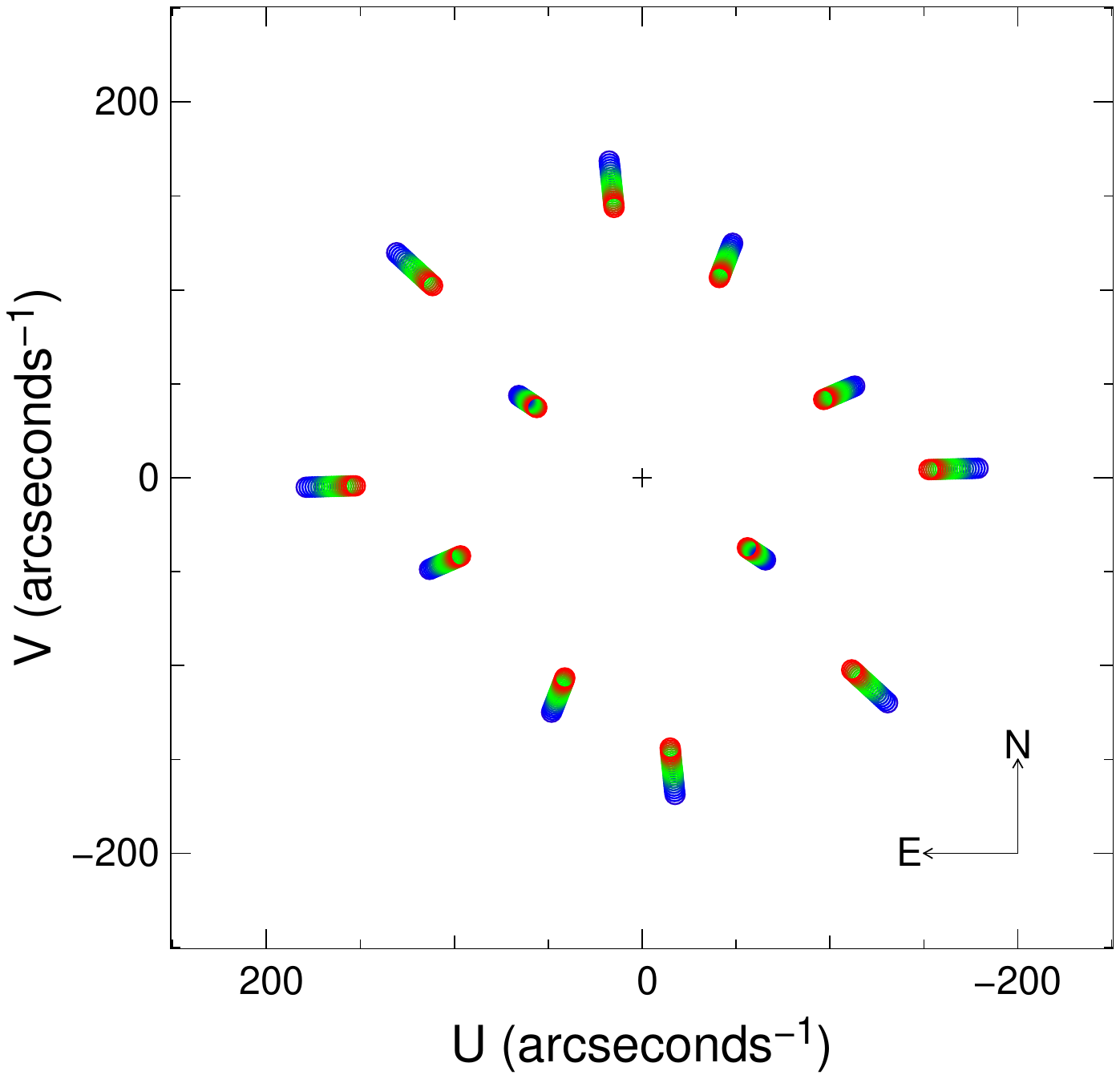}} &
\adjustbox{valign=m}{ \includegraphics[width=0.18\textwidth]{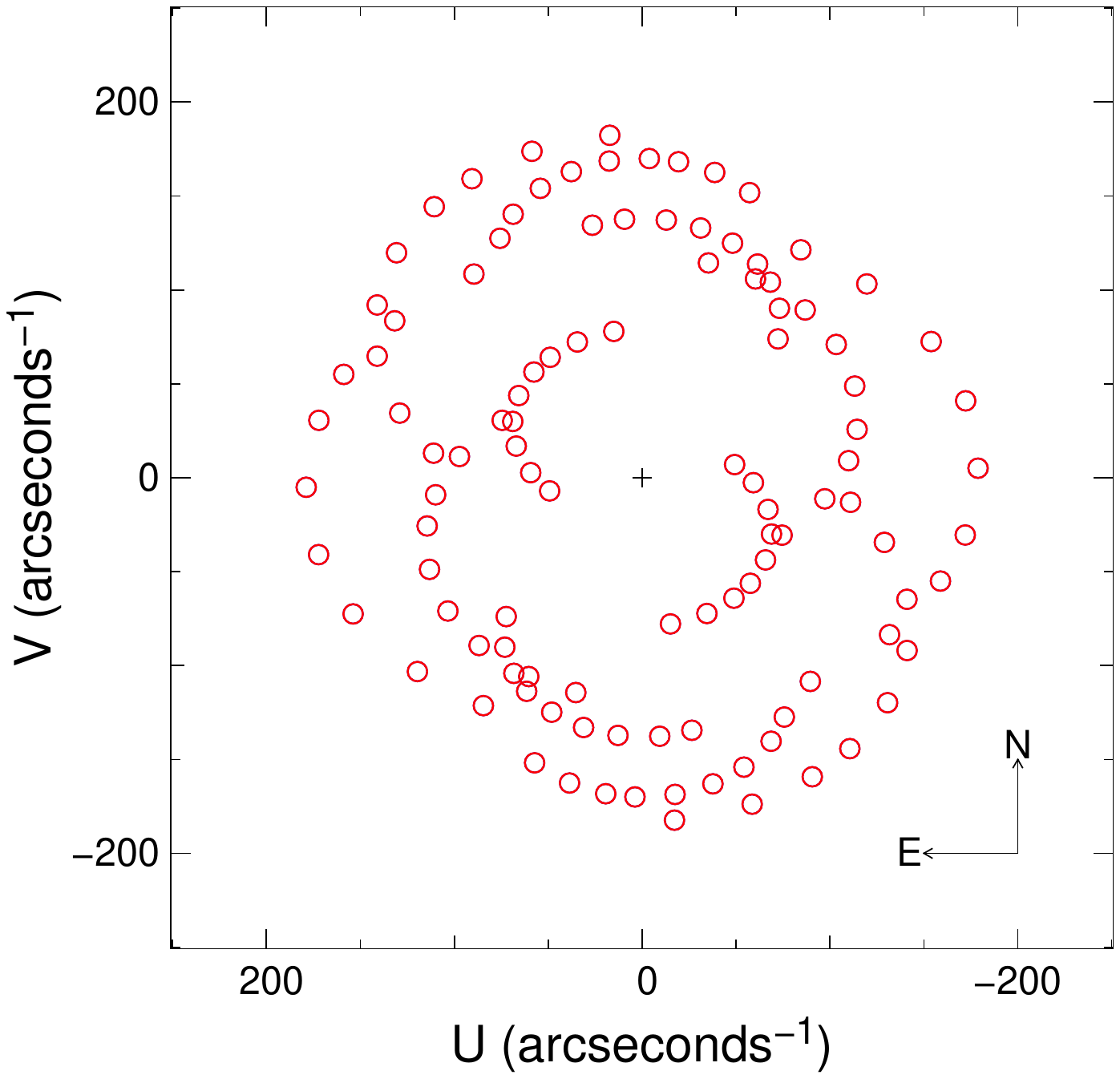}} &
 \adjustbox{valign=m}{\includegraphics[width=0.18\textwidth]{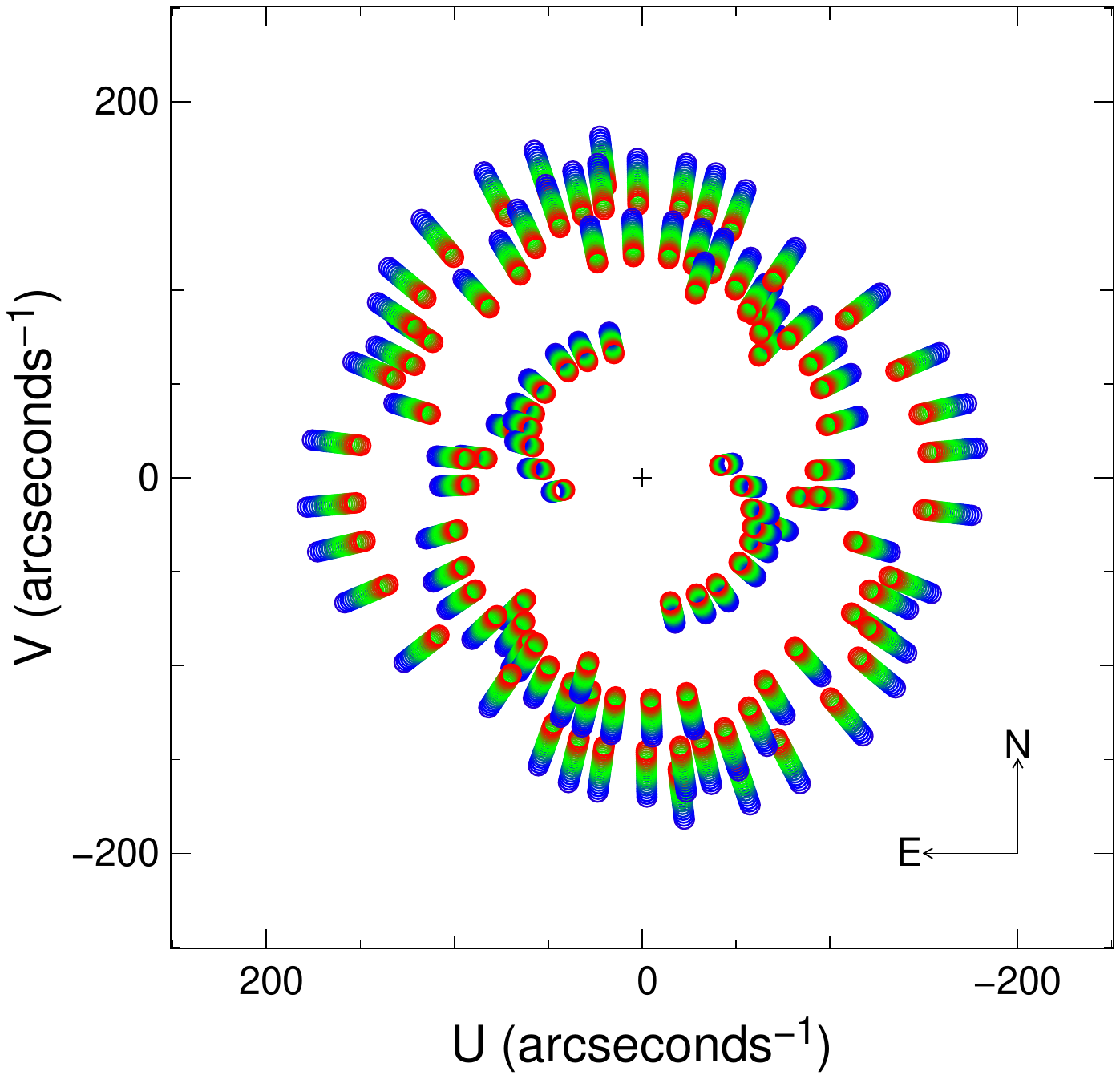}} 
  \\
\midrule
    4 T + conf &
 \adjustbox{valign=m}{\includegraphics[width=0.18\textwidth]{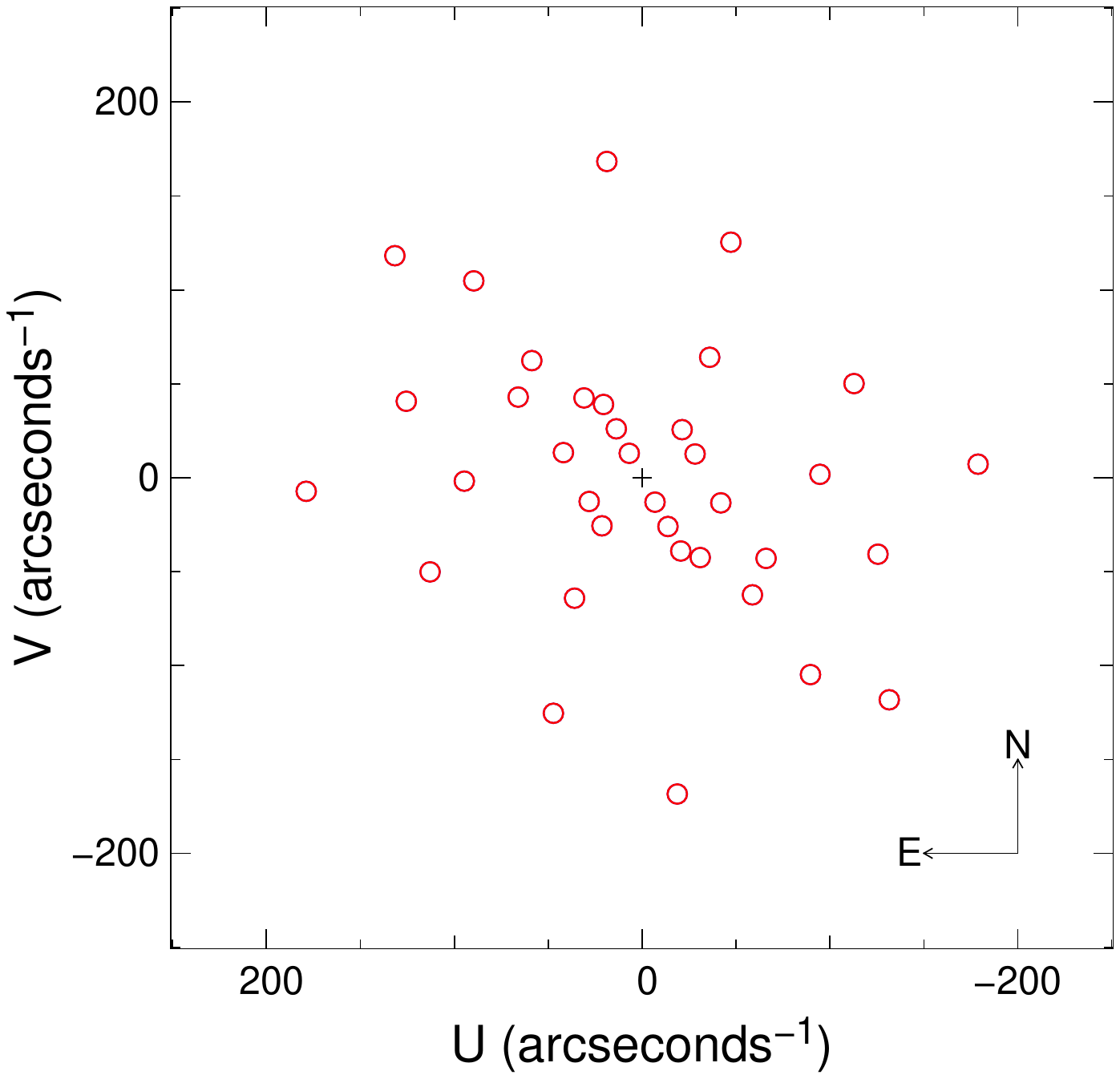}} &
 \adjustbox{valign=m}{\includegraphics[width=0.18\textwidth]{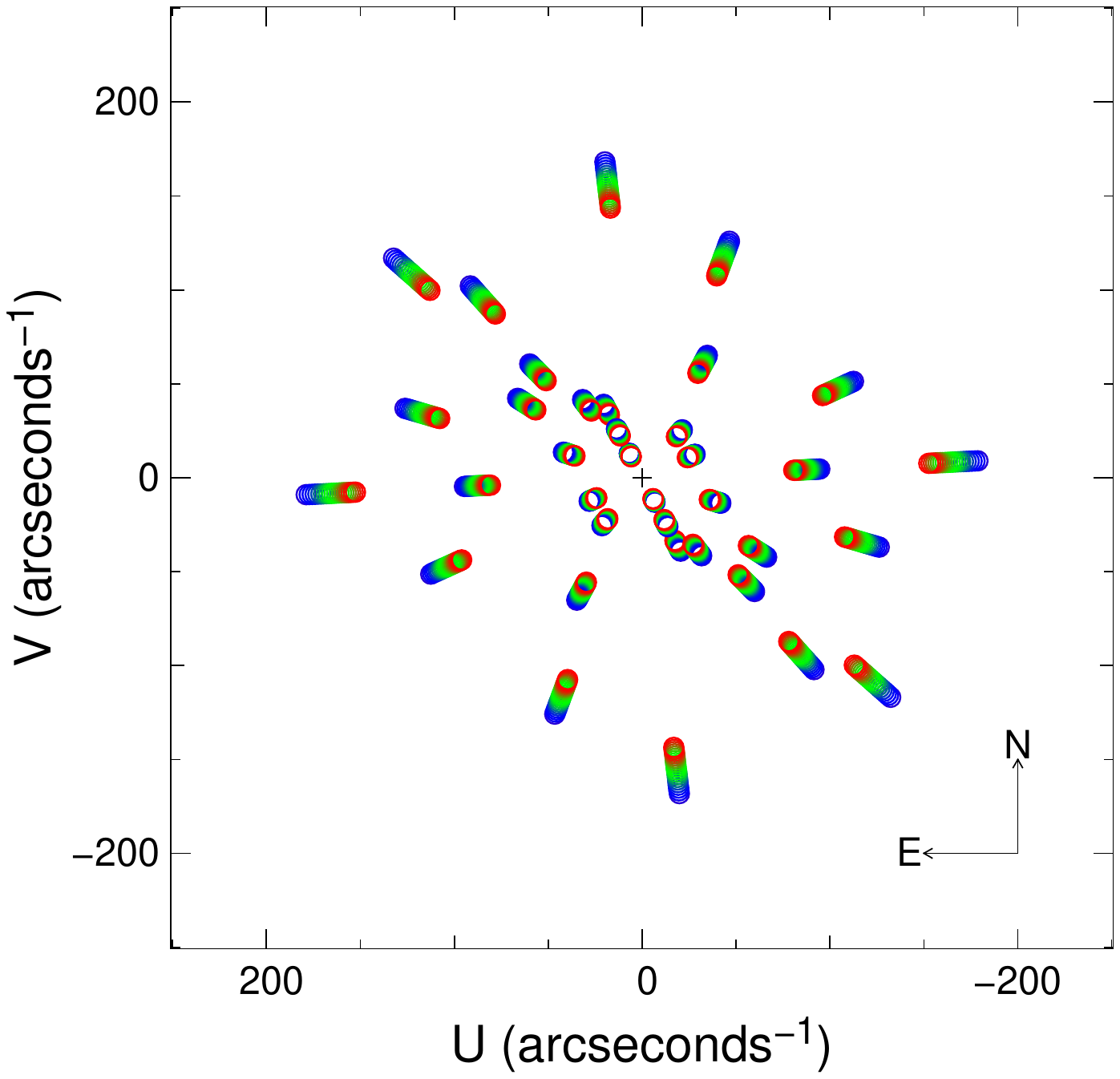}} &
\adjustbox{valign=m}{ \includegraphics[width=0.18\textwidth]{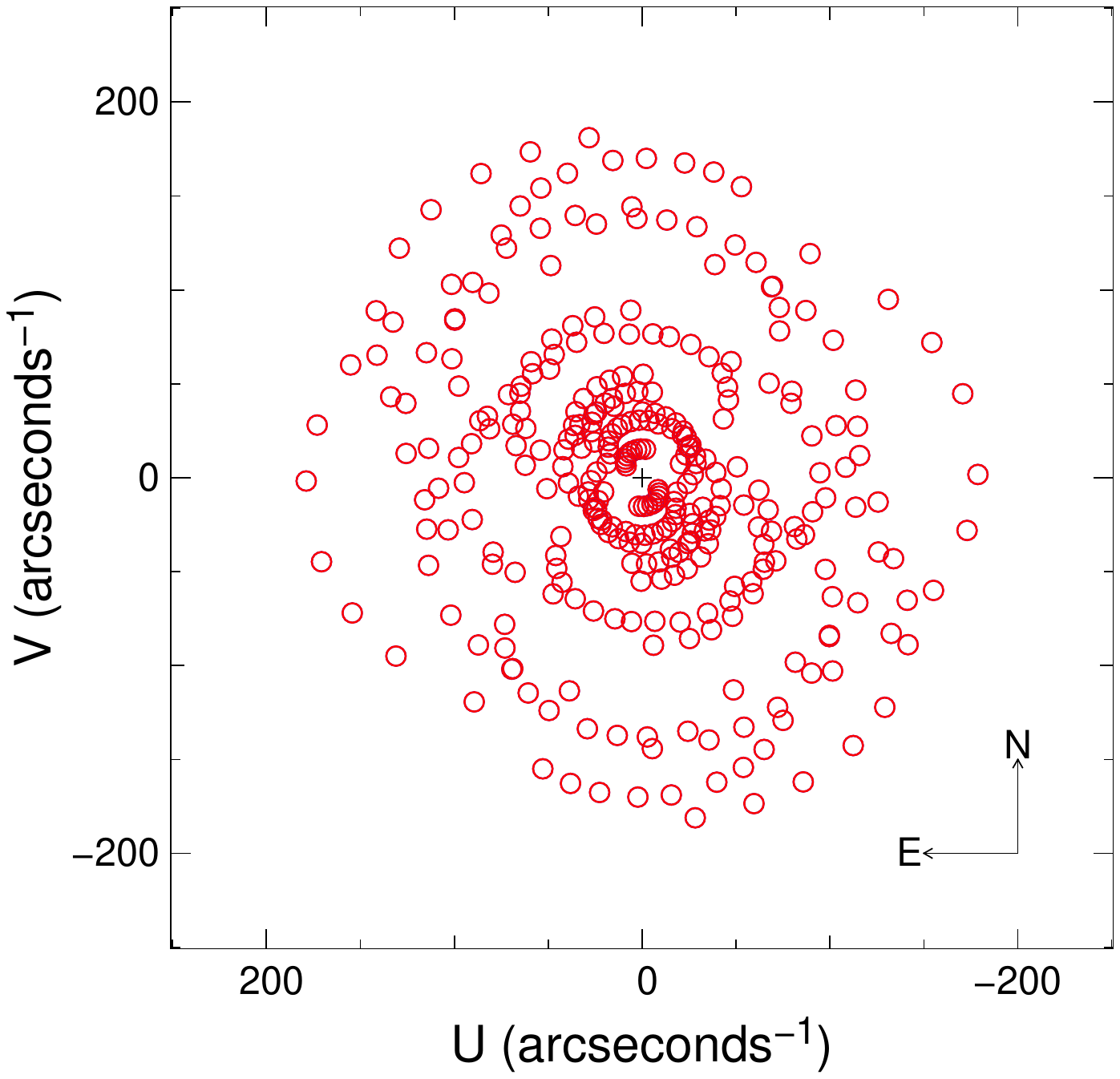}} &
\adjustbox{valign=m}{ \includegraphics[width=0.18\textwidth]{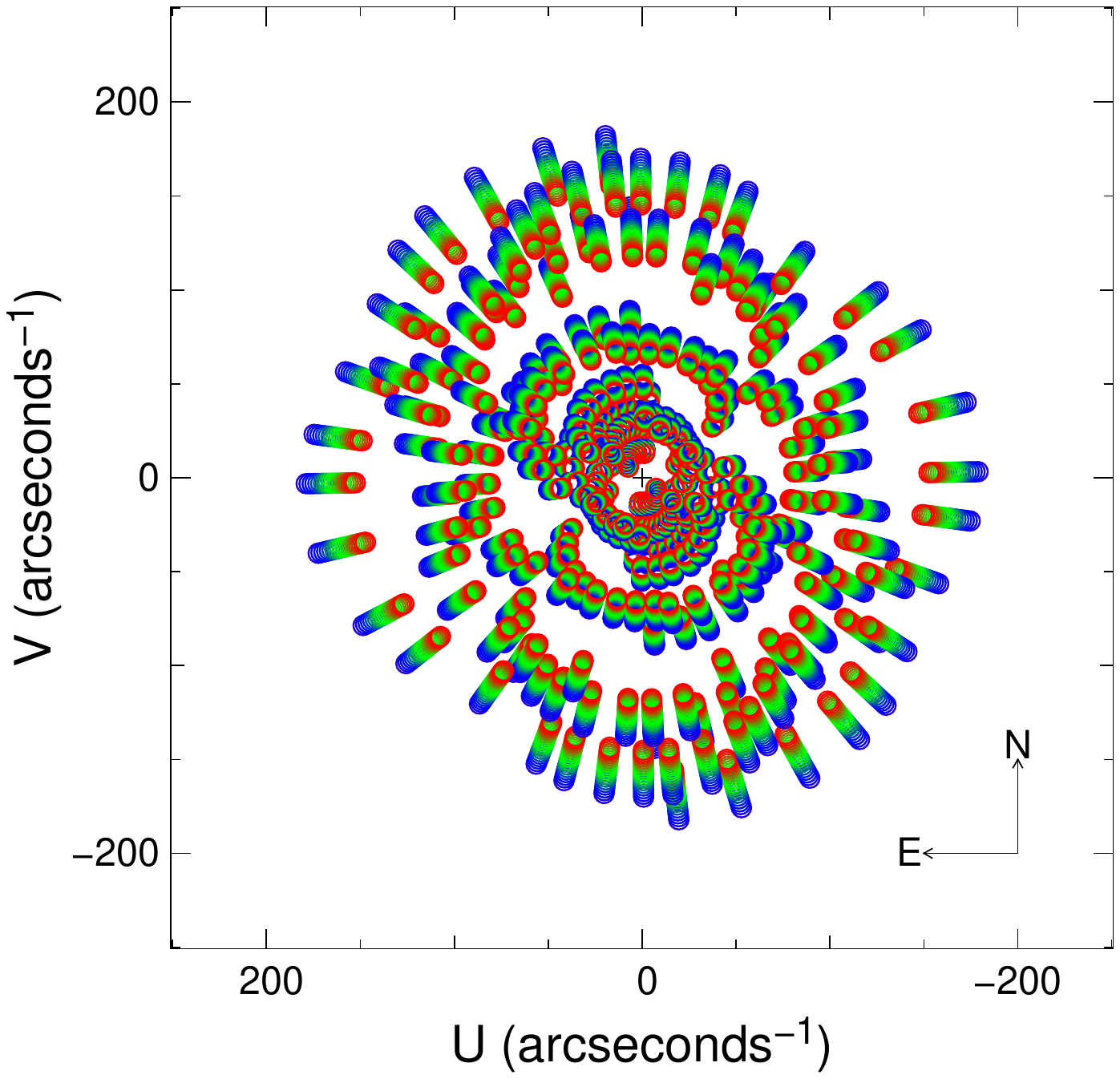}}\\
\bottomrule
  \end{tabular}
\end{sidewaystable}

\subsection{The phase problem}

\begin{figure}[htbp]
\vspace{-1cm}
  \centering
\begin{center}
\begin{minipage}[t]{.3\linewidth}
\vspace{10pt}
  \includegraphics[width=1.4\textwidth,origin=t,angle=0]{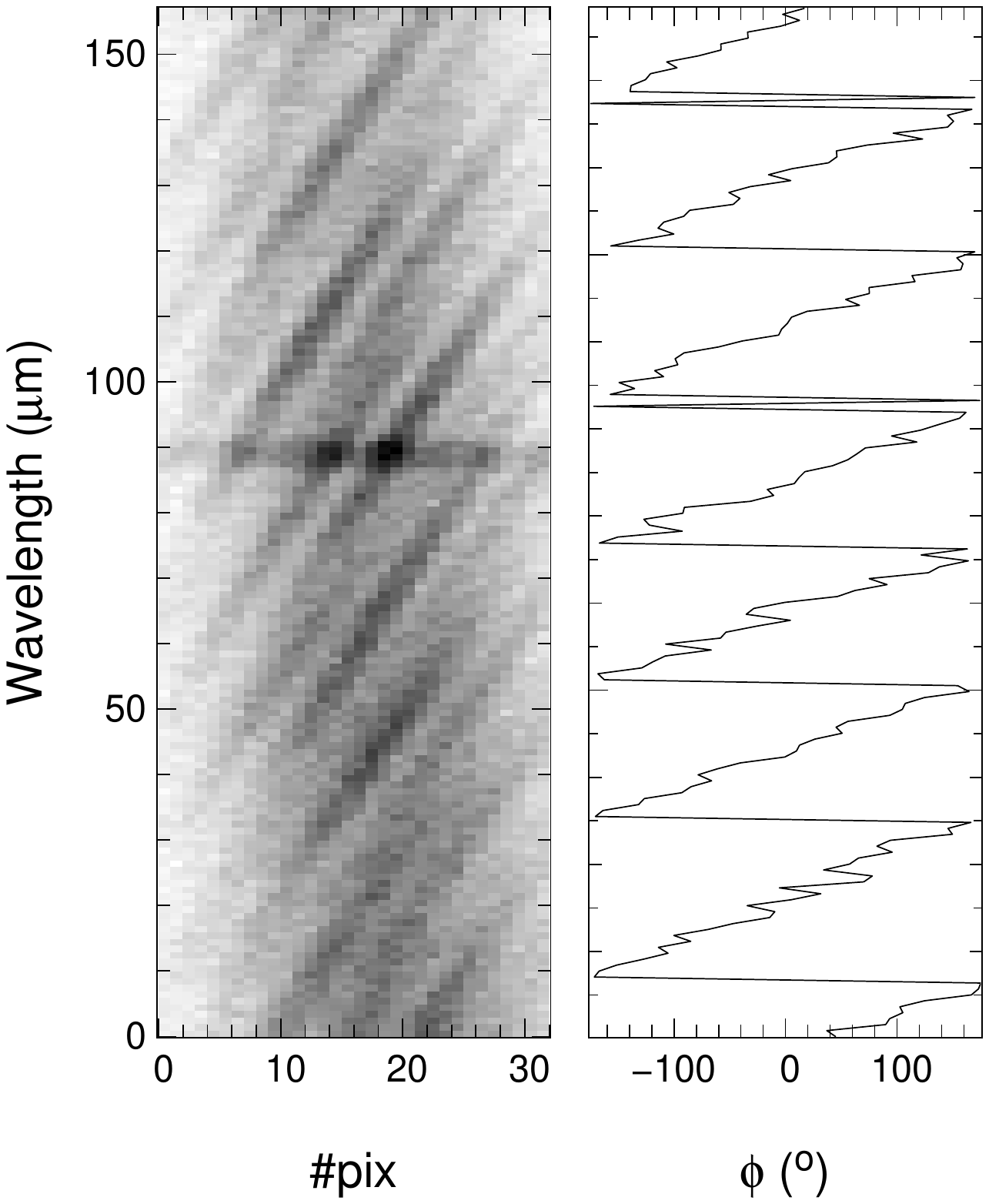}
\end{minipage}%
\hspace{.5cm}
\begin{minipage}[t]{.58\linewidth}
\vspace{0pt}
\centering
  \includegraphics[width=1\textwidth,origin=t,angle=0]{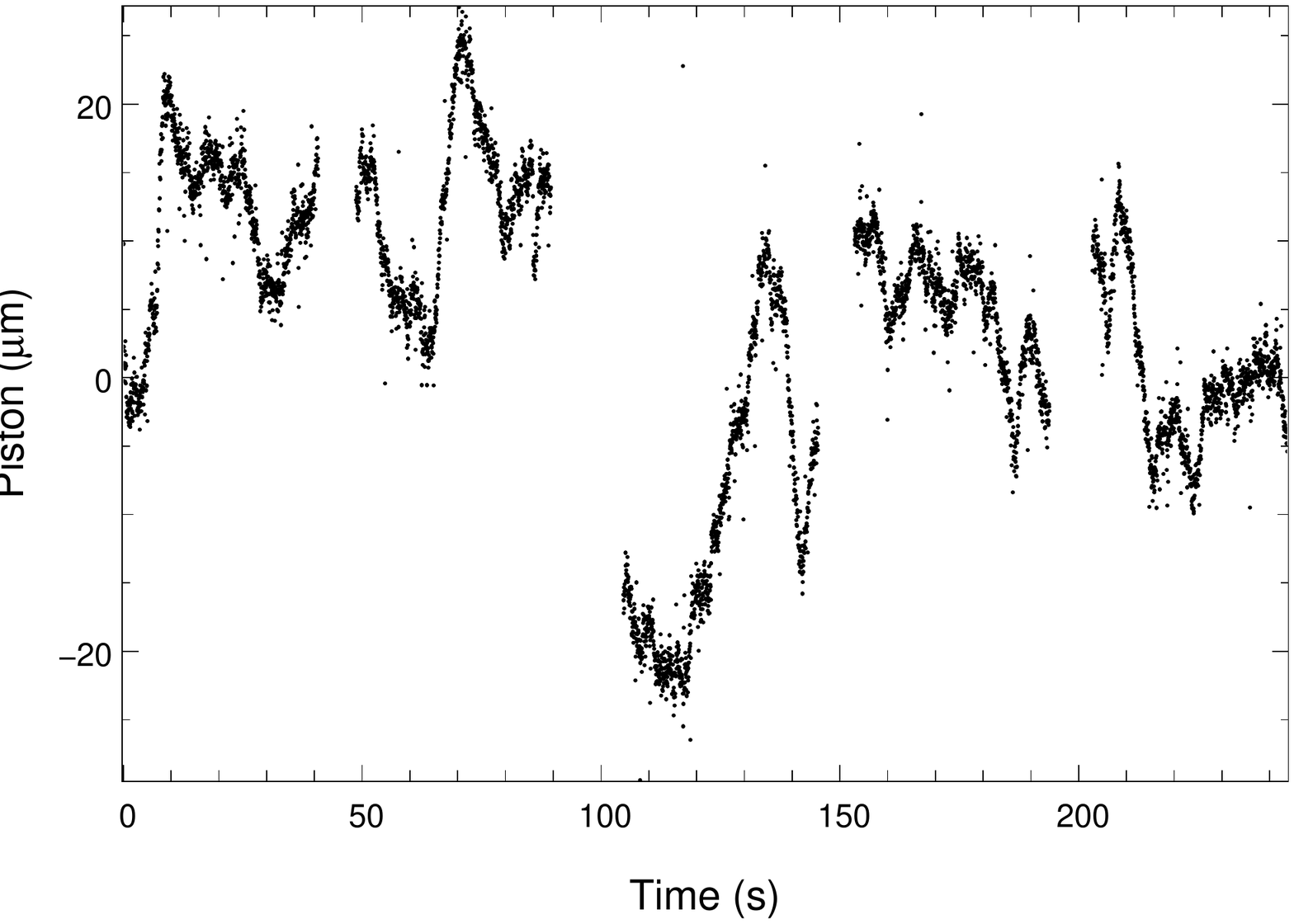}
\end{minipage}
\end{center}
\vspace{-3cm}
  \caption{{\bf From left to right:} example {\sc amber} spectrally dispersed
  fringes, the corresponding phase, and a time-sequence of
  OPD from Tatulli \etal\ (\cite{2007AandA...464...29T}). Reproduced with permission.} \label{fig:phaseAtm}
\end{figure}

So, we have a way to measure the complex visibility. However, the
actual measurement of the fringes is affected by a series of effects
we detail here, and we explain a set of workarounds on how to measure
the amplitude $\mu$ and phase $\phi$ of the object of interest. The
effects can be classified into visibility attenuation factors $A
\leq 1$ and phase factors $\phi$. The following list is ordered by
decreasing magnitude on the observables:

\begin{itemize}
\item The atmospheric turbulence adds a phase term $\phi^p$ between
  the telescopes, sometimes called ``atmospheric piston'' because it
  comes from a change in the optical path difference $\delta(t)$
  between the two telescopes. This term varies as $\phi^p_\lambda(t) =
  2 \pi \delta(t) / \lambda$ as a first approximation (see an
  example phase shape in Fig.~\ref{fig:phaseAtm}).
\item Atmosphere also puts higher-order terms (like the tip/tilt
  effect) which will affect the instantaneous flux $I_1$ and $I_2$. It
  can also affect other terms due to the speckle
  pattern (Mourard \etal\ \cite{1994AandA...288..675M}) but today most of the combiners
  use optical fibers to wash out this effect, so we will not detail it
  here.
\item A finite exposure time can also be of trouble, as it transforms
  the phase term variations ${\sigma_\phi}^p$ into contrast variations
  $A({\sigma_\phi}^p) = \me^{-\sigma^2_\phi}$
  (Tatulli \etal\ \cite{2007AandA...464...29T}).
\item Chromatic longitudinal dispersion makes the optical path delay
  $\delta(t)$ dependent of wavelength $\delta_\lambda(t)$. This effect
  is explained in details in Tubbs \etal\ (\cite{2004SPIE.5491..588T}) and Vannier \etal\ \cite{Vannier2006a}.\label{item:chromaticOPD}
\item Polarisation effects, either in the beam feeding or inside the
  instrument can make the contrast time-variable and even kill it. The
  contrast variation due to polarisation is noted:
  $A(\Delta_\pi)$. This is due to a difference of speed propagation
  between the two linear polarisations (birefringence effect) due to
  asymmetric setups or birefringent materials in the instruments
  (e.g. optical fibres). One can then extinct one of the polarisation
  to avoid this effect by using a linear polarizer, or an elegant
  solution is to introduce a birefringent plate with relevant
  properties to compensate for this effect (Lazareff \etal\ \cite{Lazareff2012}).
\end{itemize}

In addition to these effects, more fundamental effects like photon
noise $\sigma_\phi$ or detector noise $\sigma_{\rm det}$, grouped
in \emph{additive noises} $b$, need to be taken into account.  To
summarize, one can include all these effects into the interferometric
equation~\ref{eq:interferogramme} and consider an arbitrary number of
telescopes $Ntel$, which can be written as:

\begin{eqnarray}\nonumber
  I &=& \sum_{j=1}^{\rm Ntel} I_j \\\nonumber
&+& \sum_{k=
   1}^{\rm Ntel-1}\sum_{j=
    k+1}^{\rm Ntel} I_j I_k
  A({\sigma_\phi}^p) A(\Delta_\pi) A(\delta) \mu^{\rm obj}_{j,k} \cos \left(
  \frac{2 \pi}{\lambda}(x + \delta) + \phi^{\rm obj}_{j,k} \right)\\
&+& b
  \label{eq:interferogrammeComplet}
\end{eqnarray}

where $j,k$ are telescope indices; $x$ is a space coordinate;
$\lambda$ is the wavelength; $\mu^{\rm obj}_{j,k}$ and $\phi^{\rm obj}_{j,k}$ are
the object's visibility and phase; $\delta$ is the atmospheric optical
path difference (OPD), varying with time (see
Fig.~\ref{fig:phaseAtm}). All the beam intensity terms $I_j$ and $I_k$
depend on space $x$, wavelength $\lambda$, and time $t$;
$A({\sigma_\phi}^p)$ is an attenuation factor coming from the finite
exposure time of each frame, and depends on the atmospheric
conditions, or the fringe tracker performances; $A(\delta)$ is an
attenuation factor dependent of the spectral resolution of the
instrument and the value of the OPD $\delta$; $A(\Delta_{j,k})$ is an
attenuation factor depending on the polarization state of both
contributiong beams; finally, $b$ is a zero-mean noise.

The consequence of the phase problem is that the object phase cannot
be measured directly. The following section provides some guidelines
on how to cope with this fact.


\section{Observables}\label{sec:observables}

The goal of \emph{observables} is to extract the relevant information,
i.e. $\mu_o$ and $\phi_o$ from the somewhat perturbed
equation~\ref{eq:interferogrammeComplet} compared to
eq.~\ref{eq:interferogramme}.

\subsection{Squared visibility}

One way to minimise the turbulence effect of the phase on the
visibility is to take profit of the developments made for speckle
interferometry in the 70's
(Labeyrie \cite{1970AandA.....6...85L}). Indeed, the modulus of the
instantaneously-measured complex visibility can be computed and
averaged over time, minimising the effects of phases fluctuations. To
simplify the calculations, one can remove the square root of the
modulus, hence calculating squared modulus. With the Fourier method,
this gives:

\begin{equation}
  \widetilde{\mu^2} = \left\|\frac{N_{\rm bases}{\rm FT}_{f_{\rm
        i,j}}\left[I(x,\lambda) \right]}{{\rm
      FT}_{0}\left[I(x,\lambda) \right]} \right\|^2
\label{eq:squaredVis}
\end{equation}

Here one can see that most of the phase effects wipe out from this
visibility estimate, but there are a number of issues related to it:
\begin{itemize}
\item Additive noises can lead to biases (see how to treat them
      in Perrin \cite{2003AandA...400.1173P}, and also in this book:
      ten Brummelaar \cite{Brummelaar2015}),
\item The multiplicative terms $A({\sigma_\phi}^p)$, $A(\delta)$, and
      $A(\Delta_\pi)$ in eq.~\ref{eq:interferogrammeComplet} do not
      wipe out from the estimator, making calibration a very acute
      issue for this type of estimator (See for example
      Millour \etal\ \cite{Millour2008}).
\end{itemize}

In terms of error estimate, the squared visibility method has a
well-documented bibliography and I would suggest reading one of these
papers: Tatulli \etal\ (\cite{2007AandA...464...29T}), Petrov \etal\ (\cite{2007AandA...464....1P}), or ten
Brummelaar \cite{Brummelaar2015}).

\subsection{Differential visibility, coherent visibility}

Another way to minimize turbulence on the visibility measurement (the
terms $\frac{2 \pi}{\lambda}\delta$ and $A({\sigma_\phi}^p)$) is to
wipe it out explicitely.

In other words, as shown in Fig.~\ref{fig:phaseAtm}, the phase as a
function of $\lambda$ exhibits a ``slope'', directly related to
OPD. This provides us a way to measure $\widetilde{\delta}$.  The
coherent flux $\widetilde{\gamma_{i,j}(0)}$ of
eq.~\ref{eq:coherentdegree} can be corrected from the associated phase
term, and its real part averaged.

\begin{equation}
\widetilde{\mu} = \Re\left[ \widetilde{\gamma_{i,j}(0)} \times \exp^{-\frac{2 \i \pi}{\lambda}\widetilde{\delta}}\right]
\label{eq:linearVis}
\end{equation}

This estimator of visibility is calculated by the {\sc midi} pipeline
(Koehler \etal\ \cite{koehler2008}) and is planned to be implemented
in the {\sc matisse} instrument too. It is often called ``coherent
visibility'' or ``linear visibility''. Many aspects are not treated
here (like e.g. time-averaging) as these are pure signal processing
aspects (you can refer to Papoulis \cite{2002prvs.book.....P} to see
how time-averaging of the quotient of two random variables can be
done) and they would be too long to describe here.

To overcome the $A({\sigma_\phi}^p)$ term, supposedly
wavelength-invariant, it may be convenient to divide the above
estimate of visibility by its wavelength-average. This provides a
visibility measurement whose average value is 1 and whose variations
with respect to wavelengths are kept. It is then called ``differential
visibility'' as its variations are only relevant relative to a virtual
reference wavelength (also called ``reference channel'').

\begin{equation}
\widetilde{\mu}^{\rm diff} = \frac{\widetilde{\mu}}{<\widetilde{\mu}>_\lambda}
\label{eq:diffvis}
\end{equation}

Two flavors of differential visibility may be computed:
\begin{itemize}
\item normalizing the squared visibility of eq.~\ref{eq:squaredVis} by its wavelength-average.
\item normalizing the linear visibility of eq.~\ref{eq:linearVis} along $\lambda$.
\end{itemize}

The first method was used in the early times of {\sc amber}. The
second method is the one proposed today in the {\sc amber} data
reduction software, and also on {\sc vega}. It will be proposed also
for {\sc matisse}. I would suggest reading the papers Millour
(\cite{2006EAS....22..379M}) and Mourard et
al. (\cite{2009AandA...508.1073M}) for more information.

\subsection{Closure phase}

The phase $\phi^{\rm obj}_{j,k}$ of the object is usually considered as lost
when going through the atmosphere. However, a phase measurement out of
3 telescopes was invented for radio-astronomy
(Jennison \cite{1958MNRAS.118..276J}), called ``closure phase''. This
closure phase has extremely interesting properties in that it washes
out the atmospheric disturbances from the phases.

Let us call $1, 2, 3$ the three telescopes of interest. The object's
phases are, according to the ZVC theorem, linked with each baselines
$\phi_{1,2}^{\rm obj}$, $\phi_{2,3}^{\rm obj}$, and $\phi_{3,1}^{\rm
  obj}$. On the other hand, the atmospheric disturbances affect the
phase of the wavefront prior each \emph{telescope}, hence the
atmospheric phases can write $\phi^{\rm atm}_1$ $\phi^{\rm atm}_2$
$\phi^{\rm atm}_3$. Baseline-wise, the phases can be expressed then:

\begin{eqnarray}
\Phi_{1,2} & = & \phi_{1,2}^{\rm obj} + \phi^{\rm atm}_2 - \phi^{\rm atm}_1 \\
\Phi_{2,3} & = & \phi_{2,3}^{\rm obj} + \phi^{\rm atm}_3 - \phi^{\rm atm}_2 \\
\Phi_{3,1} & = & \phi_{3,1}^{\rm obj} + \phi^{\rm atm}_3 - \phi^{\rm atm}_1 \\
\end{eqnarray}

When summing the phases from the three baselines, the atmospheric
disturbances disappear and the summed phase becomes:

\begin{equation}
\Psi_{1,2,3} = \phi_{1,2}^{\rm obj} + \phi_{2,3}^{\rm obj} + \phi_{3,1}^{\rm obj}
\label{eq:computeclosure}
\end{equation}

\begin{figure}[htbp]
\vspace{-4cm}
  \centering
 \includegraphics[width=1.2\textwidth,angle=-0]{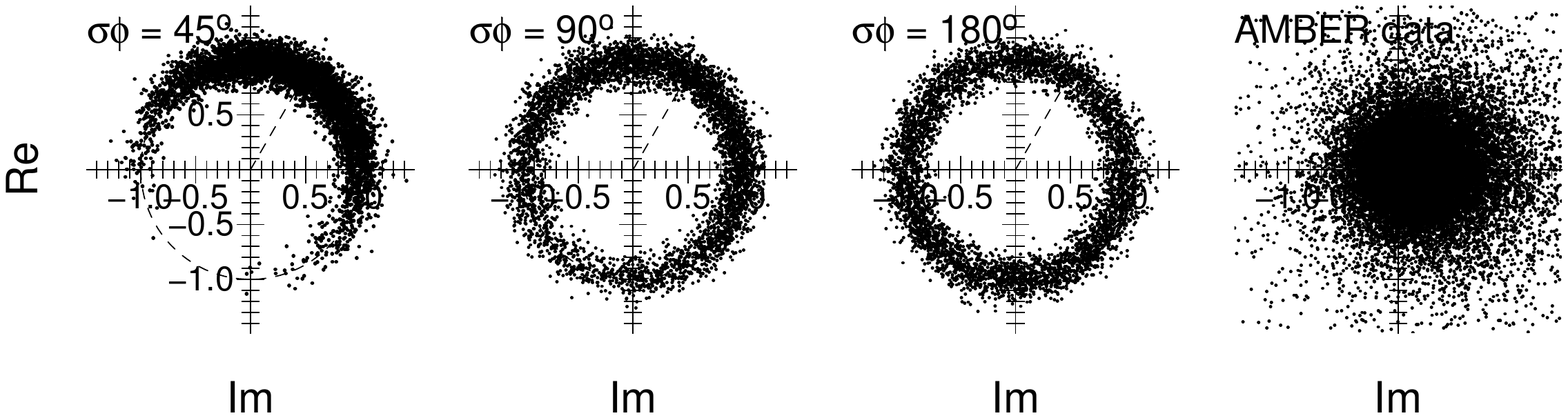}\\
\vspace{-7cm}
 \includegraphics[width=1.2\textwidth,angle=-0]{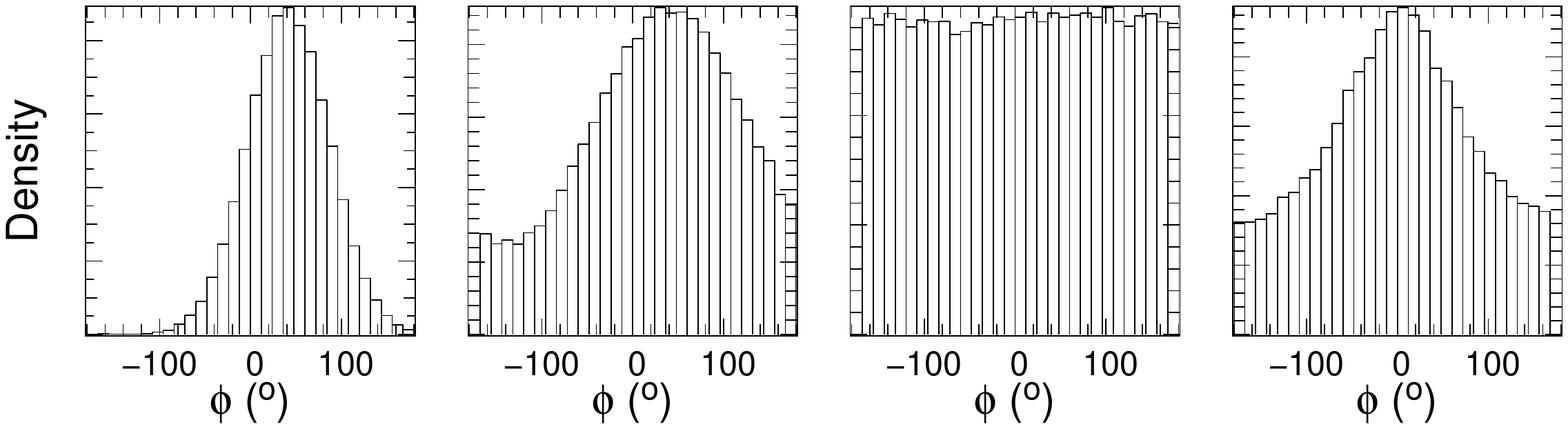} 
\vspace{-4cm}
\caption{Illustration
 of the phase wrapping effect, showing the transition of phases
 histograms from almost Gaussian to a nearly uniform distribution. The
 top part shows the real and imaginary parts of the complex values,
 while the lower part shows the corresponding phase histograms. The
 three first panels are synthetic data, while the last one is actual
 very high signal-to-noise ratio {\sc amber}
 data.}  \label{fig:phasewrapping}
\end{figure}

This is called the closure phase. However, one cannot simply sum the
phases, as the phase noise is usually very large compared to $60^{o}$
(or 1 rad). The phases histogram can therefore be very far from a
pristine Gaussian distribution, often close to an uniform
distribution. The corresponding phase-wrapping effect is illustrated
in Fig.~\ref{fig:phasewrapping}.

The way to compute it as an observable is to compute the
\emph{bispectrum} directly in the complex plane, average it
and then take the phase:

\begin{equation}
\widetilde{\Psi_{1,2,3}} = \arg{\left<{\rm FT}_{f_{\rm
      1,2}}\left[I(x,\lambda) \right] \times {\rm FT}_{f_{\rm
      2,3}}\left[I(x,\lambda) \right] \times {\rm FT}_{f_{\rm
      3,1}}\left[I(x,\lambda) \right]\right>_t}
\end{equation}

When the three visibilities on the three baselines have similar
values, the closure phase noise is just $\sqrt{3}$ higher than the
true phase noise. However, this is not true at all when the three
baselines have very different visibility amplitudes (for example when
one baseline is exactly in a zero of visibility). In such a
case, the closure phase noise is approximately equal to the highest
phase noise from the three baselines. One can refer to
Chelli \etal\ (\cite{Chelli2009b}) for more information.

For details on what means the closure phase, and how to interpret it,
please read section~\ref{sect:interpreting_data}.

\subsection{Differential phase}

\subsubsection{How to get it}

``Differential phase'' can mean many things: the phase difference
between the two telescopes (what has been called ``phase'' in this
paper), the phase difference between two offset positions (also
referred as ``phase referencing''), the phase difference between two
separated beams (Aime \etal\ \cite{1986JOSAA...3.1001A}), or the phase
difference between two adjacent wavelengths, introduced by Beckers
(\cite{1982AcOpt..29..361B}). We will not talk about the three first
phases, but explain here the last one which had quite some
applications in the last decade for optical interferometry.

If we take a look to the phase term in
eq.\ref{eq:interferogrammeComplet}, we note it is made primarily of
two components:

\begin{equation}
\phi_{i,j}(\lambda) = \phi^{\rm obj}_{i,j}(\lambda) + \frac{2 \pi
  \delta(t)}{\lambda} + o\left(\frac{1}{\lambda}\right)
\end{equation}

These two components are the object's phase $\phi^{\rm
obj}_{i,j}(\lambda)$, basically fixed as a function of time (if there
is no baseline smearing) but possibly varying as a function of
wavelength (for example inside an emission line, or as a function of
spatial frequency), and the OPD term $\frac{2 \pi \delta(t)}{\lambda}$
strongly varying as a function of time. All higher-order terms (like
the water vapor term) are contained within the term
$o(\frac{1}{\lambda})$.

If one is able to estimate properly the OPD term, it can be subtracted
from the individual phase measurements, and the remaining phase
variations as a function of wavelength can be averaged. To avoid
wrapping effects in presence of noise, this procedure must be done in
the complex plane, by computing a cross-spectrum:

\begin{equation}
W^{\rm no P} = \widetilde{\gamma_{i,j}(0)} \times \me^{-2\i\pi\frac{\delta(t)}{\lambda}}
\label{eq:removeOPD}
\end{equation}

We note here that we remove only the achromatic OPD effect, but as
mentionned in page~\pageref{item:chromaticOPD}, other effects can also
affect the phase. They are present essentially as a phase offset plus
higher order terms. The phase offset can be removed by computing
phase differences between the current wavelength (called ``work
channel'') and another wavelength (called ``reference channel'',
bearing many similarities to the one used for differential
visibility), i.e.:

\begin{equation}
\widetilde{\phi^{\rm diff}} = \arg \left< W^{\rm no P}(\lambda_{\rm
  work}) W^{\rm no P}(\lambda_{\rm ref}) \right>_t
\label{eq:interspectrum}
\end{equation}

One can note here that equations~\ref{eq:removeOPD} and
\ref{eq:interspectrum} can be swapped in the process without any
issue.

The reference channel can be computed by taking one wavelength, or by
averaging several wavelengths. The most used method is to compute the
reference channel with all but one of the wavelengths (to avoid
introducing a quadratic bias). The differential phase can be related
to the object phase according to the following equation:

\begin{equation}
\widetilde{\phi^{\rm obj}_{i,j}} = \frac{N_\lambda-1}{N_\lambda} \phi^{\rm diff}_{i,j} + \frac{\alpha}{\lambda} + \beta
\label{eq:phiobjphidiff}
\end{equation}

where $N_\lambda$ is the number of wavelengths in the reference
channel, and ``1'' is the one wavelength in the work channel. The
small multiplicative factor $\frac{N_\lambda-1}{N_\lambda}$ has to be
taken into account due to the definition of the reference channel, as
detailed in Millour (\cite{2006PhDT........46M}) page 91. Usually, this
factor is negligible, as most today spectro-interferometric
instruments have 100's of channels, but it may be large-enough for
wideband instruments to be considered. The two terms $\alpha$ and
$\beta$ are lost in the process, but may be recovered by using the
self-calibration method (see Sect.~\ref{sect:selfcal}).

\section{Interpreting interferometric data}
\label{sect:interpreting_data}

What has been described in the previous sections is how to obtain
interferometric data, but no word has been written about how to
interpret these data. The goal of this section is to see how to
compare interferometric measurements with a model, the image
reconstruction part being treated in a separated article (Young \&
Thi\'ebaut \cite{Young2015}, this book). We divide this section in two
sub-sections: qualitative interpretation of data (``first sight''
interpretation) and quantitative interpretation, with a few examples
of implementations.

\subsection{First-sight interpretation}

\subsubsection{Description of observables}

Interferometer data usually come as Optical Interferometry Flexible
Image Transport System (OIFITS) data files, which are based on the
FITS format. For more information on the OIFITS specifications, see
Pauls \etal\ (\cite{2004SPIE.5491.1231P,2005PASP..117.1255P}). Most of the
time, an OIFITS file contains squared visibilities, plus optionally
closure phases and/or differential visibilities and differential
phases, depending on the instrument used. 

These different observables provide already some information on the
object. Table \ref{tab:aveclesmains} presents some simple examples of
the considerations you can make at ``first sight''. For example, a
visibility close to 1 is measured with a 130\,m baseline in the
near-infrared for an object much smaller than 2\,mas, i.e.  smaller
than the angular resolution
$\theta_{\mathrm{res}}\simeq\lambda/B$. 

More generally,

\begin{description}
\item[The visibility] value indicates whether the object is resolved
  (low visibility) or not (visibility close to 1).
\item[The differential visibility] is a relative measurement of the object
     size along wavelengths. A lower differential visibility in an
     emission line indicates an emitting region larger than in the
     continuum.
\item[The phase] is sensitive to astrometric position of a given
     source. As such, it provides both information on the asymmetry
     (skewness of the intensity distribution) of the object and its
     photocenter (astrometry).
\item[The closure phase] gives the information whether the object is
     asymmetric (skewness): a zero or $\pi$ closure phase may
     correspond to a symmetric object (but not in 100\% of the cases),
     whereas a non-zero closure phase (modulo $\pi$) indicates for
     sure an asymmetric object. The astrometric position is lost in
     the closure phase signal.
\item[The differential phase] is sensitive to the astrometric position
     at one wavelength relative to another wavelength. Therefore,
     astrometric shifts in emission lines can be measured with it, or,
     if a large chunk of wavelength is available, it allows one to
     scan phase across spatial frequencies for an achromatic object
     (like a binary star for example). It contains also the
     information of the closure phase wavelength-variations, the only
     relevant information coming from closure phase being then its
     wavelength-average value.
\end{description}

A few example of qualitative features seen on the above observables
and their signification are provided in Table~\ref{tab:aveclesmains}.

\begin{sidewaystable}
\caption{Qualitative information which can be retrieved from
  interferometric observables.}\label{tab:aveclesmains} \centering
  \begin{tabular}{||c|c|c|c|c||}
    \hline
    \hline
    Observable & Value or features & Qualitative information & Model-fitting guidelines & Illustration \\
     &  &  &  &  Table~\ref{tab:zoologie_observables}\\
    \hline
    Visibility & Close to 0 & Object $\diameter\gg1.22 \lambda/B$ rad & Add a resolved component & (a) \\
    & Close to 1 & Object  $\diameter\ll\lambda/B$ rad & Uniform disk size & (e) \\
    & Cosine shape & Binary star! & Use a binary model & (i) \\
    \hline
    Diff. vis. & $\searrow$ in an emission line & Bigger emission & Add an emitting envelope & (c) \\
    & $\nearrow$ in an emission line & Smaller emission & Add an inner region/disk  & (g) \\
    & $\searrow$ in an absorption line & Central object in absorption & Absorbing central object & \\
    & $\nearrow$ in an absorption line & Shell in absorption & Absorbing shell  & \\
    \hline
  Clos. phase & $\neq0^{\circ}$ and $180^{\circ}$ & Asymmetric object & Add a point source & (f) \\
     & $=0^{\circ}$ or $180^{\circ}$ & Object possibly symmetric & - & (b) \\
    \hline
    Diff. phase & Sine shape & Binary star! & Use a binary model & (k) \\
     & ``S'' shape in a line & Object is rotating! & Use a kinematic model & (h) \\
     & ``V'' shape in a line & Asymmetric object in line & See Weigelt \etal\ (\protect\cite{2007AandA...464...87W}) & (d) \\
     & ``W'' shape in a line & A bipolar outflow? & See Chesneau \etal\ (\cite{Chesneau2011})  & (j) \\
    \hline
    \hline
  \end{tabular}
\end{sidewaystable}

\begin{sidewaystable}
\caption{The optical long-baseline interferometric ``observables zoo'', showing all the different cases one can face with current spectro-interferometric instruments, illustrated with actual published interferometric data. The letters link to the Table~\ref{tab:aveclesmains}. Reproduced with permission.}

\label{tab:zoologie_observables} \centering
  \begin{tabular}{||c|c|c|c||}
 \hline
 \hline
 Visibility & Closure phase & Diff. Vis. & Diff. Phase \\
 \hline
(a) Resolved source & (b) Zero & (c) $\searrow$ in emission line & (d) ``V'' shape\\
Benisty \etal\ (\cite{2010AandA...511A..74B}) & Monnier \etal\ (\cite{2006ApJ...647..444M}) & Stee \etal\ (\cite{2012AandA...545A..59S}) & Weigelt \etal\ (\cite{2007AandA...464...87W}) \\
\includegraphics[width=0.25\textwidth,height=0.25\textheight,angle=0,origin=c]{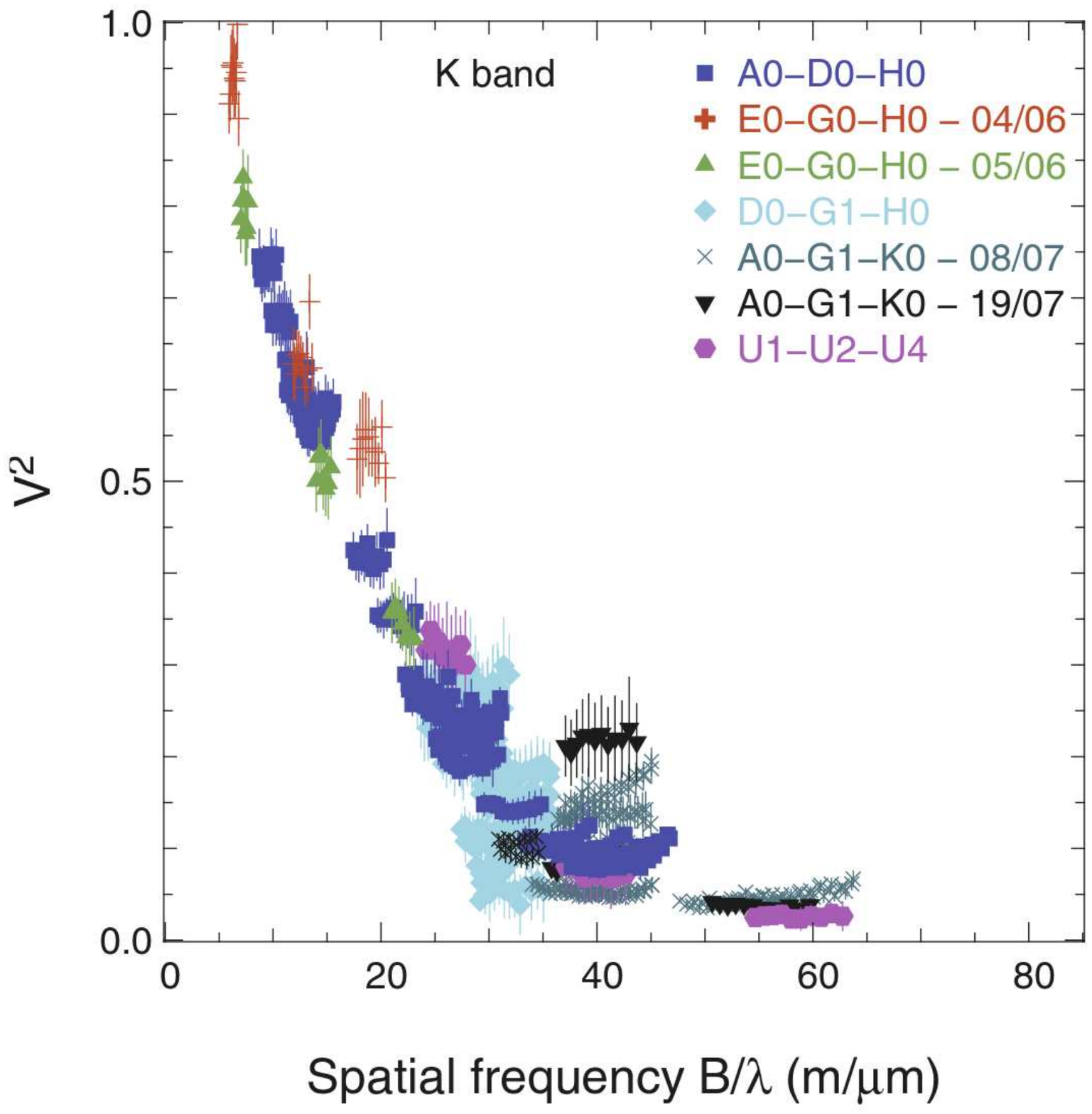} &
\hspace{-1cm}\includegraphics[width=0.3\textwidth,height=0.3\textheight,angle=0,origin=c]{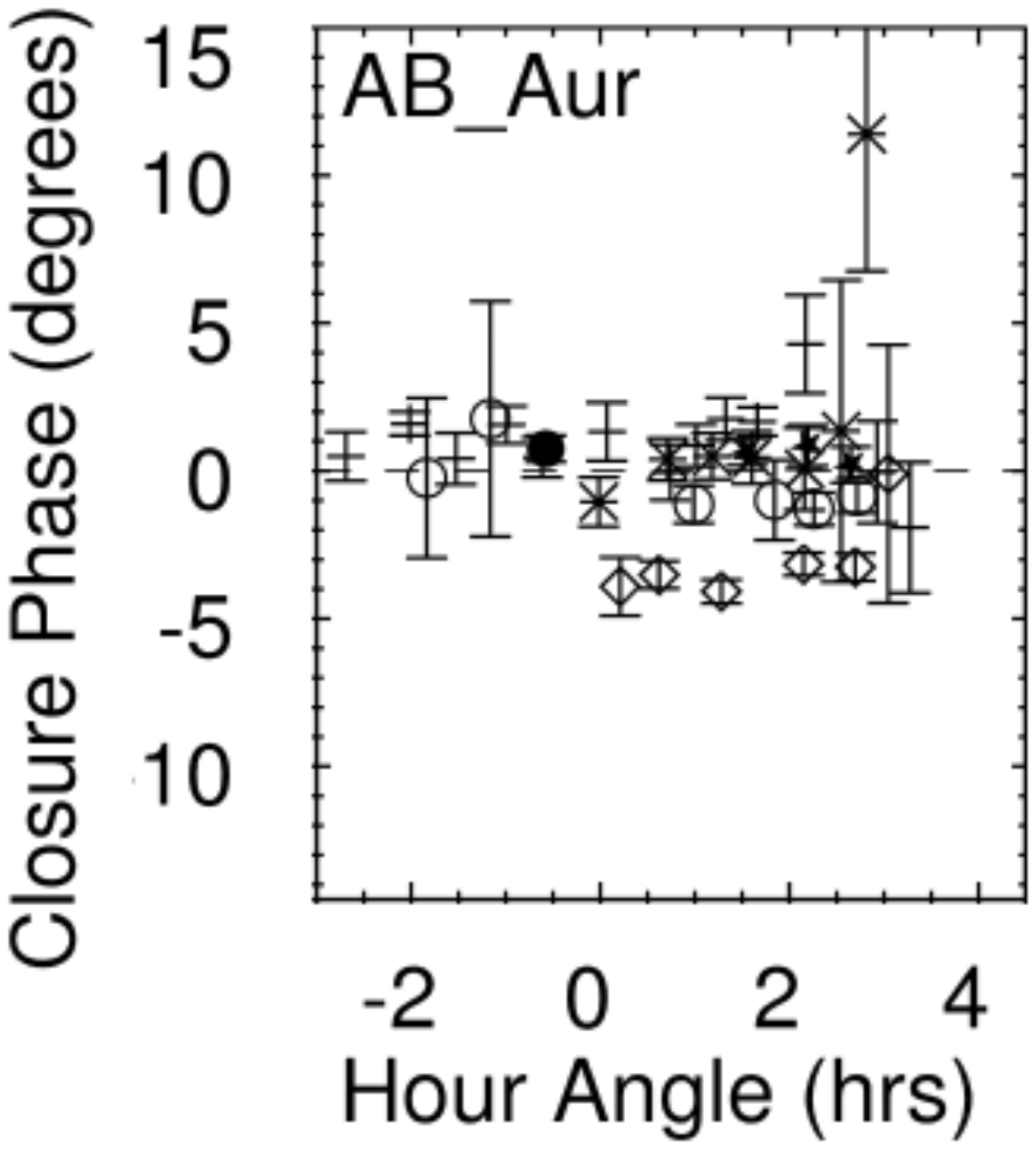} &
\hspace{-1cm}\includegraphics[height=0.25\textwidth,width=0.25\textheight,angle=-0,origin=c]{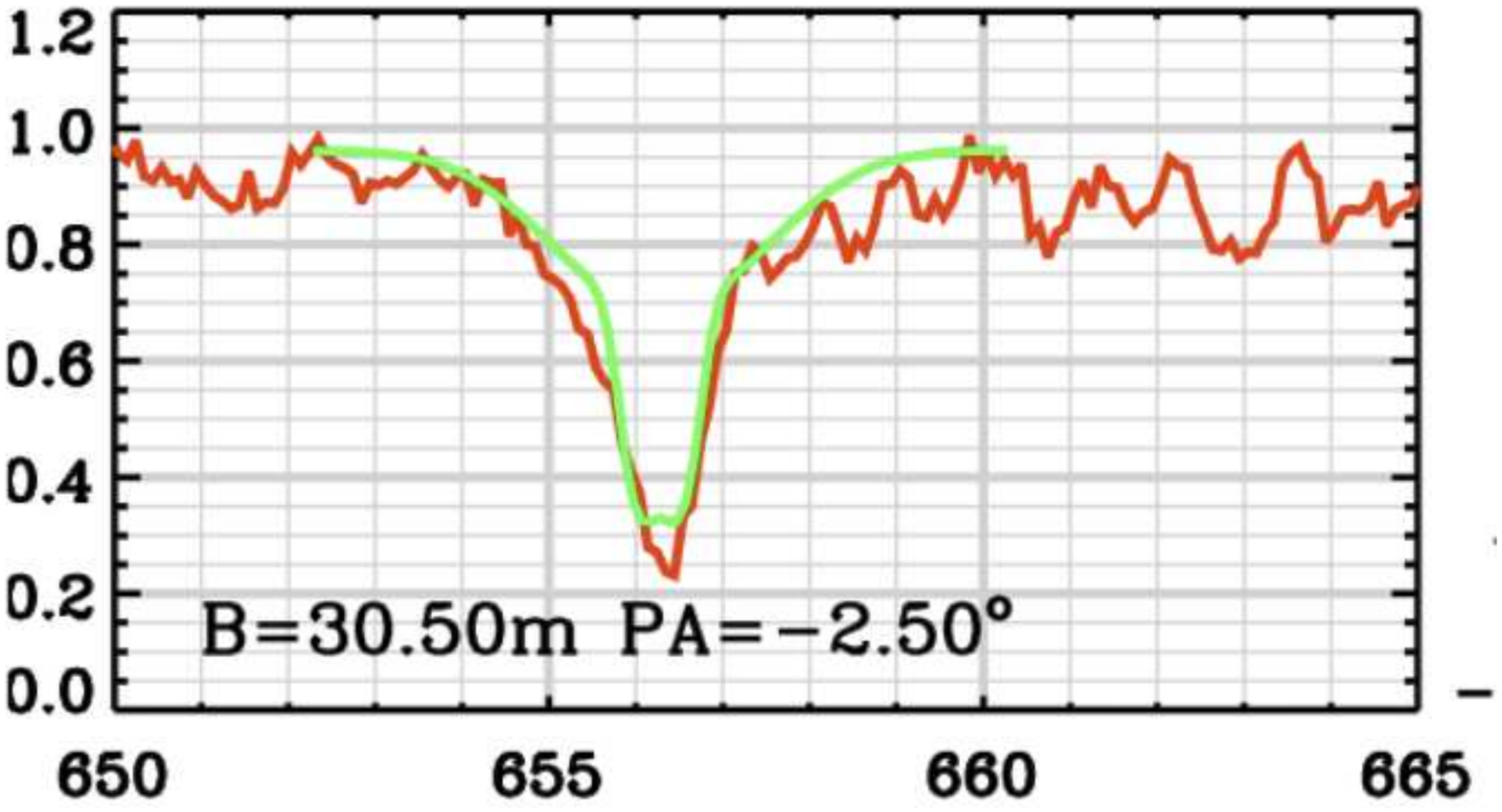} &
\hspace{-1cm}\includegraphics[width=0.2\textwidth,height=0.3\textheight,angle=0,origin=c]{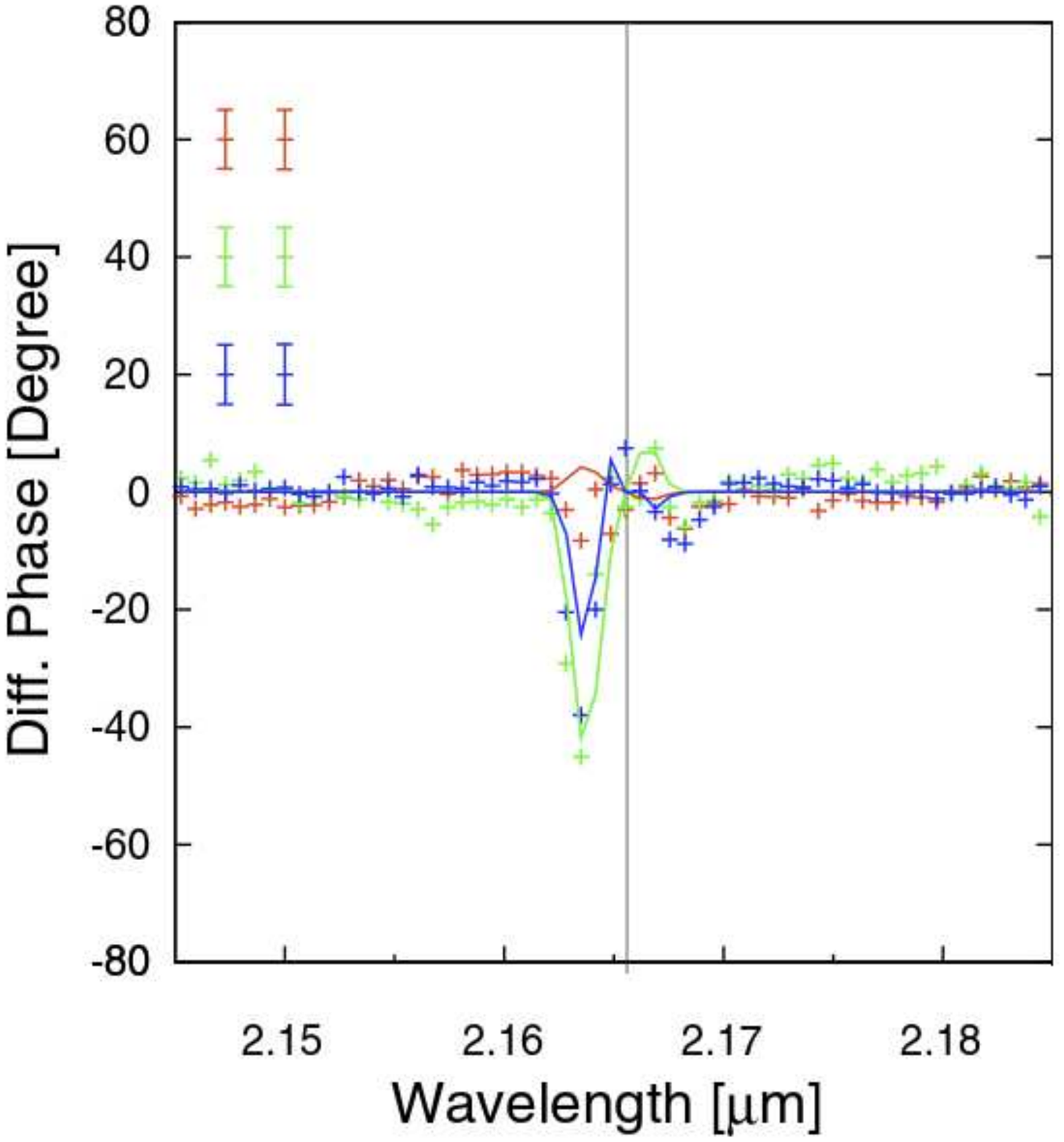} \\
(e) Unresolved source & (f) Non-zero & (g) $\nearrow$  in emission line & (h) ``S'' shape\\
Demory \etal\ (\cite{2009AandA...505..205D}) & Monnier \etal\ (\cite{2006ApJ...647..444M}) & Kraus \etal\ (\cite{2008AandA...489.1157K}) & Meilland \etal\ (\cite{2012AandA...538A.110M}) \\
 \includegraphics[width=0.25\textwidth,angle=-0,origin=c]{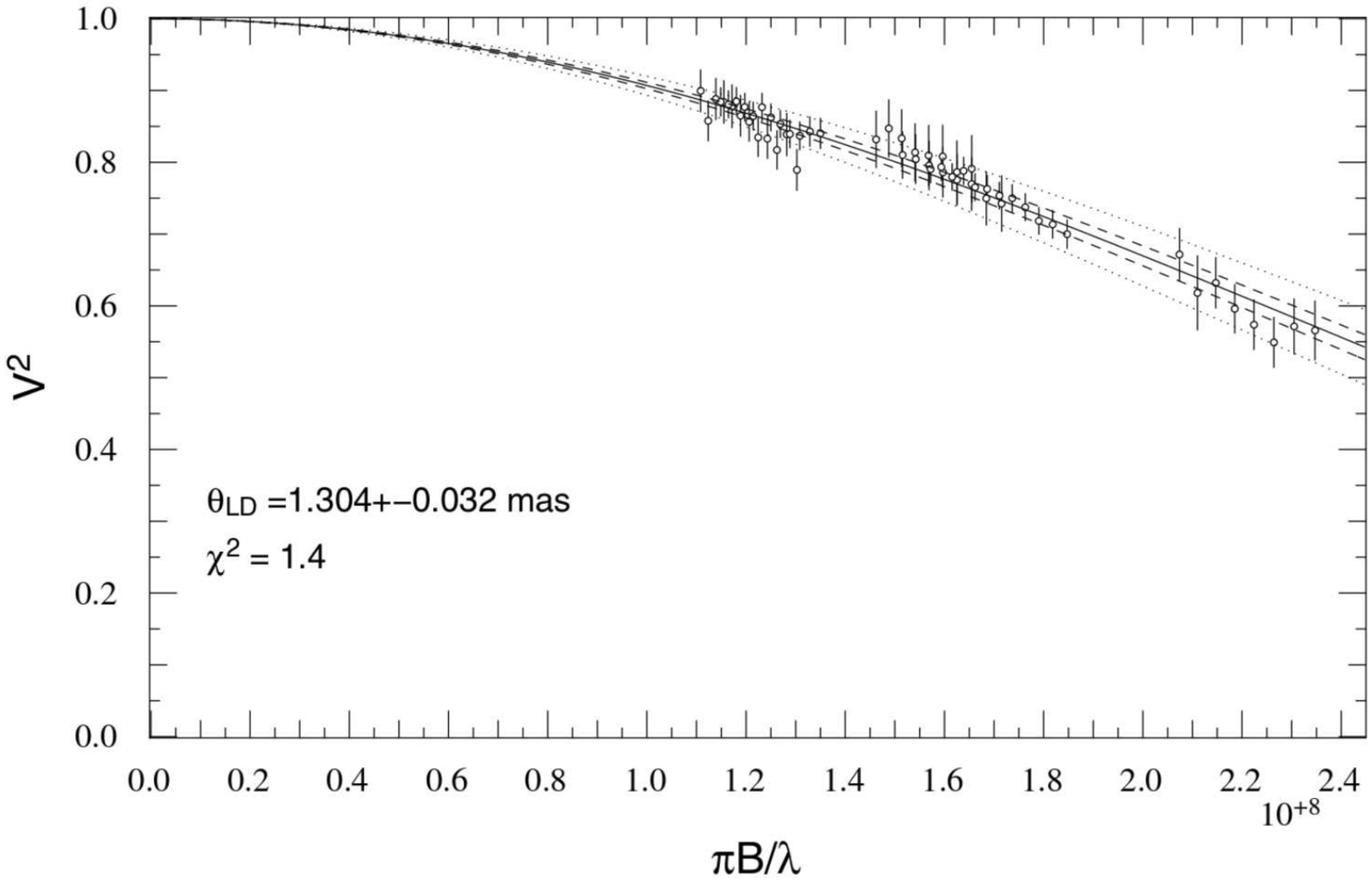} &
\hspace{-0.5cm}\includegraphics[width=0.25\textwidth,height=0.25\textheight,angle=0,origin=c]{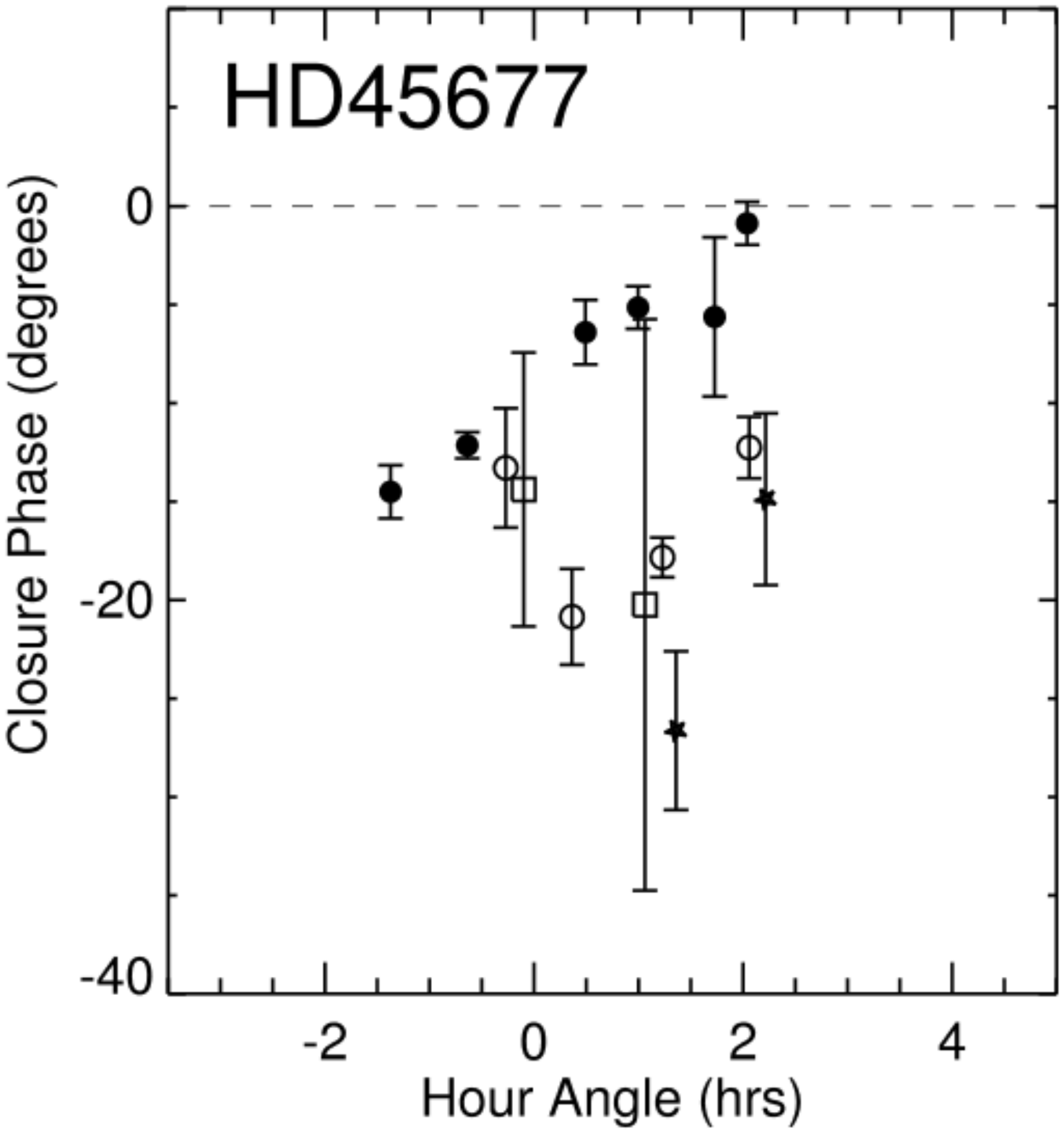} & 
\hspace{-1cm}\includegraphics[width=0.25\textwidth,height=0.25\textheight,angle=-0,origin=c]{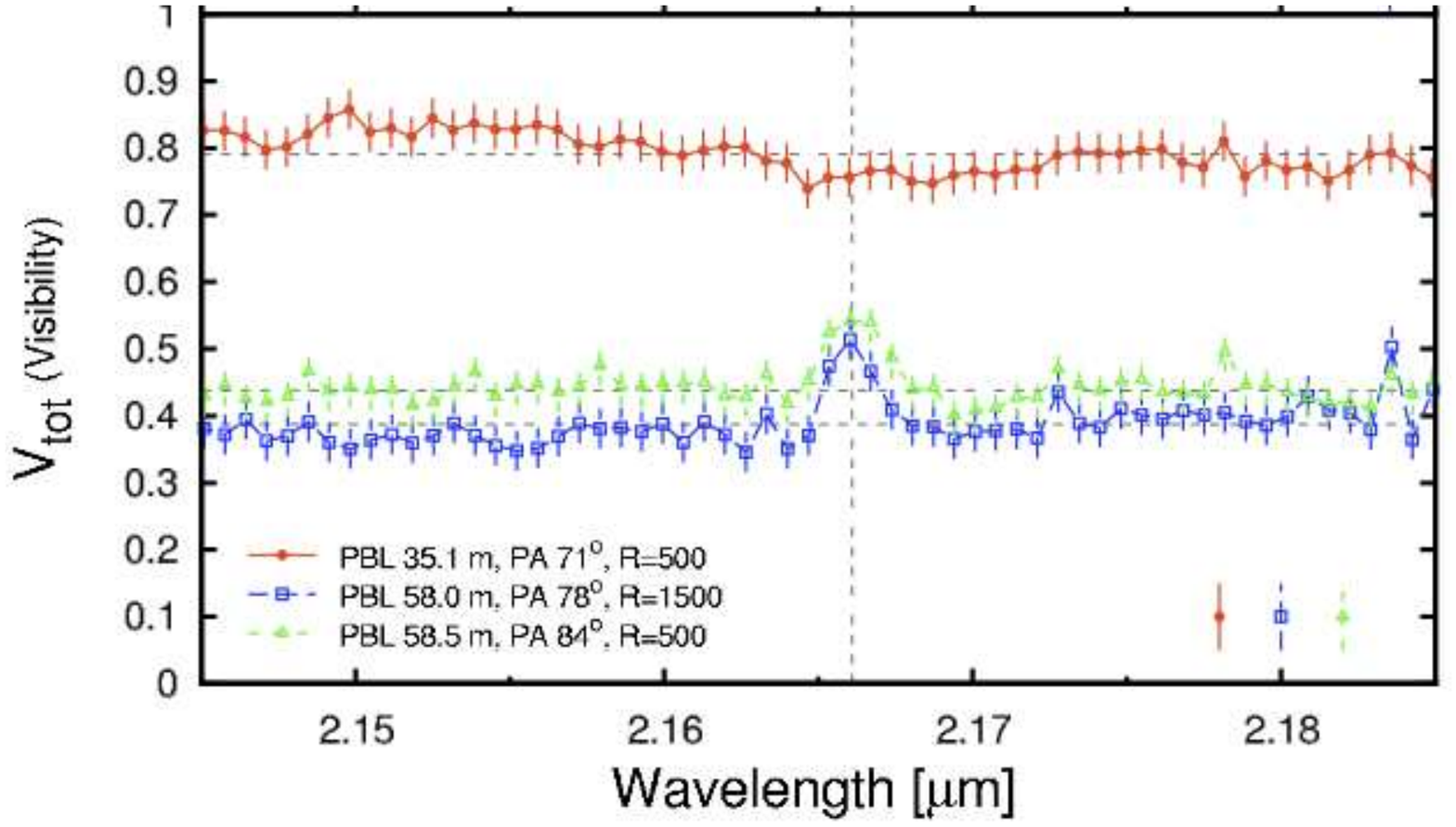} &
\hspace{-1cm}\includegraphics[width=0.3\textwidth,height=0.2\textheight,angle=-0,origin=c]{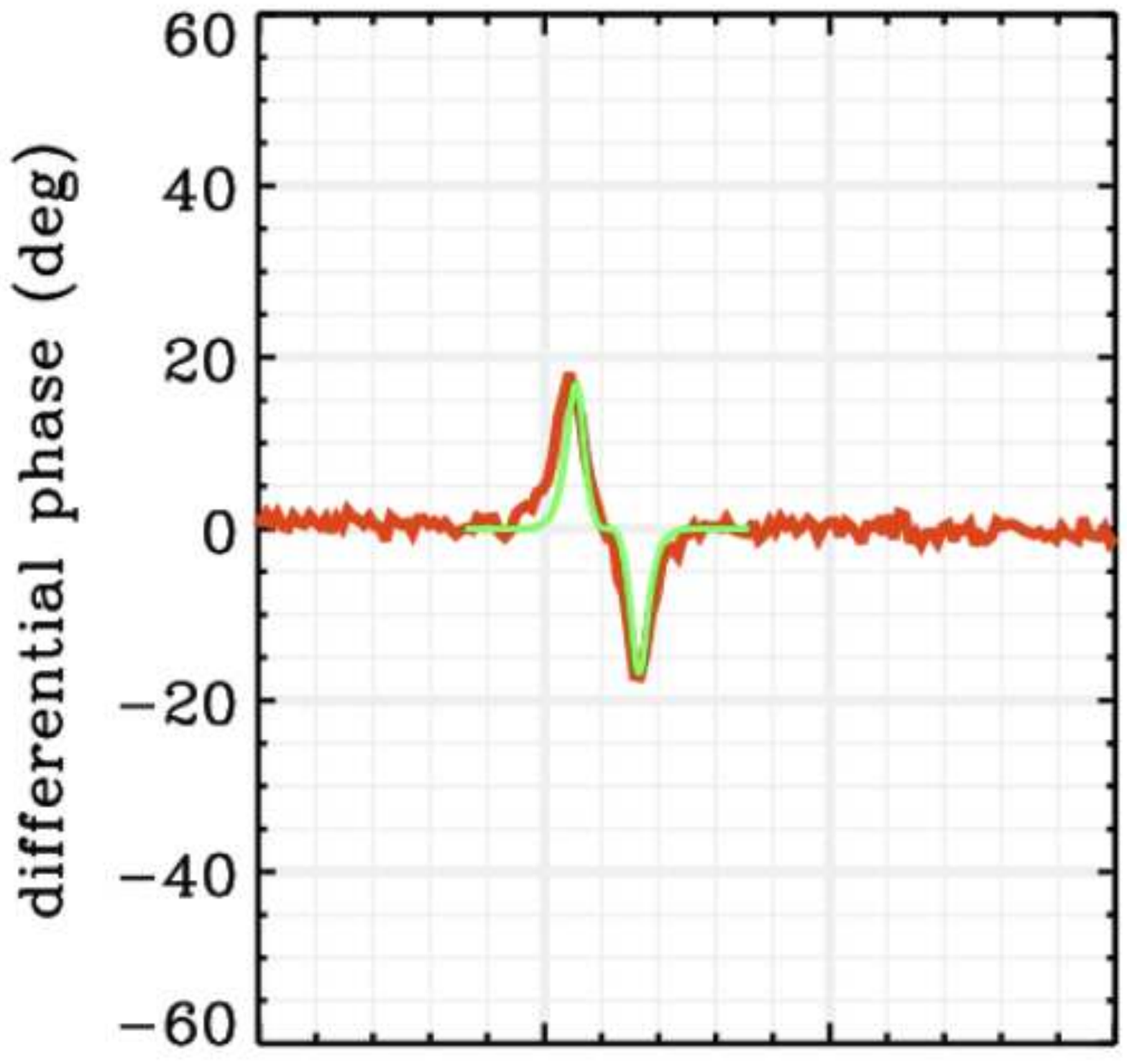} \\
\cline{3-3}
(i) Cosine shape &  &  \multicolumn{1}{c}{(j) ``W'' shape} & (k) Sine shape\\
Meilland \etal\ (\cite{2011AandA...532A..80M}) & & \multicolumn{1}{c}{Chesneau \etal\ (\cite{Chesneau2011})} & Meilland \etal\ (\cite{2011AandA...532A..80M})  \\
\includegraphics[width=0.25\textwidth,angle=-0]{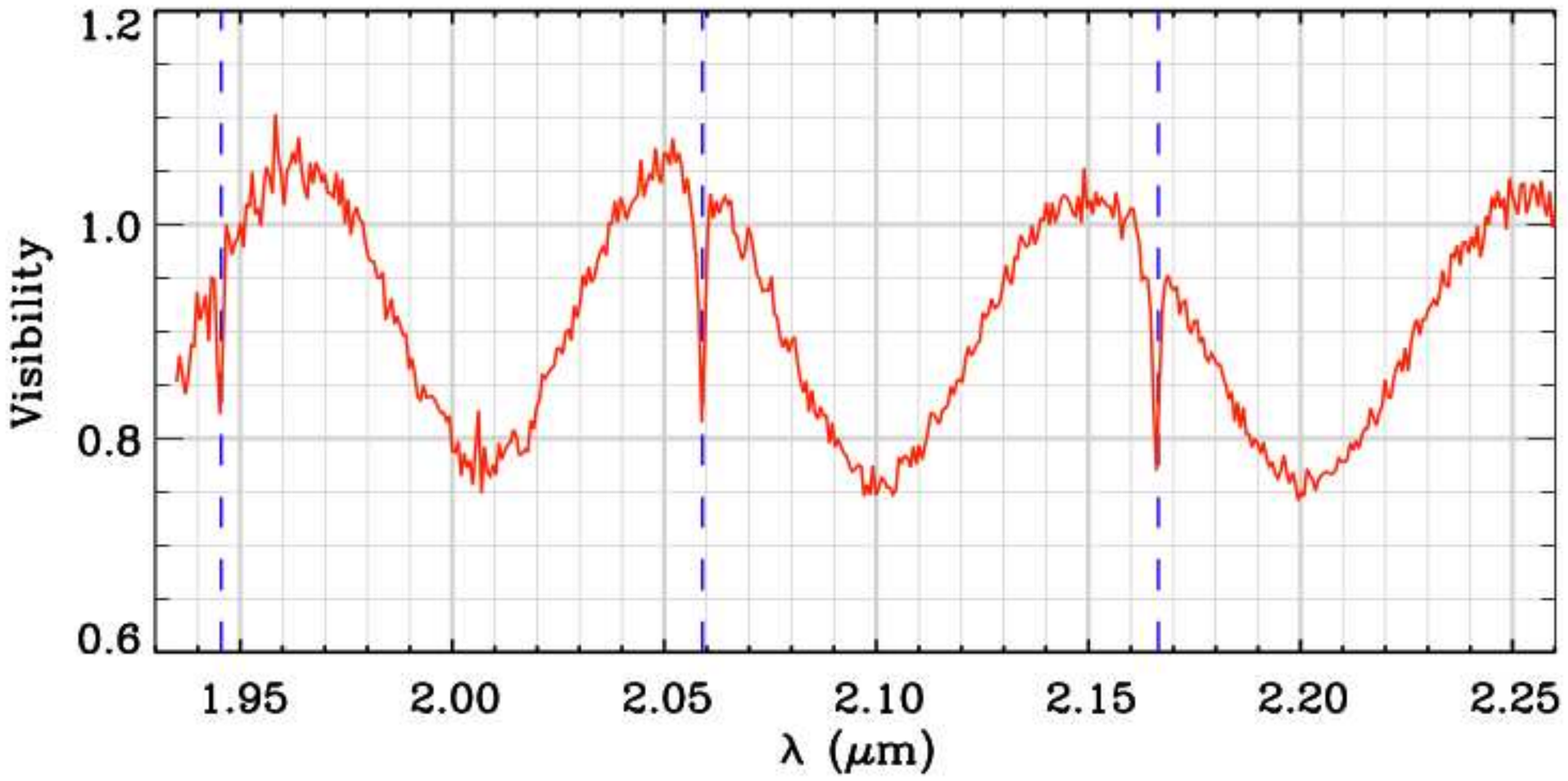} & & 
\multicolumn{1}{c}{\hspace{-1cm}\includegraphics[width=0.3\textwidth,height=0.2\textheight,angle=0]{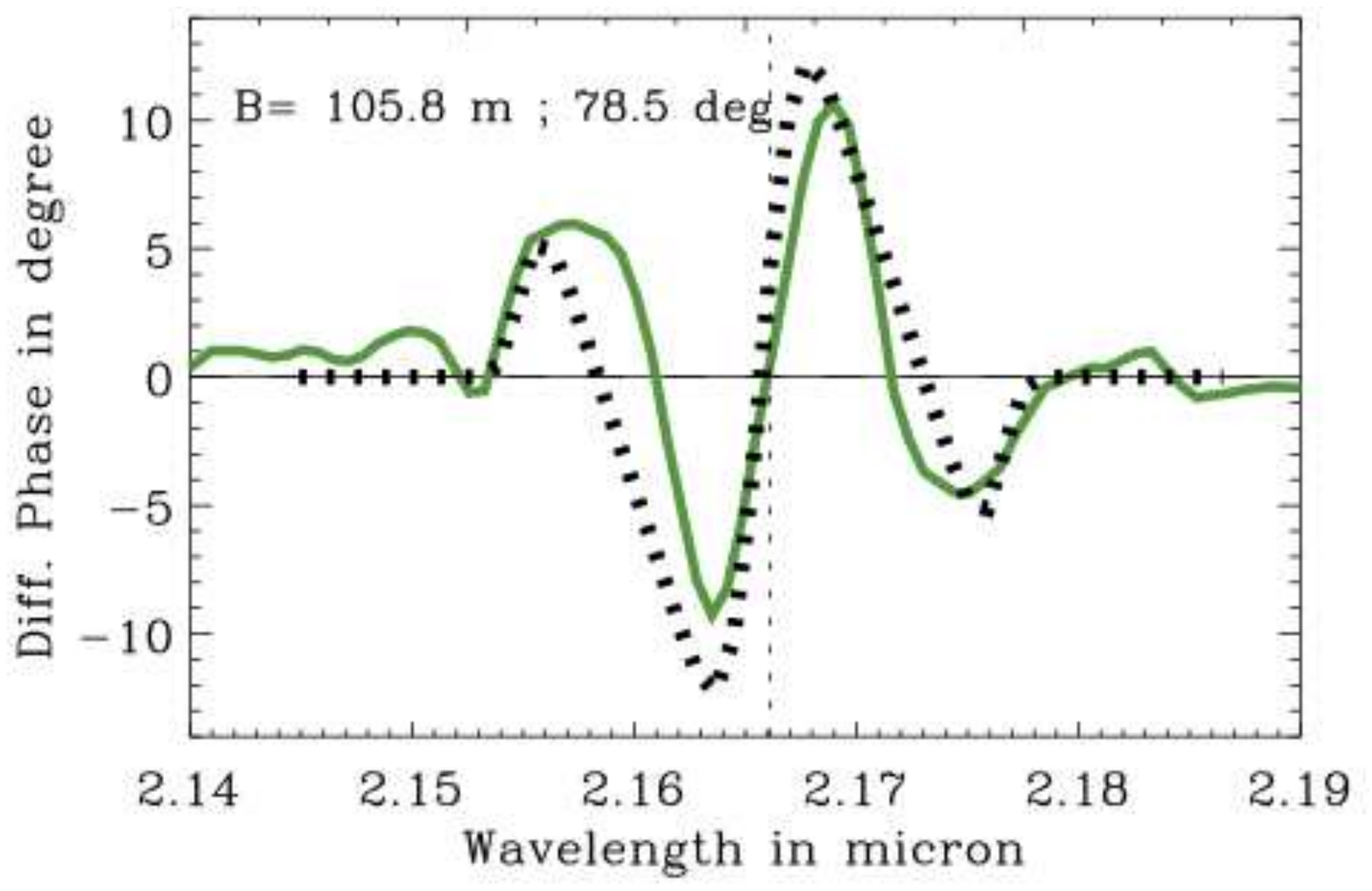}} &
\hspace{-1cm}\includegraphics[width=0.25\textwidth,angle=-0]{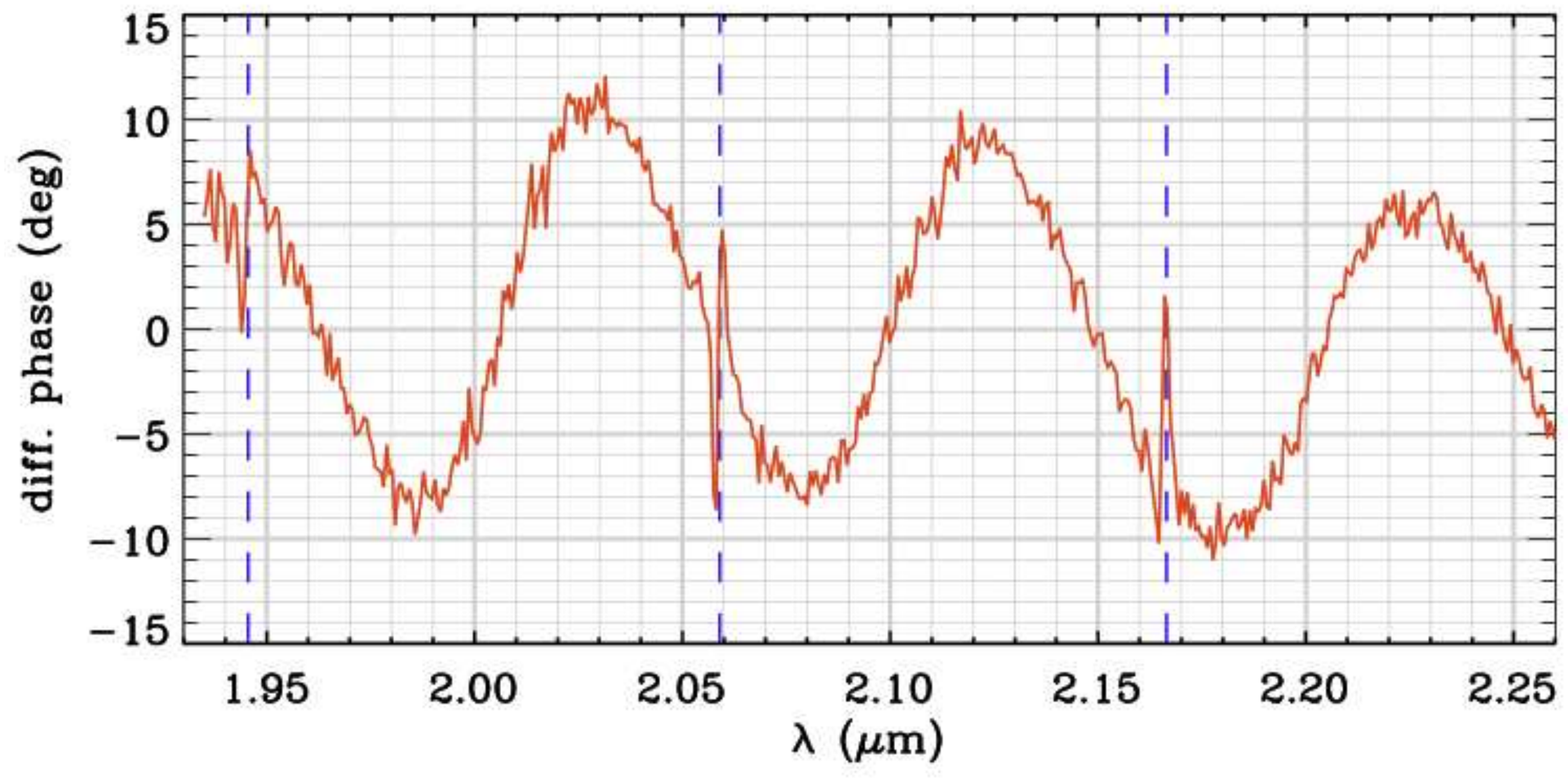}\\
 \hline 
\hline
 \end{tabular}
\end{sidewaystable}

\subsubsection{What can I do without a model?}

Already a significant amount of things! The visibility, closure phase
and differential phase provide already a large number of information
about the geometry of an object.

In practice, two quantitative information can be extracted from the
visibilities and differential phases without a model:

\begin{itemize}
\item Size estimate with the visibilities (if the visibility is larger
      than 20\%), using simple ad-hoc models (like uniform disks:
      Demory \etal\ \cite{2009AandA...505..205D}), Gaussian disks:
      Tristram \etal\ \cite{2007AandA...474..837T}, or even rings:
      Kishimoto \etal\ \cite{2011AandA...527A.121K},
\item Photocentre $p$ variations with the differential phase, or
      spectro-astrometry, using the simple relation $\phi = 2 \pi p B
      / \lambda$ when an object is unresolved.
\end{itemize}

However, one needs to bear in mind that these quantitative estimates
are valid \emph{only} when the object is barely resolved. For more
details, please read Lachaume (\cite{2003AandA...400..795L}). In all
other cases, on needs to use a model-fitting tool like \texttt{LITpro}
(Tallon-Bosc \etal\ \cite{Tallon-Bosc2008}), \texttt{fitOmatic} (Millour et
al. \cite{Millour2009a}), or \texttt{SIMTOI} (Kloppenborg et
al. \cite{Kloppenborg2012})

\subsection{Quantitative Interpretation: model-fitting}

When one has a model $m_i$ of an object, with parameters $p$, and
wishes to compare it with the data acquired, one needs to quantify how
the model matches the data. This match is perfect when the synthetic
observables computed from the model $m_i(p)$ are equal to the
observations, described by the measurements $x_i$:

\begin{equation}
\forall i, x_i = m_i(p)
\label{eq:datamodel}
\end{equation}

However, this never happens due to the presence of noise. One way to
best match the model to the data is to compute squared differences
between $m_i(p)$ and $x_i$, taking into account the noise $\sigma_i$:

\begin{equation}
\chi^2 = \sum_i \frac{(x_i - m_i(p))^2}{\sigma_i^2}
\label{eq:chi2}
\end{equation}

Minimizing this quantity by changing the parameters values $p$ provide
a plausible solution to the eq.~\ref{eq:datamodel}. This minimization
is made by an \emph{optimization} algorithm, and the whole process is
called ``model-fitting''. 

Specific aspects of long-baseline interferometry, like the use of
wrapped phases (see Fig.~\ref{fig:phasewrapping}), heterogeneous data
noise models (Schutz \etal\ \cite{2014AandA...565A..88S}), or highly
non-convex $\chi^2$, need to be taken into account. This is why
specific software have been developed to cope with it. They are
basically made of an \emph{OIFITS reader} combined with a
simplified \emph{instrument model}, and they make use of
different \emph{optimizers} going from simple descent algorithms up to
simulated annealing, or MCMC methods.

The whole process is described for the specific case
of \texttt{LITpro} in Tallon-Bosc \etal\ (\cite{Tallon-Bosc2008}). You
can also have a look to the practice sessions in this book
using \texttt{LITpro} (Domiciano \cite{2015Domiciano}).

The model itself $m_i(p)$ may be a simple analytic model (``toy''
model, described in sect.~\ref{sec:analytic_model}), very fast to
compute, or an image coming from a more advanced model (described in
Sect.~\ref{sec:generic_model}).

\subsubsection{Analytic models}
\label{sec:analytic_model}

The space distribution of light may be described by simple analytical
functions, as is the associated visibility. Due to the Zernicke \&
van-Cittert theorem, these visibility functions are the Fourier
transforms of the image functions. The most common analytical
functions are given in table~\ref{tab:analytic} and provide already a
quite complete overview of what is used nowadays to interpret
interferometric data.

\begin{table}[htbp]
  \caption{Analytical models of different shapes and their associated
  visibility function. The following parameters are used: $\diameter
  $ represents either the diameter for a ring or uniform disk, or
  FWHM. $r$ or $\vec{x}$ represent the angles in the image
  plane. $\rho = \frac{B}{\lambda}$ or $\vec{\rho}$ are the spatial
  frequencies.}  \label{tab:analytic}
  \begin{tabular}{|m{2.cm}cc|}
    \hline
    Shape & Brightness distribution & Visibility \\
    \hline
    \hline
    \vspace{0.1cm}
    Point source & $\delta(\vec{x})$ & $1$
    \vspace{0.1cm}\\
    \hline
    Background & $I_0$ & 
    $\left\{
    \begin{array}{ll}
      1 & \mbox{if } \rho = 0\\
      0 & \mbox{otherwise} \\
    \end{array}
    \right.$
    \vspace{0.1cm}\\
    \hline
    \vspace{0.1cm}
    Binary star & $I_0 \left[ \delta(\vec{x}) + R
      \delta(\vec{x} - \vec{x_0}) \right]$ &
    $\sqrt{\frac{1+R^2+2 R \cos\left(\frac{\vec{\rho} \cdot
          \vec{x_0} }{ \lambda}\right)}{1 + R^2}}$\\
    \hline
    \vspace{0.1cm}
    Gaussian &
    $I_0 \sqrt{\frac{4 \ln \left( 2\diameter \right)}{\pi}} \times \me^{- 4 \ln 2
      \frac{r^2}{\diameter^2}}$ & $\me^{- \frac{\left( \pi \diameter
        \rho \right)^2}{4 \ln 2}}$ \\
    \hline
    \vspace{0.1cm}
    Uniform disk & 
    $
    \left\{
    \begin{array}{ll}
      \frac{4}{\pi \diameter^2} & \mbox{if } r < \frac{\diameter}{2}\\
      0 & \mbox{otherwise} \\
    \end{array}
    \right.
    $
    & $  \frac{2 \mbox{J}_1\left(\pi \diameter \rho \right)}{\pi \diameter
      \rho}$\\
    \hline
    \vspace{0.1cm}
    Ring & $\frac{1}{\pi \diameter} \delta\left(r -
    \frac{\diameter}{2}\right)$ & $\mbox{J}_0(\pi \diameter \rho)$
    \\
    \hline
    \vspace{0.1cm}
    Exponential & $\me^{-k_0 r}$, $k_0 \ge 0$ & $\frac{k_0^2 }{1+k_0^2 \rho^2}$ \\
    \hline
    \vspace{0.1cm}
    Any circular object & $I(r)$ & $2 \pi
    \int_0^\infty I(r) \mbox{J}_0(2 \pi r \rho) r
    dr$ \\
    \hline
    \vspace{0.1cm}
    Pixel (image brick)&
    $
    \left\{
    \begin{array}{ll}
      \frac{1}{l L} & \mbox{if } x < l \mbox{ and }  y < L\\
      0 & \mbox{otherwise} \\
    \end{array}
    \right.
    $
    & $\frac{\sin(\pi x l) \sin(\pi y L)}{\pi^2 xy lL}$
    \\
    \hline
    \vspace{0.1cm}
    Limb-darkened disk & 
    $
    \left\{
    \begin{array}{ll}
      I_0 [1 - u_\lambda(1 - \mu)] & \mbox{if } r < \frac{\diameter}{2}\\
      \mu = cos(2r / \diameter) & \\
    \end{array}
    \right.
    $
    &  
    $
    \left\{
    \begin{array}{ll}
      \frac{
        \left[ 
          \alpha \frac{\mbox{J}_1(x)}{x} +
          \beta \sqrt{\pi/2} \frac{\mbox{J}_{3/2}(x)}{x^{3/2}} 
          \right]^2
      }{
        \left( \frac{\alpha}{2}+\frac{\beta}{3} \right)^2} 
      \\
      \alpha = 1 - u_\lambda \\
      \beta = u_l\lambda \\
      x = \pi \theta_{\rm LD} \frac{B}{\lambda}\\
    \end{array}
    \right.
    $
    \\
    \hline
  \end{tabular}
    \vspace{0.1cm}
with $f$, $g$ analytical functions like the ones described in
table~\ref{tab:analytic}, $x$, $y$ are coordinates in the image plane,
$u$ and $v$ are spatial frequencies, and $\alpha$ and $\beta$ are
zoom \& shrink factors.
\end{table}

All these analytical models can be added together to produce combined
models. This is feasible thanks to the linearity of the Fourier
Transform, and other properties described below:

\subsubsection{Generic properties of Fourier Transform}
\label{sec:generic_model}

Here I describe the generic properties of the Fourier transform, which
can be found in any book treating FT. 
These properties are widely used to combine or to modify, stretch,
distort, the simple analytical models provided above:

\begin{description}
\item[\textbullet\  linearity (addition):] $FT[f + g] = FT[f] + FT[g]$,
\item[\textbullet\ translation (shift):] $FT[f(x-x_0,y-y_0)] =
  FT[f](u,v) \times \me^{2 i \pi, (u x_0 + v y_0)}$,
\item[\textbullet\ similarity (zoom and shrink):] $FT[f(\alpha x, \beta y)] = \frac{1}{\alpha\beta}FT[f](\frac{u}{\alpha},\frac{v}{\beta})$,
\item[\textbullet\ convolution (``blurring''):] $FT[f \otimes g] =
  FT[f] \times FT[g]$,
\item[\textbullet\ $\infty$ limit (``small'' details):] $FT[f] \stackrel{\infty}{\longmapsto} 0$,
\item[\textbullet\ $0$ limit (``large'' details):] $FT[f] \stackrel{0}{\longmapsto} 1$.
\end{description}

These Fourier transform properties can also be used to combine more
advanced models. This was for example the case of Millour \etal\
(\cite{Millour2009c}), where the authors combined a series of ring
models to build the pseudo-3D toy-model of a spiral nebula.

\subsubsection{Advanced models}
\label{sec:advanced_model}

For more advanced models, which do not have a simple analytical
expression, one can produce a pixellized map of the model and
Fourier-transform it. Most model-fitting software allow, or are
planned to allow to Fourier-transform maps of otherwise computed
models. This is for example the case
of \texttt{ASPRO}, \texttt{fitOmatic}, and will be in a near-future in
the distributed version of \texttt{LITpro}.

\subsection{Image reconstruction}

Image reconstruction is treated in details in Young \&
Thiebaut (\cite{Young2015}) later in this book. I invite the
reader to take a look to that article.

\subsection{The advent of differential phase}

Differential phase has changed the panorama of possibles in
long-baseline interferometry. Millour (\cite{2006PhDT........46M})
presented a theoretical approach to explain the potential of
differential phase to bring new information, independent of a model,
to model-fitting and image reconstruction. As this work in in French,
I translate most of the related content here to the English reader:

\subsubsection{potential in model-fitting}

I was interested here to quantify the information brought by
differential phases in addition to the one brought by closure
phases. I took inspiration from the demonstration of Lachaume
(\cite{2003AandA...400..795L}) and considered first a N-point-sources
model, but resolved by the interferometer. These sources are described
by $2N_{\rm s}-2$ parameters for position (global centroid is unknown
and all sources coordinates are described relative to the first one),
and $N_\lambda N_{\rm s}$ fluxes, i.e.

\begin{equation}
N_{\rm param} = 2N_{\rm s}-2+N_\lambda N_{\rm s}
\label{eq:nparams}
\end{equation}

No other hypothesis is done otherwise than observing several sources
at many wavelengths simultaneously. Great.

Accounting for the number of observables will help us quantify the
maximum number of sources that can be modeled $N_{\rm s\,max}$. This
maximum is given by zeroing the degrees of freedom of the
model-fitting problem (i.e. modelling $N_{\rm s}$ chromatic
point-sources and comparing them to the interferometer data). These
degrees of freedom $\mathds{D}$ are simply the difference between the
number of independent observations $N_{\rm obs}$ and the number of
parameters of the model $N_{\rm param}$, i.e. $\mathds{D} = N_{\rm
obs} - N_{\rm param}$. Zeroing it is simply writing the equation:

\begin{equation}
N_{\rm obs} = N_{\rm param}
\label{eq:degreefreedom}
\end{equation}

The users of interferometers can face four specific cases:

\paragraph{Full access to the complex visibility:}

This is for example the case in radio-astronomy, or the dream of every
single optical interferometrist. In this case, one has access to
$ \frac{N_{\rm tel} (N_{\rm tel}-1)}{2} N_{\lambda} $ visibilities,
the same number of phases, and $N_\lambda$ measured fluxes (the
spectrum). We therefore have:

\begin{equation}
N_{\rm obs} = N_{\rm tel} (N_{\rm
tel}-1) N_{\lambda} + N_\lambda
\label{eq:obsvisibilityphase}
\end{equation}

Putting equations \ref{eq:nparams} and \ref{eq:obsvisibilityphase} into
equation \ref{eq:degreefreedom}, we get:

\begin{equation}
  N_{\rm s\,max} = N_{\rm tel}(N_{\rm tel}-1)\frac{
    N_{\lambda}}{(N_\lambda + 2)} + 1
\label{eq:smaxobsvisibilityphase}
\end{equation}

We see here that the maximum number of sources is roughly proportional
to the number of baselines (i.e. square the number of telescopes), but
not to the number of wavelengths. On the wavelength side, the increase
of the number of sources has an asymptotic behavior.

\paragraph{Visibility only:}

This is the case with 2-telescopes instruments like {\sc vinci}, {\sc classic},
{\sc fluor} or {\sc midi}. In this case, one has access to $ \frac{N_{\rm tel}
(N_{\rm tel}-1)}{2} N_{\lambda} $ visibilities and $N_\lambda$
measured fluxes (the spectrum). We therefore have

\begin{equation}
N_{\rm obs} = \frac{N_{\rm tel} (N_{\rm
tel}-1)}{2} N_{\lambda} + N_\lambda
\label{eq:obsvisibilityonly}
\end{equation}

That number is roughly half the number of
eq.~\ref{eq:obsvisibilityphase}. This is expected as we measure only
half the information on the object (no phases).
Putting equations \ref{eq:nparams} and \ref{eq:obsvisibilityonly} into
equation \ref{eq:degreefreedom}, we get:

\begin{equation}
  N_{\rm s\,max} = \frac{ N_{\rm tel}(N_{\rm tel}-1)}{2}\frac{
    N_{\lambda}}{(N_\lambda + 2)} + 1
\end{equation}

\paragraph{Visibility and closure phase:}

This is still today the most common case. In such a case, one has
access to $ \frac{N_{\rm tel} (N_{\rm tel}-1)}{2} N_{\lambda} $
visibilities, $ \frac{(N_{\rm tel} -1) (N_{\rm tel}-2)}{2} N_{\lambda}
$ closure phases and still $N_\lambda$ measured fluxes. Therefore,

\begin{equation}
N_{\rm obs} = \frac{N_{\rm tel} (N_{\rm
tel}-1)}{2}  N_{\lambda} + \frac{(N_{\rm tel}-1) (N_{\rm
tel}-2)}{2}  N_{\lambda} + N_\lambda
\label{eq:obsvisibilityclosure}
\end{equation}

We therefore have the following:

\begin{equation}
  N_{\rm s\,max} = \frac{ (N_{\rm tel}-1)^2}{2}\frac{
    N_{\lambda}}{(N_\lambda + 2)} + 1
\end{equation}

The same comments as before applies here, except that the term
$(N_{\rm tel}-1)^2$ grows faster than the term $N_{\rm tel}(N_{\rm
tel}-1)$ of the previous paragraph.

\paragraph{Visibility, closure phase and differential phase:}

This is the case of {\sc amber}, {\sc matisse} and {\sc gravity}. The
observables are now $ \frac{N_{\rm tel} (N_{\rm tel}-1)}{2}
N_{\lambda} $ visibilities, $ \frac{N_{\rm tel} (N_{\rm tel}-1)}{2}
N_{\lambda} $ differential phases, $ \frac{(N_{\rm tel} -1) (N_{\rm
tel}-2)}{2} N_{\lambda} $ closure phases and still $N_\lambda$
measured fluxes. However, one has to note that closure phase and
differential phase are not independent measurements. Indeed, they are
both related to the object phase (see eq.~\ref{eq:phiobjphidiff} and
eq.~\ref{eq:computeclosure}). Therefore, one can write the relation
between the differential phase and the closure phase using both
equations:

\begin{equation}
   \Psi_{i,j,k} =
 \phi^{\rm diff}_{i,j} +
 \phi^{\rm diff}_{j,k} +
  \phi^{\rm diff}_{k,i} +
\beta_{i,j} +
 \beta_{j,k} + 
  \beta_{k,i} +
\frac{\alpha_{i,j}}{\lambda} +
\frac{\alpha_{j,k}}{\lambda} +
\frac{\alpha_{k,i}}{\lambda} 
\label{eq:phidiffclosure}
\end{equation}

The careful reader should have seen here that I discarded the small
term $\frac{N_\lambda-1}{N_\lambda}$, which can be neglected for a
large number of spectral channels. One can also note that one of the
terms $\beta_{i,j} + \frac{\alpha_{i,j}}{\lambda}$ can be fixed to
zero in order to set the two other offsets, and therefore fix the
global photocenter of the object (which remains unconstrained by
closure phases and differential phases).

This equation and the above additional constrain provide us with the
relevant information: the closure phase bring only additional data on
two offsets $\beta_{i,j}$ and wavelength-slopes
$\frac{\alpha_{i,j}}{\lambda}$ that are missed by differential
phases. All other information (wavelength variations) are contained
both in closure phase and differential phases. Therefore, the closure
phase provides $(N_{\rm tel}-1) (N_{\rm tel}-2)$ independent
observables instead of the $ \frac{(N_{\rm tel} -1) (N_{\rm
tel}-2)}{2} N_{\lambda} $ accounted just before.  Therefore, we get:

\begin{equation}
N_{\rm obs} = \frac{N_{\rm tel} (N_{\rm
tel}-1)}{2}  N_{\lambda} + (N_{\rm tel}-1) (N_{\rm
tel}-2)  + \frac{N_{\rm tel} (N_{\rm
tel}-1)}{2}  N_{\lambda} + N_\lambda
\label{eq:obsvisibilityclosurediffphase}
\end{equation}

and the maximum number of sources that can be modeled is:

\begin{equation}
  N_{\rm s\,max} = \frac{ (N_{\rm tel}-1)(N_{\rm tel}N_{\lambda}-2)}{(N_\lambda + 2)} + 1
\end{equation}

\begin{figure}[htbp]
  \centering 
\vspace{-1cm}
\includegraphics[width=0.48\textwidth]{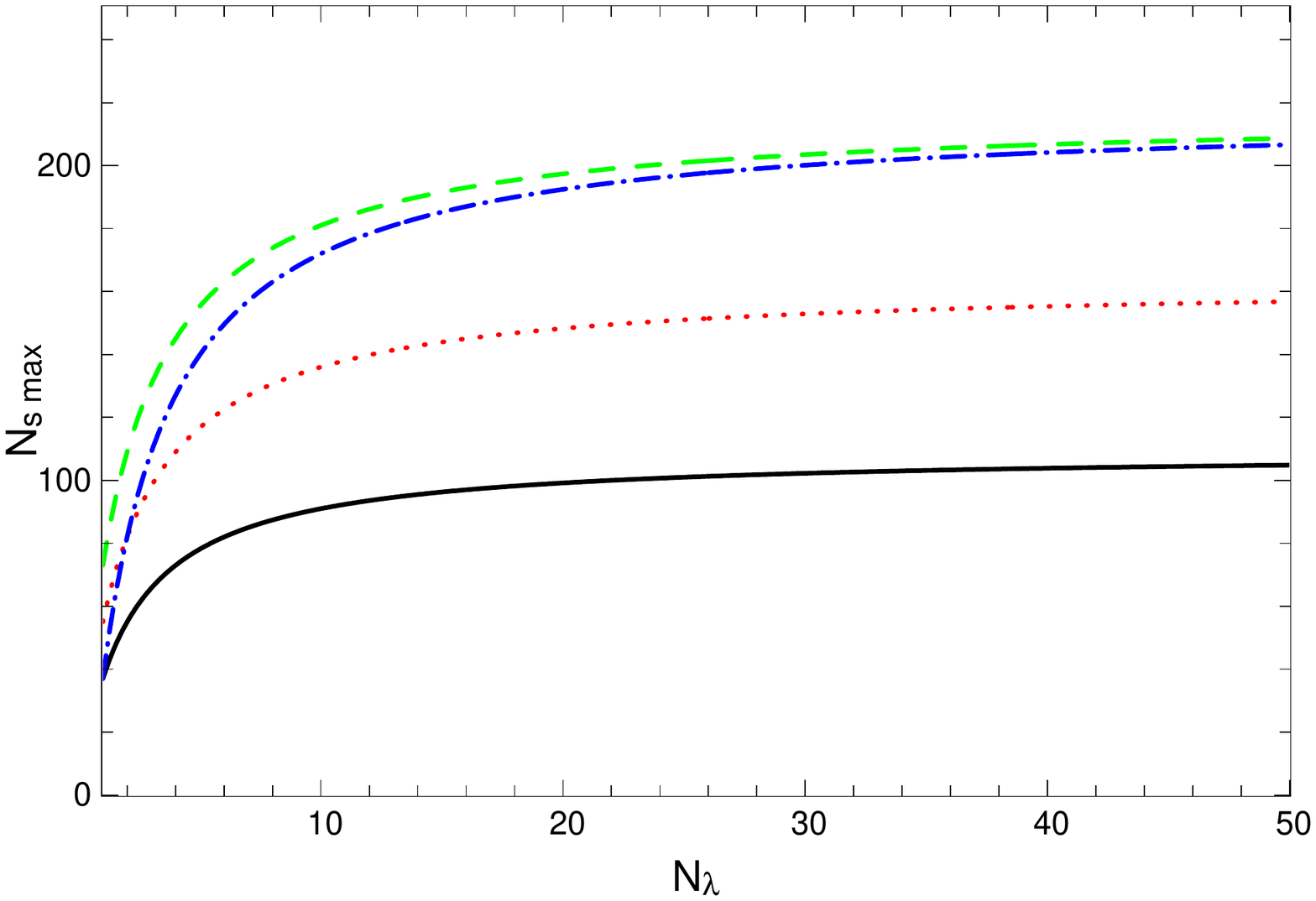} 
\includegraphics[width=0.48\textwidth]{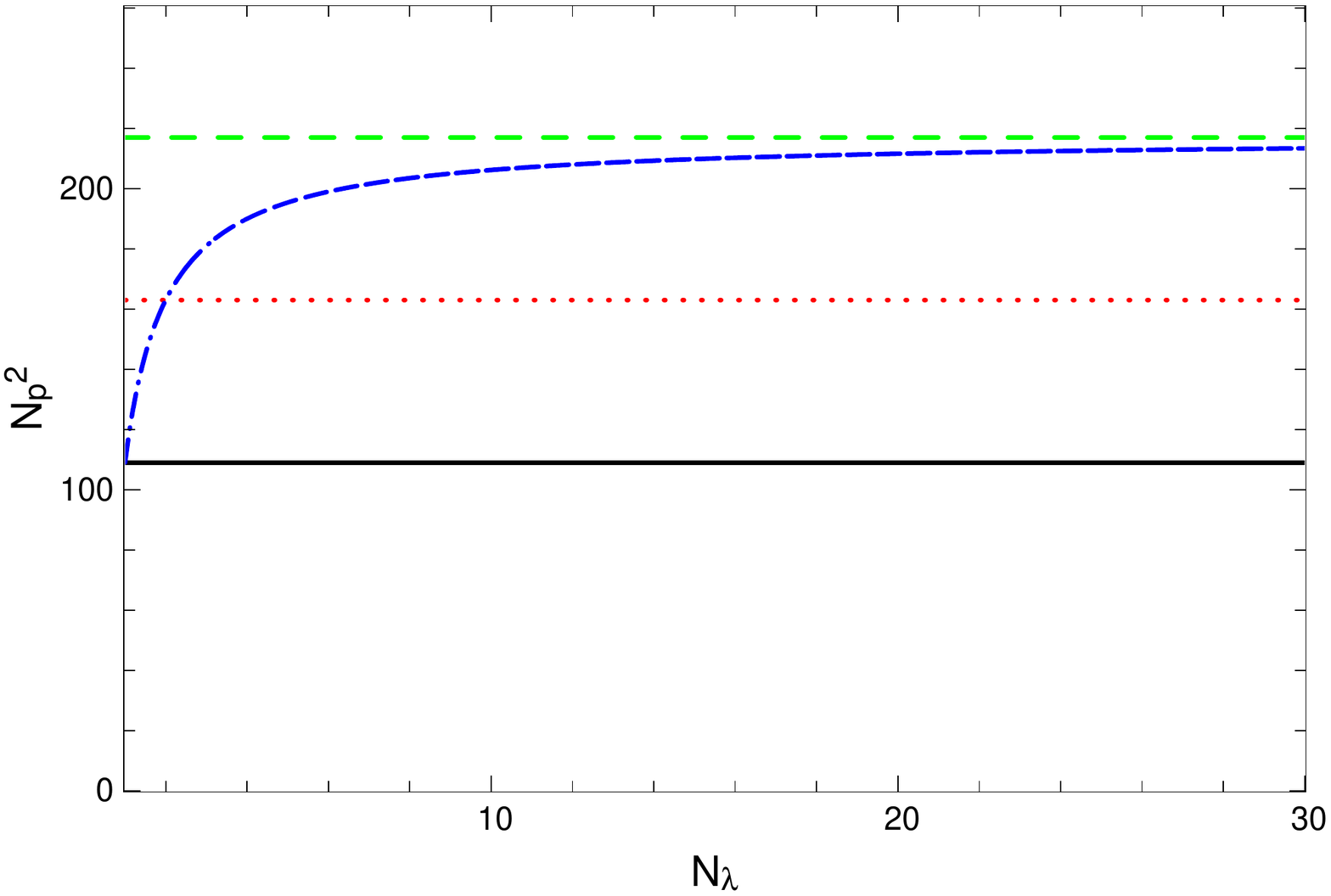} 
\vspace{-3cm}
\caption{Illustration
    of the potential of using the differential phase in model-fitting
    or image reconstruction for 3 nights of 4 telescopes observations
    (like in the case of {\sc matisse}) compared to other cases. {\bf
    Left:} modelling $N_{\rm s\,max}$ point-sources, the green dashed
    line using visibilities and phases, the black solid line is using
    visibilities alone, the red dotted line using visibilities and
    closure phases, and finally the blue dash-dotted line using
    visibilities, closure phases and differential phases. {\bf Right:}
    reconstructing a $N_{\rm p}\times N_{\rm p}$ pixels
    chromatic image. The colours are the same. }
\label{fig:phasediffmodelfittingimagerie}
\end{figure}

These different cases are illustrated in
Fig.~\ref{fig:phasediffmodelfittingimagerie}. We see that, typically
above 20 spectral channels, the use of differential phases put us in a
case almost similar as if there was a true phase measurement for
model-fitting. This was the main motivation to develop the
model-fitting tool \texttt{fitOmatic}, which can make use of chromatic
parameters.

\subsubsection{in image reconstruction}
\label{sect:selfcal}

The wavelength-differential phase was not considered in imaging until
Millour (\cite{2006PhDT........46M}), Schmitt \etal\ (\cite{2009ApJ...691..984S}) and Millour \etal\ (\cite{Millour2011}). Indeed, differential phase provide a corrugated
phase measurement (as described in eq.~\ref{eq:phiobjphidiff}), which,
in theory, can be incorporated into a self-calibration algorithm, in a
very similar way as what is done in radio-interferometry (Pearson \&
Readhead \cite{1984ARAandA..22...97P}).

As early as \cite{2003RPPh...66..789M}, J. Monnier anticipated
\emph{``revived activity }[on self-calibration]\emph{ as more interferometers with
‘imaging’ capability begin to produce data.''}
And indeed, the conceptual bases for using differential phases in
image reconstruction were laid in Millour
(\cite{2006PhDT........46M}): the same reasoning as in the previous
section can be applied, except that an image is made of pixels whose
positions are pre-defined. The only unknown information is therefore
the spectrum of each pixel. If $N_{\rm p}$ is the image size ($N_{\rm
p}=128$ for a 128 x 128 image), the number of unknown $N_{\rm param}$ is
equal to:

\begin{equation}
N_{\rm param} = N_{\rm p}^2 N_\lambda
\label{eq:nparamsimage}
\end{equation}

and the maximum number of pixels that can be reconstructed is:

\paragraph{Full access to the complex visibility:}
  \begin{equation}
    N_{\rm p} = \sqrt{N_{\rm tel} (N_{\rm tel}-1) + 1} 
    \label{eq:image_visi_phi}
  \end{equation}
\paragraph{Visibility only:}
  \begin{equation}
    N_{\rm p} = \sqrt{\frac{N_{\rm tel} (N_{\rm tel}-1)}{2} + 1}
    \label{eq:image_visi_only}
  \end{equation}
\paragraph{Visibility and closure phase:}
  \begin{equation}
    N_{\rm p} = \sqrt{(N_{\rm tel}-1)^2 + 1}
    \label{eq:image_visi_clos}
  \end{equation}
\paragraph{Visibility, closure phase and differential phase:}
  \begin{equation}
    N_{\rm p} = \sqrt{ \frac{(N_{\rm tel}-1) (N_{\rm tel} N_\lambda - 2)}{N_{\lambda}} + 1}
    \label{eq:image_visi_clos_phidiff}
  \end{equation}

Fig.~\ref{fig:phasediffmodelfittingimagerie} shows the same behavior as
for model-fitting: typically above 20 spectral channels, the use of
differential phases put us in a case almost similar as if there was a
true phase measurement, i.e. as if there was no atmosphere in front of
the interferometer, given that one is able to take profit of the
information contained in the differential phase.

The Schmitt \etal\ (\cite{2009ApJ...691..984S}) paper was a first
attempt to use differential phases in image reconstruction. They
considered that the phase in the continuum was equal to zero, making
it possible to use the differential phase (then equal to the phase) in
the H$\alpha$ emission line of the $\beta$ Lyr system. They were able
this way to image the shock region between the two stars at different
orbital phases.

\begin{figure}[htbp]
\vspace{-2cm}
  \centering \includegraphics[width=1. \textwidth,angle=-0]{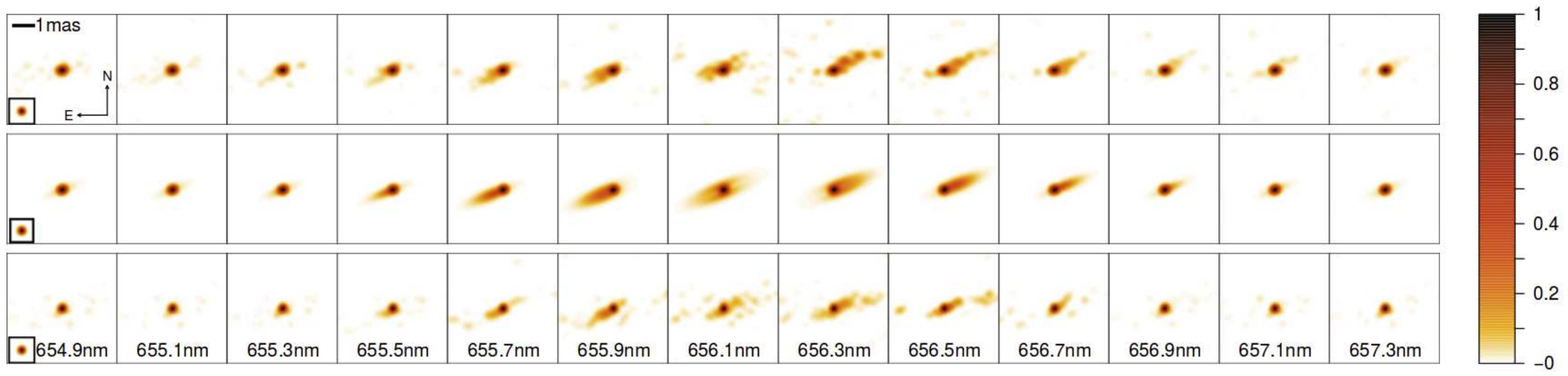}
\vspace{-3cm}
 \caption{Image
    from Mourard \etal\ (\cite{Mourard2014}), showing the kinematics
    of the $\phi$ Per Be star disk through the H$\alpha$ emission
    line. The top row shows the reconstructed images, the middle row a
    best-fit kinematics model and the lower row a reconstruction based
    on the model, for comparison. Reproduced with permission.}  \label{fig:mourardpaper}
\end{figure}

The paper Millour \etal\ \cite{Millour2011} went one step further, by
using an iterative process similar to radio-interferometry
self-calibration (Pearson \& Readhead \cite{1984ARAandA..22...97P}) in
order to reconstruct the phase of the object from the closure phases
and differential phases. This way, they could reconstruct the image of
a rotating gas+dust disk around a supergiant star, whose image is
asymmetric even in the continuum (non-zero phase). This method was
subsequently used in a few papers to reconstruct images of supergiant
stars (Ohnaka \etal\ \cite{Ohnaka2011, 2013AandA...555A..24O}). A more
recent work (Mourard \etal\ \cite{Mourard2014}) extended the method to
the visibilities, in order to tackle the image reconstruction
challenges posed by visible interferometry, lacking the closure phases
and a proper calibration of spectrally-dispersed visibilities. The
image-cube reconstructed with this technique in Mourard \etal\
(\cite{Mourard2014}) is shown in Fig.~\ref{fig:mourardpaper}.

Of interest are also new developments made on the core image
reconstruction algorithms to include the differential phases into the
process (Schutz \etal\ \cite{2014arXiv1407.1885S}, or
Soulez \etal\ \cite{2014ipco.conf..255S}). These new algorithms are
very promising and they must be confronted sooner or later to real
datasets.

\section{Interferometry hardware}

Combining telescopes which are hundreds of meters apart is a difficult
matter, which needs a set of functions described below and shown in
Fig.~\ref{fig:schemaInterf}:

\begin{enumerate}[(a)]
\item Telescopes, to collect the light, mostly defined by their
      diameter $D$,
\item Set of periscopes (sometimes grouped in ``switchyards'') to shape,
      collimate, and feed the beam through light tunnels, defined by a
      fixed length $\Delta_{\rm fixed}$,
\item Delay lines to compensate for the variable delay due to pointing
      the telescope and the atmosphere's effects, defined by a
      time-variable optical path length $\Delta_i(t)$,
\item Combiner instrument to effectively produce the interference
      pattern.
\end{enumerate}

We will try not to repeat the numerous descriptions of how to build an
interferometer. 
We just describe here what makes an interferometer working:

\subsection{Telescopes}

Interferometry telescopes are, in principle, no different from
``regular'' astronomical telescopes. However they differ on three
aspects we will detail in the following subsections: they must be
{\it smart, tough and large}.

\subsubsection{``smart''}

A ``smart'' telescope is a reliable one. Indeed, a single telescope
needs to be operational (i.e. not undergoing technical failures or
maintenance) most of its time.

\begin{figure}[htbp]
\vspace{-2cm}
  \centering
  \begin{tabular}{lr}
    \includegraphics[width=0.48\textwidth]{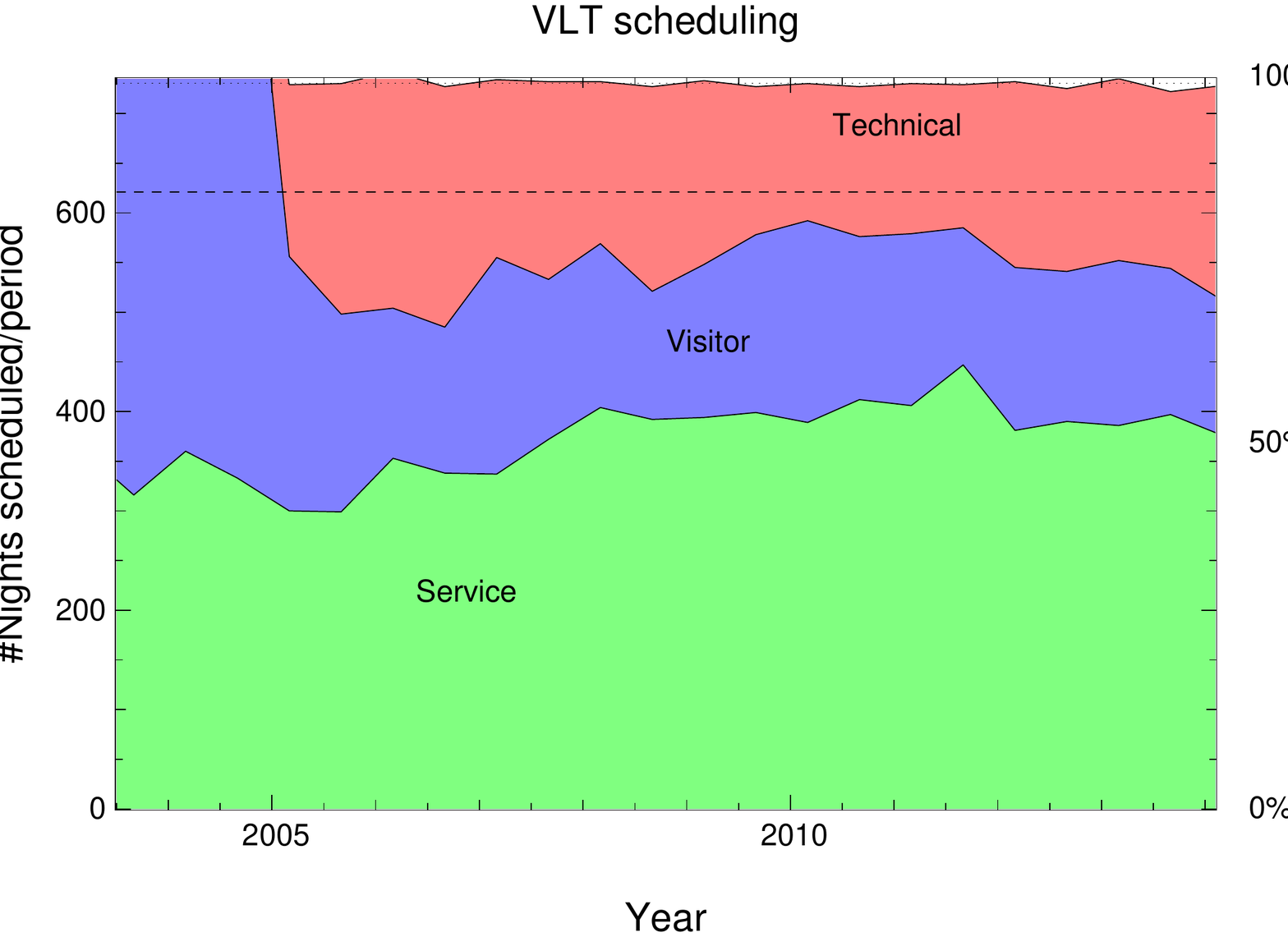} &
    \includegraphics[width=0.48\textwidth]{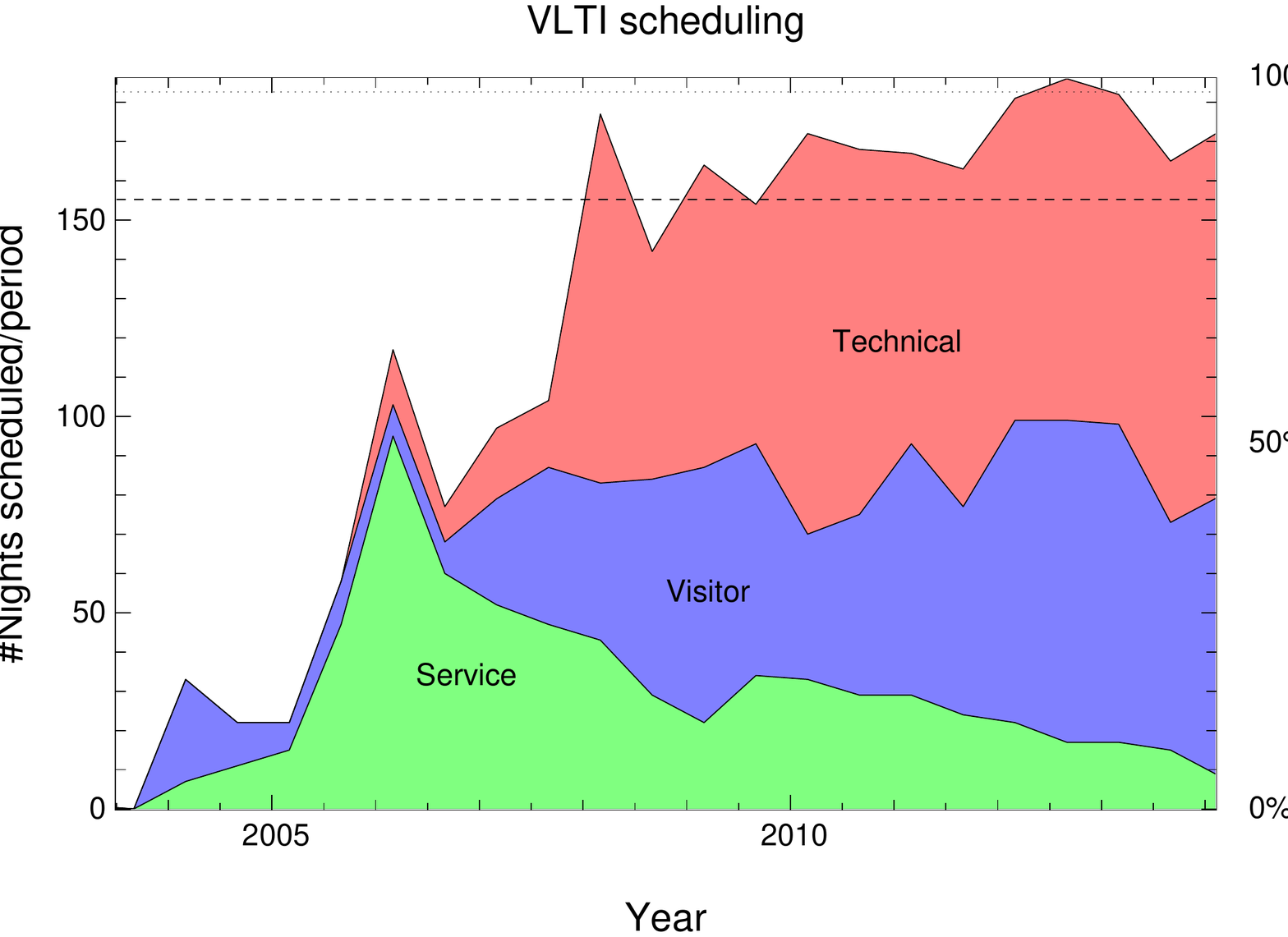} \\
  \end{tabular}
\vspace{-3cm}
  \caption{{\bf Left:} VLT scheduling taken from the ESO telescope
    schedule$^1$, including all 4 Unit Telescopes (hence \# of nights
    multiplied by 4). Different colours represent different times on
    the telescopes. The dashed line represent the clear skies night,
    while the dotted line represent the total number of nights over
    one observing period (6 months). {\bf Right:} The same plot for
    the {\sc vlti}.}
  \label{fig:schedule}
\end{figure}

For example, if we take the ESO telescope
schedule\footnote{\url{http://archive.eso.org/wdb/wdb/eso/sched_rep_arc/form}}
for the Unit Telescopes on the {\sc vlti}, the Observatory confidence in their
telescopes is given by the ratio of observing nights scheduled by the
available clear skies observing nights (Lombardi \etal\ \cite{2009MNRAS.399..783L}: 310
  nights/years at Paranal). This is illustrated by
Figure~\ref{fig:schedule} where we provide the number of scheduled
nights (visitor or service mode) versus the average number of clear
nights at Paranal. ESO usually accounts for reliable telescopes 89\%
of the clear sky time.

On the other hand, the {\sc vlti} scheduling, illustrated in
Figure~\ref{fig:schedule} tells us that the ESO observatory uses 50\%
of the available clear sky nights, after a learning curve on the
interferometer between 2003 and 2006. This is well explained if we
consider all 4 telescopes are used for interferometry. The probability
of having all 4 telescopes online in a given night is just
$0.89^4\equiv63\%$ of the available time. The difference between the two
numbers (50\% and 63\%) comes from the numerous additional sub-systems
needed by the {\sc vlti} to operate (delay lines, fringe tracker,
instrument, etc.).

A word on the ``technical'' time plotted here: the cumulative time,
including technical one, exceeds the number of clear sky nights, since
some technical activities do not need the telescope open. We also note
here that the {\sc vlti} technical time was not fully taken into account in
the scheduling until 2008, leaving the impression that the {\sc vlti} was
``idle'' most of the time.

\subsubsection{``tough''}

A ``tough'' telescope is a stable one. ``stable'' means the telescope
do not transmit vibrations to the instrument. Usual instruments at the
focus of telescopes are sensitive (at first order...) to transverse
vibrations (i.e. ``tip/tilt'' vibrations). This puts some requirements
on the tip/tilt pointing and stability accuracy (See for example a
study in Altarac \etal\ \cite{Altarac2001}).

Unfortunately, an interferometer is sensitive {\it both} to transverse
and longitudinal vibrations (a.k.a. ``OPD'' or ``piston''
vibrations). Both of the large telescopes interferometers are subject
to such vibrations as they were not designed in the first time to be
used in an interferometer (Hess \etal\ \cite{2003SPIE.4837..342H}, Millour \etal\ \cite{Millour2008}). To
overcome these effects, an active dampening system had to be
integrated into both facilities (Hess \etal\  \cite{2003SPIE.4837..342H}, Lizon \etal\ \cite{2010SPIE.7739E.138L}, Poupar \etal\ \cite{2010SPIE.7734E.101P}, Spaleniak \etal\ \cite{2010SPIE.7734E.126S}).

On smaller telescopes facilities, vibrations have also been
investigated but this effect has a much smaller amplitude
(Merand \etal\ \cite{2001AAS...198.6104M}).

It is worth to note that the VLT instruments are themselves affected
by vibrations (Sauvage, private communication), and vibrations
assessment are a part of the ELTs design.

\subsubsection{``large''}

``Large'' telescopes means large collecting area, means more sensitive
interferometer. However, one needs to bear in mind that the gain in
sensitivity is true {\it only} for a constant strehl ratio of the
telescope PSF, simply because the overall effective transmission of
the system, when using optical fibres, is multiplied by the strehl
ratio. This is why very large telescopes interferometers ({\sc vlti}
\& Keck) have been equipped with adaptive optics (Arsenault \etal\ \cite{2003Msngr.112....7A}).

\subsection{Feed through}

The light is fed by a series of mirrors from the telescope to the
delay lines building. This is where a large part of the light
propagation occurs in the interferometer and where potentially several
issues can happen to the beam. Three possibilities exist today to
transport the beam:
\begin{itemize}
\item through air (e.g. in {\sc vlti} and Keck-I),
\item through vacuum (e.g. in {\sc iota}, {\sc chara}, {\sc npoi}),
\item through fibers (developed for the OHANA
  project, Woillez \etal\ \cite{2014ipco.conf..175W}).
\end{itemize}

The air transportation is the simplest to setup with just tunnels and
relay optics to be installed (no bulky vacuum tubes and
pumps). However, the air introduces chromatic longitudinal dispersion
when large delays are compensated, which affects the fringe signal and
is not easy to overcome for high-precision measurements
(Tubbs \etal\ \cite{2004SPIE.5491..588T},
Vannier \etal\ \cite{Vannier2006a}).

\begin{figure}[htbp]
  \centering
  \includegraphics[width=0.9\textwidth]{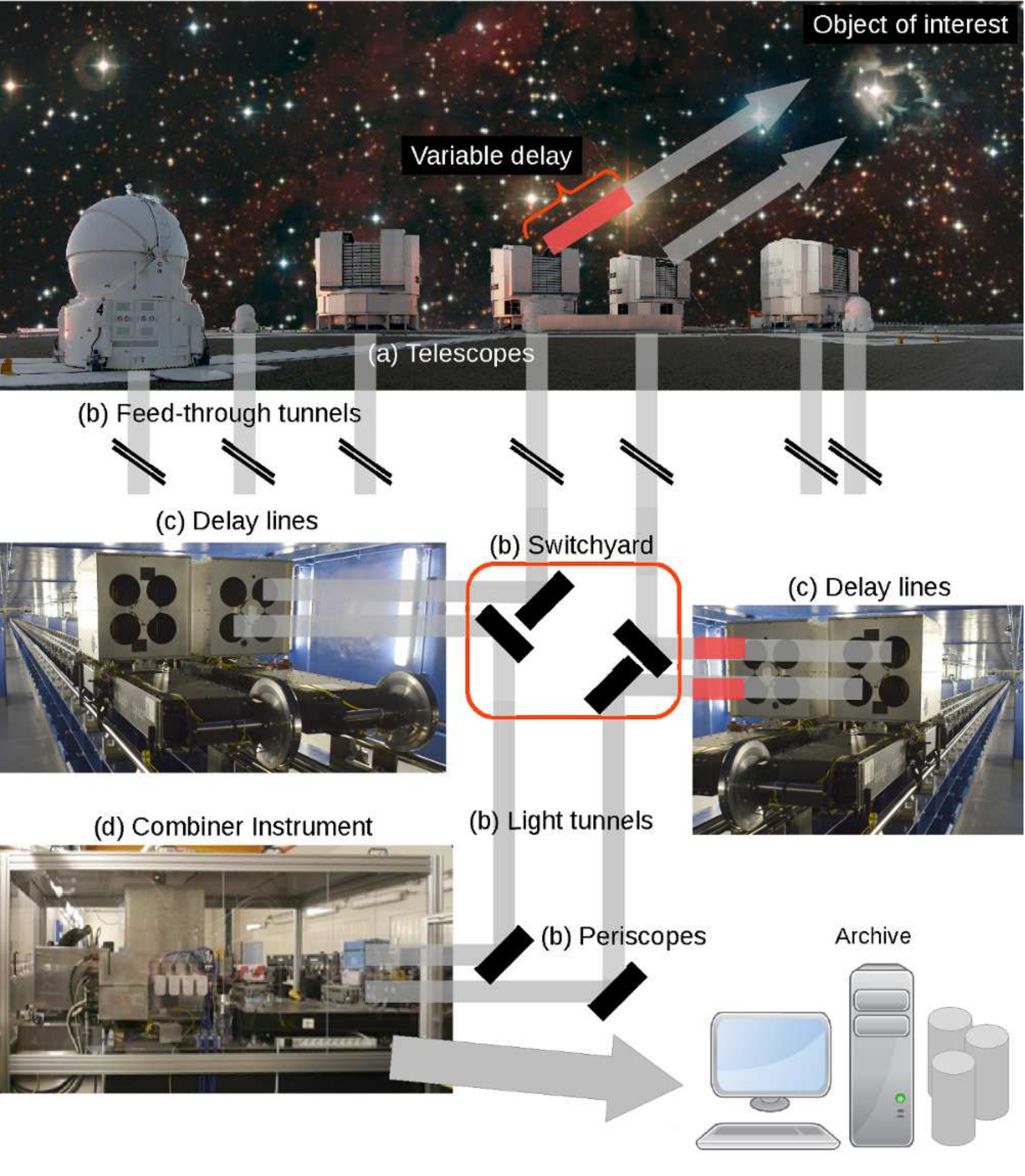}
  \caption{How to combine beams from separated telescopes illustrated
    with the {\sc vlti}. The collected beams are first collimated and fed
    into tunnels by a set of mirrors in the telescopes, put into delay
    lines to compensate for the delay introduced by pointing and other
    effects (in red), then fed into the interferometric instruments
    that records the data.}
  \label{fig:schemaInterf}
\end{figure}

\subsection{Delay lines}

Delay lines are a set of movable mirrors which compensate for the
optical delay induced by the pointing of the telescope. These are
supposedly simple at first glance but one needs to consider the
required precision (less than 1\micron), and range of motion (half the
largest telescope baseline, i.e. it can be hundreds of meters). These
optical systems are in no way simple to build and operate, as the
mirrors-bearing carriage has to provide sub-micron position accuracy
on hundreds of meters with a continuous motion of a few centimetres
per second...

Several technical solutions have been implemented, which all have
advantages and drawbacks: the delay lines can be in the air (like on
{\sc vlti} or Keck-I) or in vacuum (like on {\sc iota}), or partially
in vacuum and partially in air (like in {\sc chara}, {\sc npoi}). They
can be one stage ({\sc vlti}) or two stages ({\sc iota}, Keck-I, {\sc
  chara}, {\sc npoi}) with a long-stroke fixed delay (easier to
manufacture) and short-stroke moving delay.

\subsection{Combiner}

The last element of the interferometer is the Combiner. It is
basically a Michelson or a Fizeau interferometer plugged-in to a very
sophisticated video camera with some degree of spectral dispersion and
a feedback loop to stabilise the fringes.
The fringe combination can be done in different ways and we refer the
reader to Berger (\cite{Berger2015}, this book) for further details.

All these sub-systems are shown in the illustration
Fig.~\ref{fig:schemaInterf}.

\section{Optical interferometry in 2015}

\subsection{{\sc vlti}}

\begin{figure}[htbp]
\vspace{-2cm}
  \centering
  \includegraphics[width=0.7\textwidth]{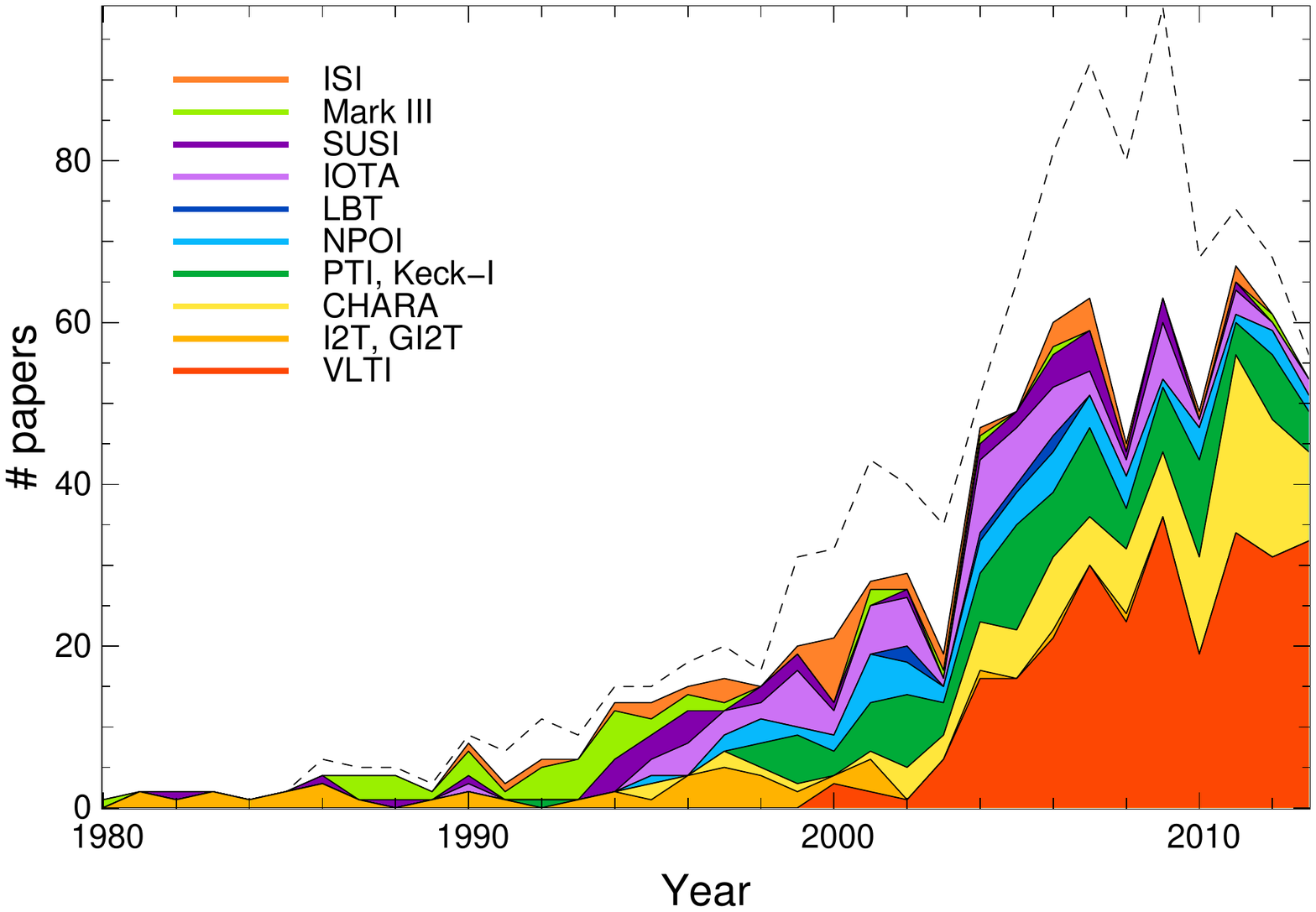}
\vspace{-4cm}
  \caption{Publications related to long-baseline interferometry per
    year (Data from the JMMC \url{http://apps.jmmc.fr/bibdb/}). The
    dash line represent the total number of publication whereas the
    colors are per facility (i.e. observations papers only). We can
    see a steep increase after 2002 with the advent of {\sc vlti}, and a
    decrease after 2010 due to the closure of several facilities
    (Keck-I, IOTA, ...).}
  \label{fig:pubInterf}
\end{figure}

The {\sc vlti} is the only large-aperture interferometer in operation
today. With its four 8-meter class telescopes, supplemented by four
movable 2-meter class telescopes (see Fig~\ref{fig:schemaInterf}), it
offers versatility and sensitivity at the same time. It saw its first
light in 2001 (Glindemann \etal\ \cite{2001Msngr.104....2G}) with the
{\sc vinci} instrument, and has since seen its capabilities
increasing: 2 recombined telescopes in 2001
(Kervella \etal\ \cite{2003SPIE.4838..858K}), mid-infrared with {\sc
midi} (Leinert \etal\ \cite{2003Ap&SS.286...73L}), 3 telescopes and a
high spectral resolution in 2004 with {\sc amber}
(Petrov \etal\ \cite{2007AandA...464....1P}) and 4 telescopes in 2010
with {\sc pionier} (Le Bouquin \etal\ \cite{2011AandA...535A..67L},
but lacking a high or medium spectral resolution, and just open to the
general community since 2015). The {\sc vlti} is noticeably the most
productive interferometric facility in the world (see
Fig.~\ref{fig:pubInterf}). Applying for observing time is open to any
professional astronomer, and its archive is public after typically one
year of ownership by the PI. The next generation instruments {\sc
matisse} and {\sc gravity} will offer to the wide community four
telescopes and, for the first time, real imaging capabilities.

\subsection{{\sc chara}}

{\sc chara} is a six-1-meter-class telescopes interferometer, which is
funded and operated by the Georgia State University. It has developed
in the 2000s a collaborative framework allowing teams from all over
the world to install and operate instruments on the facility. As a
result, {\sc chara} is the interferometer with the most number of
currently-operated instruments, with {\sc classic}, {\sc climb}, {\sc
mirc}, {\sc pavo}, {\sc vega}. {\sc chara} has the longest operating
baselines (300\,m) and works in the visible range, making it the
sharpest telescope on Earth.

\subsection{{\sc npoi}}

{\sc npoi} is a six-telescope interferometer jointly operated by the
Lowell Observatory and the US Navy. It consists of small apertures
movables siderostats for imaging as well as fixed siderostats
dedicated to astrometry. Its recent developments include a 6-telescope
visible instrument {\sc vision} and the commissioning of new longer
baselines. It is currently limited by the small apertures, but the
developement plan includes the installation of meter-class telescopes
in the future.

\subsection{The legacy}

The long-term durability of the data acquired by the interferometers
make it a goldmine for future astronomers. Therefore some
observatories have made an effort to archive the obtained data and to
make it public for future use. This is e.g. the case for the ESO
science archive facility\footnote{available at
  \url{http://archive.eso.org}} which provides raw datasets from all
the open {\sc vlti} instruments plus, more recently, data from the visitor
instrument {\sc pionier}.

Since ESO provides only raw datasets, a community effort is being made,
led by JMMC\footnote{available at \url{http://www.jmmc.fr/oidb.htm}},
to provide a reduced database called OIDB. It will provide in a
near-future reduced datasets which have been published, in order to
make them accessible for future use.

An effort has also been conducted to make the legacy Keck-I and PTI
instruments data avilable through the PTI \& Keck Public
Database\footnote{available
at \url{http://irsa.ipac.caltech.edu/data/NExScI_PTI_KI/}}. Future
users can access freely these data and use them in a publication
provided they follow the publishing guidelines available at ESO and
NExScI webpages.

\section{The future: new instruments, new possibilities}

The {\sc vlti} has been at the leading edge for optical interferometry
in the last decade. However, the two first-generation instruments,
{\sc amber} and {\sc midi} are now 10 years old, and even though they
have unmatched features (high spectral resolution for {\sc amber} and
N band for {\sc midi}), they start to show their limits. Therefore,
ESO issued a call for proposals in 2005 to build second generation
instruments. Two projects were selected: {\sc
gravity}\footnote{\url{http://www.mpe.mpg.de/ir/gravity}}, aiming at
performing micro-arcseconds astrometry on the Galactic Center in the
near-infrared, and {\sc
matisse}\footnote{\url{https://www.matisse.oca.eu}}, aiming at opening
the L band in addition to bring imaging capabilities to the {\sc vlti}
in the mid-infrared. They both come to the sky in the 5+ years from
now.

The {\sc chara} array is being fitted with adaptive optics to improve
by a large factor its performances, especially at short wavelengths,
offering new possibilities of performant instrumentation in the 5+
years to come.

In the meantime, several projects have emerged to pave the way of
future facilities: a visible interferometry prospective is being
conducted today (Stee et al. in prep.), to make emerge a new
generation instrumentation at {\sc vlti} and {\sc chara} in the 10+
years; a more general prospective is conducted by the Europan
Interferometry Initiative to direct future instruments in the same
timeline (Pott, private communication); the Planet Formation Imager
project (Kraus \etal\ \cite{2014_kraus}) aims at imaging and
characterizing an exoplanet in the 20+ coming years; finally several
bold prototypes of completely new combination schemes (le
Coroller \etal\ \cite{2015AandA...573A.117L},
Labeyrie \etal\ \cite{2001sf2a.conf..505L}, and see also the
conclusion of this book:
Labeyrie \cite{2015Labeyrie}), \emph{hypertelescopes}, are being
imagined, developed and tested to gather the technologies necessary
for the 40+ years to come.

\vspace{1cm}

{\it Acknowledgements: The author would like to thank R. Petrov, A. Meilland and G. Dalla Vedova for reading through this paper and for suggesting improvements. Thanks also to J.-F. Sauvage for interesting discussions about technical aspects on the VLT, and to J.-U. Pott for pushing the prospective on the future of interferometry.}


\end{document}